\documentclass{pinchcr}   % Comment the above and uncomment this
\usepackage{graphicx}
\usepackage{multirow}
\usepackage{amssymb}
\usepackage{amsmath}

\title{ }

\begin{document}

%\maketitle

%\dedication{This is an example of a dedication.}

%\acknowledgements

%This is an example of an Acknowledgements section which has been implemented with
%the \verb+\acknowledgements+ command.

%\tableofcontents

\maintext

%&&&&&&&&&&&&&&&&&&&&&&&&&&&&&&&&&
\chapter*{Dry, aligning, dilute, active matter: A synthetic and self-contained overview}
%&&&&&&&&&&&&&&&&&&&&&&&&&&&&&&&&&

%%%XXX check that acronyms are defined properly

%%%XXX where to mention that v0->0 limit is singular?

%%% Authors
Hugues Chat\'e$^{1,2,3}$ \& Beno\^it Mahault$^{3,4}$\\\\
$^1$Service de Physique de l'Etat Condens\'e, UMR 3680 CEA-CNRS, Universit\'e Paris-Saclay, CEA-Saclay, 91191 Gif-sur-Yvette, France \\
$^2$ Beijing Computational Science Research Center, Beijing 100094, China \\
$^3$ Department of Physics, Universal Biology Institute, Graduate School of Science, The University of Tokyo, 7-3-1 Hongo, Bunkyo-ku, Tokyo, Japan \\
$^4$ Max Planck Institute for Dynamics and Self-Organization (MPIDS), 37077 G\"ottingen, Germany\\\\
%%%

%%% Abstract
Dry, aligning, dilute, active matter (DADAM), with its many adjectives, refers to a corner of the multidimensional, fast-growing field of active matter studies. This corner, however, has both historical and theoretical importance for the entire field. 
These lectures notes first describe this particular status of DADAM. 
We then provide an overview of our current knowledge of DADAM in a synthetic and coherent manner.
This constitutes the bulk of these notes. 
For convenience, we first describe the phenomenology of simple active particle models representing the basic DADAM classes,
limiting ourselves to two spatial dimensions, where most work has been performed. 
Then we discuss the continuous hydrodynamic theories derived from these models using the Boltzmann-Ginzburg-Landau approach. 
We show in particular that they are largely faithful to the microscopic level, albeit qualitatively.
In the conclusion, we come back to the nodal situation of DADAM within active matter studies and 
provide perspectives on how one can expand DADAM knowledge into various directions, 
approaching more realistic and more complex situations in a controlled way.
%%%%

%&&&&&&&&&&&&&&&&&&&&&&&&&&&&&&&&&
\section{Introduction}
\label{intro}
%&&&&&&&&&&&&&&&&&&&&&&&&&&&&&&&&&

\subsection{Singular situation of DADAM}

After more than 20 years have passed, it is fair to say that the field of active matter physics was born in 1995. 
Not that nobody had ever worked before on out-of-equilibrium systems spending energy at the level of some local units to produce motion
or deformation ---a loose definition of active matter. 
But in 1995, in a radical move typical of statistical physics, Tam\'as Vicsek and collaborators, motivated by collective animal and
bacterial behavior, introduced a very simple model of collective motion \shortcite{vicsek1995novel}. 
In the now-famous Vicsek model, pointwise particles move at constant speed, locally aligning their velocities in the presence of noise. 
If the noise is weak enough, velocities can align on large scales. Such collective motion in the Vicsek model no more pretends to 
describe realistic situations than the Ising model does pretend to describe real magnetic materials. 
And indeed, over the years, the role played by the Vicsek model is not without resemblance to that of the Ising model in the wider
setting of statistical physics.

In fact, the Vicsek model in two dimensions can be seen as an XY model where the spins are forced to fly in the 
direction given by their orientation. Collective motion is tantamount to orientational polar order 
resulting from spontaneous breaking of rotational symmetry.
This caught the attention of John Toner and Yuhai Tu. Listening to a talk by Vicsek, they wondered whether 
such a flying XY model could display true long-range orientational order, 
in contrast to the equillibrium XY model, well-known to only display quasi-long-range order. 
In their now-famous 1995 paper, Toner \& Tu introduced a phenomenological hydrodynamic equation for (Vicsek-style) flocking, 
and showed that its ordered phase is indeed more ordered than its equilibrium counterpart, displaying true long-range order even in two space dimensions (see J. Toner's contribution to this volume and \shortcite{toner1995long,toner1998flocks,toner2012reanalysis}).

The Vicsek model and the Toner-Tu theory represent a class of systems within dry, aligning, dilute active matter, DADAM.
``Dry" since the fluid surrounding the moving active particles is absent, a legitimate assumption with objects crawling 
or sliding on a surface. ``Aligning and dilute" since the pointwise particles have no physical size, their sole interaction being alignment.
Since 1995, the field of active matter physics has grown fast and in multiple directions, 
at the interfaces with many other disciplines: animal behavior, micro-, cellular-, and multicellular biology, materials science, 
even robotics and computer science. 
The knowledge accumulated can be organized and visualized along some main axes: 
wet vs dry, dilute vs dense, alignment vs repulsion, etc.
In the cube spanned by these last 3 directions, DADAM occupies a corner, but an important one that has influence
beyond its vicinity, for any situation where elongated objects align, where `active particles' move collectively. 

%%%%%%%%%%%%%%%%%%%%%%%%%%%
\begin{figure}[h!]
	\centering
	\includegraphics[width=0.8\textwidth]{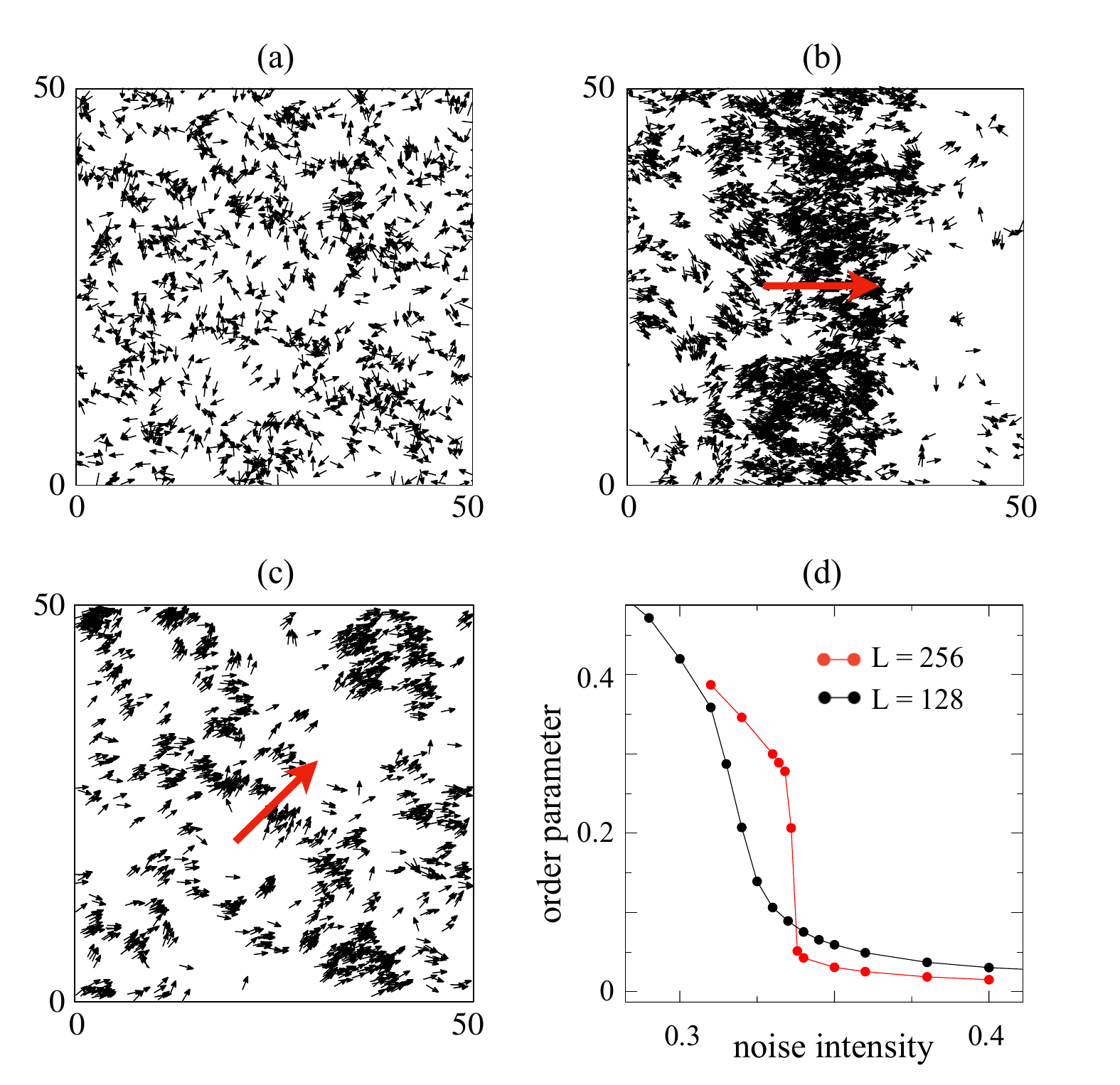}
	\caption{Canonical Vicsek model. (a-c) Snapshots of typical asymptotic configurations. 
	Each arrow represents a given particle and is oriented along its polarity.
	(a): disordered gas at low density/high noise.
	(b): high-density, high-order traveling band at intermediate noise/density.
	(c): polarly-ordered Toner-Tu liquid at low noise/high density.
	(d): variation of polar order parameter $\langle|\langle \exp(i\theta_j^t)\rangle_j|\rangle_t$ with noise strength at two different
	system sizes.
	Parameters for (a-c): $\rho_0 = \frac{1}{2}$, $\eta = 0.45$, 0.2 and 0.1. In (d), $\rho_0=2$.
	}
	\label{FIG1}
\end{figure}
%%%%%%%%%%%%%%%%%%%%%%%%%%%

\subsection{Modern viewpoint on DADAM}

DADAM, as we will see below, is arguably one of the most advanced areas of active matter physics in terms of knowledge of
the phenomena at play and of their theoretical description. The class represented by the Vicsek model is only one of a few basic 
classes (see below).
 
Another well-advanced area is that of dry systems where 
active particles only experience repulsive interactions, a situation well-known to allow, in spite of the absence of 
explicit attractive interactions, for MIPS or motility-induced phase separation \shortcite{cates2015motility}. 
The theme of phase-separation, or density segregation between a vapor and a dense `liquid', is actually central to our modern understanding of DADAM, 
but the mechanisms at play in this case are different from those leading to MIPS. 

For a long time, following the seminal works of Vicsek {\it et al.}, Toner and Tu, Ramaswamy, and others,
the problem of the emergence of collective motion, and more generally of orientational order, was approached as an order/disorder
transition: at strong noise local alignment is too weak to produce order (Fig.~\ref{FIG1}a), 
whereas decreasing the noise one sees collective motion,
i.e. order  (Fig.~\ref{FIG1}c). In 1995, and over the following years, 
Vicsek and collaborators studied the order/disorder transition numerically at moderate system sizes and number of particles, 
concluding to a continuous phase transition at a critical point representing a new universality class 
\shortcite{vicsek1995novel,czirok1997spontaneously}.

This view held until it was discovered that, at finite but larger system sizes, the transition is discontinuous  (Fig.~\ref{FIG1}d)
and that, near onset of global order, active particles spontaneously organize themselves into a number of 
high-density high-order bands traveling in a sparse disordered gas  (Fig.~\ref{FIG1}b) 
\shortcite{gregoire2004onset,chate2008collective}.
The idea of phase-separation, of some coexistence phase separating the disordered phase from a homogeneous ordered one, 
started being mentioned here and there  \shortcite{mishra2010fluctuations,marchetti2013hydrodynamics},
but it is the work of Solon \& Tailleur on the active Ising model \shortcite{solon2013revisiting,solon2015flocking} that triggered a change of perspective from
an order/disorder problem to a phase separation framework akin to a liquid-gas transition. In this framework, detailed below,
the traveling bands of the Vicsek model are characteristic of the coexistence phase separating the disordered gas from an ordered liquid.

\subsection{These lectures notes}

Most of these lecture notes focus on results obtained in two spatial dimensions (2D). 
This is the dimension of choice for dry active matter, which often consists in particles in strong interactions 
with a substrate and/or confinement, 
something common for 2D or quasi-2D systems such as motility assays, vertically vibrated granular media, or groups of walking/crawling
animals and cells.

Many of the following results rely on numerical simulations, a common situation for out-of-equilibrium systems
 involving numerous nonlinear mechanisms. All the simulations reported, be they of particle-based models or hydrodynamic partial
 differential equations, have used periodic boundary conditions, something hard
 to realize experimentally, but very useful and simpler in practice in systems of moving particles. Even though real situations 
 almost always have to deal with interactions of particles with walls, and even though interesting if not fundamental phenomena
 arise in active matter systems in the presence of walls (see, e.g., the contribution of Yariv Kafri in this volume), 
 it is better, in a first step, to set one free from walls in order to better grasp the essential bulk behavior of these systems.
 
 We have kept the reference list at some minimal level. There are two main reasons for this. One is that we present below a 
 (hopefully) coherent and self-contained account of what has emerged from series of papers reporting partial progress and sometimes contradicting each other. Sending the reader to some of these historical papers might trigger some confusion. 
 Unfortunately, there is no review paper on DADAM, and we hope that this work will fill this gap to some extent.  
There exist, nevertheless, more general review articles about active matter which, at least, 
 offer some organized access to original papers: 
 \shortcite{ramaswamy2010mechanics,marchetti2013hydrodynamics,bechinger2016active}. 
  
The rest of these lecture notes is organized in two main parts, Sections~\ref{micro} and \ref{hydro}, followed by 
discussion and perspectives in Section~\ref{discuss}. 
Section~\ref{micro}, after a definition of the 3 basic DADAM classes, presents
the phenomenology of these classes at the microscopic level, as it emerges from the study of corresponding simple Vicsek-style models. 
Section~\ref{hydro} deals with the continuous ``hydrodynamic" theories for the DADAM classes that can be derived from
the particle models studied in Section~\ref{micro}. In both Section~\ref{micro} and \ref{hydro}, we insist first on the global features
common to all 3 classes, before describing class-specific ones.
The concluding Section~\ref{discuss} first comes back to the overall good qualitative agreement obtained between
microscopic and hydrodynamic levels before providing   
perspectives on how one can expand DADAM knowledge into various directions, 
approaching more realistic and more complex situations in a controlled way.

\section{Particle-level phenomenology of the 3 basic DADAM classes}
\label{micro}

\subsection{3 basic classes and their Vicsek-style representative}

In the ``canonical" Vicsek model, point particles move at constant speed $v_0$, changing their velocity at discrete timesteps $\Delta t$ 
to align it with the local average of the velocities of neighboring particle within distance $r_0$. Setting, without loss of generality,
$\Delta t =1$ and $r_0=1$, labeling particles $j=1,\ldots,N$, 
their position ${\bf r}_j$ and velocity direction $\theta_j$ are governed, in two space dimensions, by:
\begin{subequations}
\begin{eqnarray}
{\bf r}_j^{t+1} &=& {\bf r}_j^{t} + v_0 {\bf e}(\theta_j^{t+1}) \label{eqvm1} \\
\theta_j^{t+1} &=& {\rm arg} [ \langle \exp(i\theta_k^t)\rangle_{k\sim j} ] + \eta \xi \label{eqvm2}
\end{eqnarray}
\label{eqcvm}
\end{subequations}
where ${\bf e}(\theta)$ is the unit vector along $\theta$, and $\xi$ is a uniform, delta-correlated noise over the interval $[-\pi,\pi]$, drawn independently for each particle.
There are only two main parameters, the noise strength $\eta$, and the global number density $\rho_0=N/L^2$ 
(supposing the particles evolve in a square domain of linear size $L$). 
The speed $v_0$ is usually taken to be rather large ($v_0=0.5$ in most of the following), 
a numerically advantageous choice since small speed is known to
push the observation of asymptotic behavior to larger system sizes.

Numerous variants exist: the noise can be implemented differently, e.g. the ``vectorial" noise case where  
$\theta_j^{t+1} = {\rm arg} [ \langle \exp(i\theta_k^t)\rangle_{k\sim j} + \eta \xi]$ 
where $\xi$ is now a unit-modulus, random-phase complex number. 
The updating scheme can be different, e.g. the ``backward" updating of the original Vicsek {\it et al.} paper \shortcite{vicsek1995novel}
uses $\theta_j^t$, not $\theta_j^{t+1}$ in the first equation above. Continuous time updating can be used, e.g.
$\dot{\theta_j}=K \langle \sin(\theta_k-\theta_j)\rangle_{j\sim k} + \eta \xi$. Most of the self-imposed constraints 
such as constant speed and overdamped dynamics, isotropic neighborhood, equal-weight of neighbors, 
uniform noise, can be relaxed (to some ``reasonable" extent). Another key simplification of the Vicsek model is that the particles 
have no intrinsic polarity, or, rather, that their polarity
aligns immediately and perfectly with their velocity. This may appear as an oversimplification, but it can also be relaxed, 
at the price of introducing fast dynamics aligning a physical polarity and the current velocity. 

Most of the variants sketched above do not have any qualitative influence on the collective behavior observed 
(see, however, \shortcite{chen2017fore}).
For the sake of simplicity and numerical efficiency, we adopt, in the following, 
the ``canonical" version given be Eqs.~(\ref{eqcvm}) and take it as a representative of the large class of polar
DADAM systems (hereafter called ``polar/Vicsek class''). 

\begin{figure}[h!]
	\centering
	\includegraphics[scale=0.35]{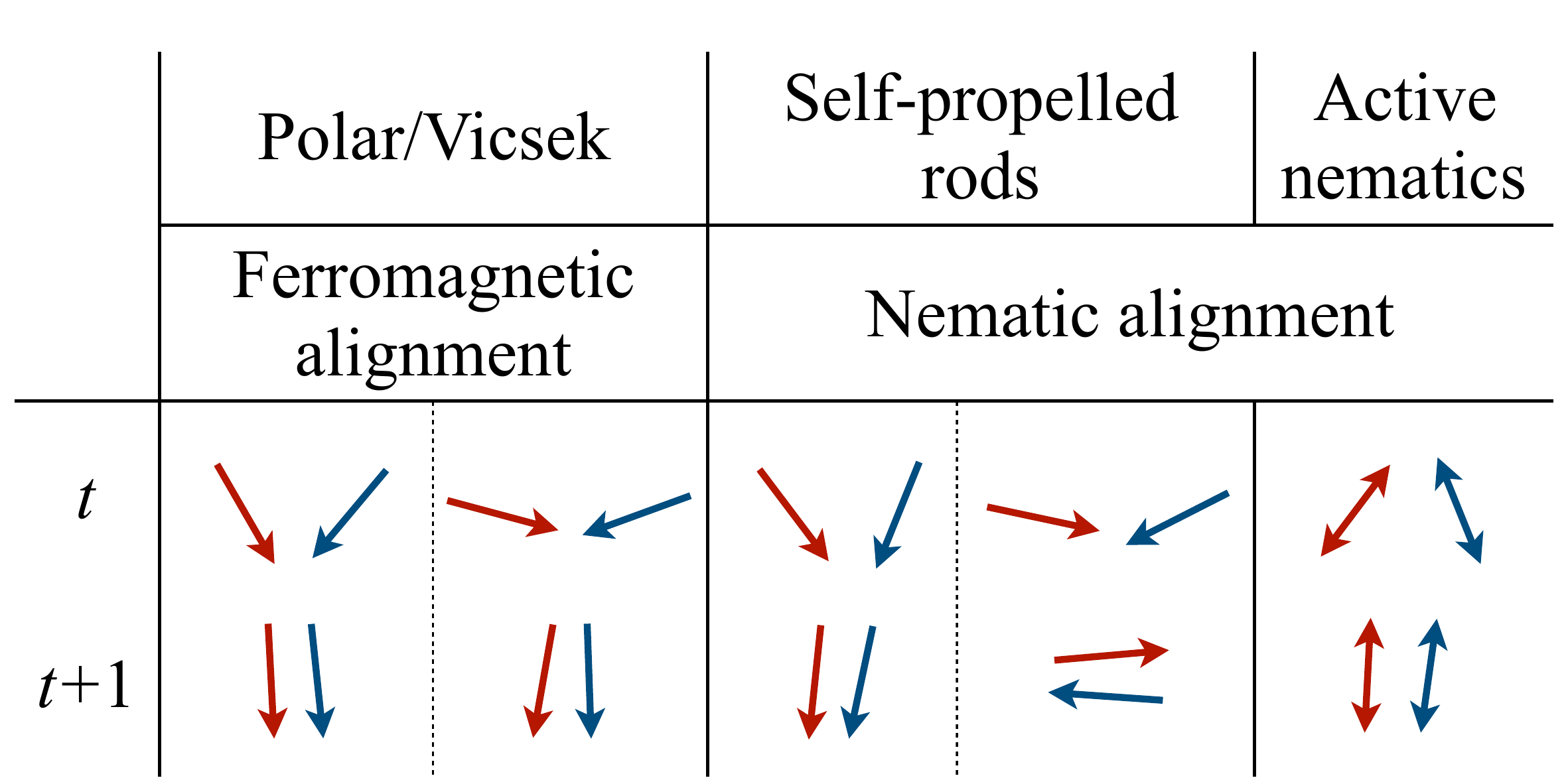}
	\caption{Vicsek-style velocity alignments for the 3 DADAM classes.
	In the active nematics case, the fact that particles can move in both directions given by their axis 
	with equal probabilities is represented by the double-headed arrows.}
	\label{fig:types_alignment}
\end{figure}

Two other basic DADAM classes can be defined within the ``Vicsek framework''.
Qualitative changes, defining new classes of DADAM, are obtained when changing the nature of the individual particle motion 
and/or the symmetry of the alignment (see Fig.~\ref{fig:types_alignment}).
Changing the ferromagnetic alignment of Eq.~(\ref{eqvm2}) to nematic alignment in which
neighboring particles at obtuse angle anti-align (as when elongated objects collide), one obtains a Vicsek-style self-propelled ``rods''
model.
Introducing spontaneous velocity reversals at finite rate $\alpha$ for the individual particle motion prevents the 
emergence of any polar order when ferromagnetic alignment is used. In the case of nematic alignment, however,
this defines a new class, usually called ``active nematics''. 
(Note that Vicsek-style active nematics are fundamentally different from the wet active nematics 
although they do share some characteristics. See, e.g. \shortcite{Doostmohammadi2018active}.)

%%%%%%%%%%%%%%%%%%%%%%%%%%%
%%%%%%%%%%%%%%%%%%%%%%%%%%%
%%%XXX renverser "X" et "y" dans cette figure? --> Cela déforme le tableau (les configurations avec alignement initial obtus prennent de la place)...

The canonical Vicsek-style models representing the 3 classes can be defined by the following equations governing
the positions ${\bf r}_i$ and velocities ${\bf v}_i=v_0 {\bf e}(\theta_i)$ of particles:
\begin{subequations}
\begin{eqnarray}
	{\bf r}_i^{t+1} &=& {\bf r}_i^t +\varepsilon(t) {\bf v}_i^{t +1} \label{eqvm3}\\
	{\bf v}_i^{t+1} &=& \left( {\mathcal R}_\eta \circ \vartheta \right) \langle {\bf v}^t \rangle_{i} \label{eqvm4}
\end{eqnarray}
with $\langle {\bf v}^t \rangle_{i}$ equals to
\begin{equation}
\langle {\bf v}^t \rangle_i^{\rm ferro} = \sum_{j\sim i} {\bf v}_j^t \quad {\rm or} \quad 
\langle {\bf v}^t \rangle_i^{\rm nema} = \sum_{j\sim i} {\rm sign}[{\bf v}_i^t\cdot {\bf v}_j^t  ] {\bf v}_j^t \; . \label{eqvm5}
\end{equation}
\label{eqvm}
\end{subequations}
In Eq.~(\ref{eqvm4}) $\vartheta$ is an operator returning unit vectors ($\vartheta({\bf u})={\bf u}/\|{\bf u}\|$),
and ${\mathcal R}_\eta{\bf v}$ rotates the vector ${\bf v}$ by a random angle drawn from a uniform distribution inside 
an arc of length $2\pi\eta$ (a cap of surface $2\pi(1-\cos(\eta))$  in 3D)
centered on ${\bf v}$.
In the polar and self-propelled rods cases, $\varepsilon=1$, while in the active nematics case where velocity reversals occur, $\varepsilon=\pm1$ and 
changes sign with probability $\alpha$. In the  self-propelled rods and the active nematics cases, nematic alignment coded by $\langle {\bf v}^t \rangle_i^{\rm nema}$ is used, while the polar, Vicsek class uses ferromagnetic alignment $\langle {\bf v}^t \rangle_i^{\rm ferro}$. 
(Note that using this ferromagnetic alignment and $\varepsilon=1$ is fully equivalent to Eqs.~(\ref{eqcvm}).)

\begin{table}[b]
%\tableparts
\centering 
{
\caption{The 3 basic DADAM classes: definition and main properties of their liquid and coexistence phases in 2D.
$^{\rm a}$Here active nematics regroups ``fast active nematics" (described at hydrodynamic level by diffusive 
Equations~(\ref{hydro_an})) 
and rods with a finite velocity reversal rate $\alpha$. Polar rods here refers to the case $\alpha=0$.  
$^{\rm b}$Single band occupying a fraction of space exists, but is unstable to long-wavelength longitudinal instability. 
This instability is superseded by transversal break up instability at high velocity reversal rates.
$^{\rm c}$Part of coexistence phase near liquid binodal consists in never-settling disordered evolution of nearly-parallel bands, with long-range polar order but no smectic order.
(LRO: long-range order. QLRO: quasi-long-range order.)}
\label{TABLE}
}
\vspace{6pt}
{\small
\begin{tabular}{@{}llllllll@{}}
\hline 
 &  \multicolumn{3}{c}{basic features}  & \multicolumn{2}{c}{coexistence phase} & \multicolumn{2}{l}{ordered liquid} \vspace{0.1cm}\\
\multirow{2}{*}{Class} & particle  & \multirow{2}{*}{alignt.} & linear     &  \multirow{2}{*}{basic object} &  \multirow{2}{*}{global pattern} &  global  &  GNF  \vspace{-0.1cm}\\
          & motion   &		     &instab. &  &  &  order &     exp.     \vspace{0.2cm} \\
Polar & \multirow{2}{*}{polar} & \multirow{2}{*}{ferro.} & \multirow{2}{*}{$\parallel$} & quantized & smectic /& 
\multirow{2}{*}{LRO} & \multirow{2}{*}{$\gtrsim 1.6$}  \vspace{-0.1cm} \\
/Vicsek & & & & traveling band & irreg. bands$^{\rm c}$ & & \vspace{0.2cm}\\
Active & \multirow{2}{*}{apolar} & \multirow{2}{*}{nema.} & \multirow{2}{*}{$\perp$} & macroscopic  & \multirow{2}{*}{band chaos} & \multirow{2}{*}{QLRO} & \multirow{2}{*}{$\gtrsim 1.6$} \vspace{-0.1cm}\\
nematics$^{\rm a}$ & & & & band$^{\rm b}$ &  & & \vspace{0.2cm} \\
Polar & \multirow{2}{*}{polar} & \multirow{2}{*}{nema.} & \multirow{2}{*}{$\perp$} & macroscopic  & \multirow{2}{*}{band chaos} & \multirow{2}{*}{LRO?} & \multirow{2}{*}{$\gtrsim 1.6$} \vspace{-0.1cm}\\
rods$^{\rm a}$ & & & & band$^{\rm b}$ &  & & \vspace{0.1cm} \\
\hline
\end{tabular}
}
\end{table}
%%%%%%%%%%%%%%%%%%%%%%%%%%%

The first 3 columns of Table~\ref{TABLE} summarize the defining features 
of the 3 basic DADAM classes represented by the Vicsek-style models defined above.
We now turn to a description of the collective properties of each of these models, 
and in particular of their phase diagram in the density/noise strength $(\rho_0,\eta)$ main parameter plane.

%%%%%%%%%%%%%%%%%%%%%%%%%%%
\begin{figure}[h!]
	\centering
	\includegraphics[scale=0.6]{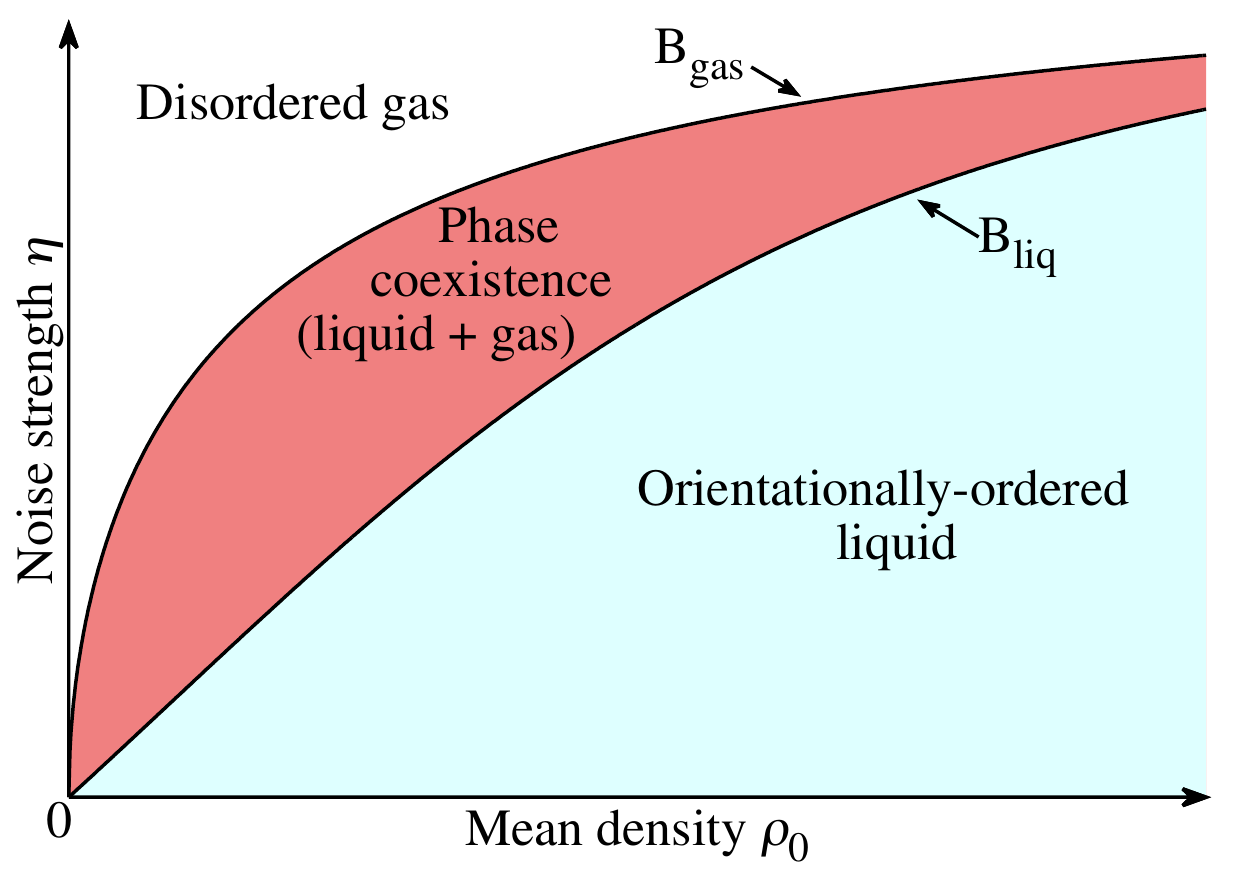}
	\caption{Schematic phase diagram of the Vicsek-style representatives of all 3 basic DADAM classes 
	in the global density/noise strength $(\rho_0,\eta)$ main parameter plane. 
	The microscopically-disordered gas is separated from the ordered liquid by a coexistence phase 
	where liquid regions move in a residual sparse gas. 
	The two binodals ${\rm B}_{\rm gas}$ and ${\rm B}_{\rm liq}$ meet at the origin, and converge to each other
	in the infinite density limit.
	 }
	\label{FIG2}
\end{figure}
%%%%%%%%%%%%%%%%%%%%%%%%%%%

\subsection{Properties common to all 3 classes}

The 3 basic DADAM classes being defined by different symmetries, their collective properties have 
important differences. To start with, the symmetry of the alignment directly rules what type of orientational
order can emerge {\it locally}: ferromagnetic alignment may lead to polar order, 
while nematic alignment allows for both nematic and polar local order.
Nevertheless, the 3 classes share many properties. 
In this section, we present these common properties, before turning to a more detailed description of the features 
specific to each class. A summary is given in Table~\ref{TABLE}.

\subsubsection{Phase diagram}

The phase diagrams of the 3 Vicsek-style models representing the 3 basic classes defined above have been 
established from careful numerical simulations.
They are all organized in the same way (Fig.~\ref{FIG2}): for high noise strength $\eta$ and/or low global density $\rho_0$, 
the persistent random walks of
individual particles cannot synchronize globally under the aligning interactions, 
and the system remains in a microscopically-disordered ``gas" state. 

%%%%%%%%%%%%%%%%%%%%%%%%%%%
\begin{figure}[h!]
	\centering
	\includegraphics[width=0.85\textwidth]{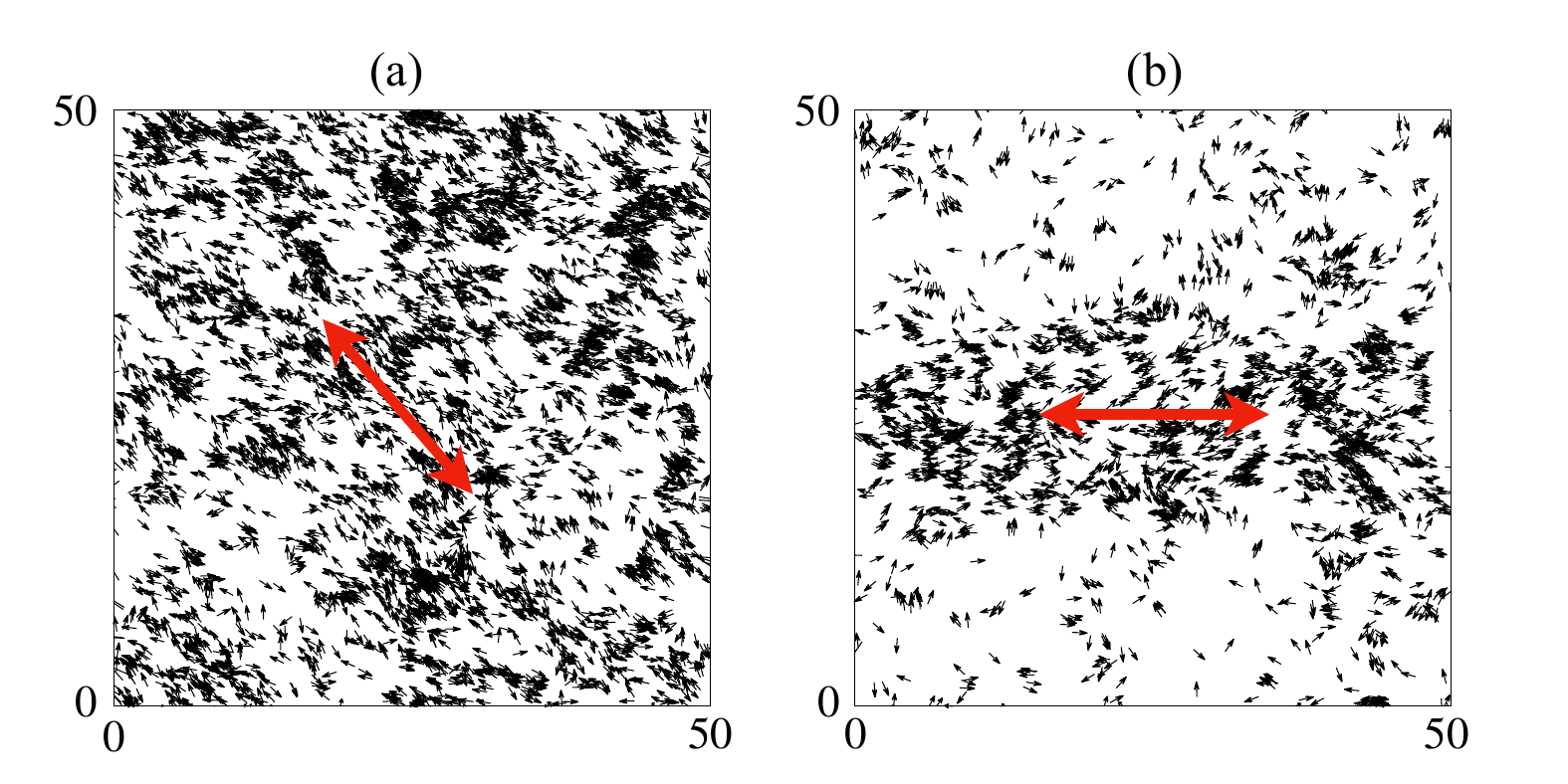}
	\caption{Typical snapshots of the nematic liquid (a) and a nematic band (b) for the Vicsek-style self-propelled rods model
	(defined by Eqs.~(\ref{eqvm}) at moderate system size.
	Each arrow represents a given particle and is oriented along its polarity, the large double-headed red arrows
	indicate the direction of the global nematic order.
	Parameters: $\rho_0 = 1$ \& $\frac{1}{2}$, $\eta = 0.15$.
	 }
	\label{FIG:NEMA}
\end{figure}
%%%%%%%%%%%%%%%%%%%%%%%%%%%

For low-enough $\eta$ and/or a high-enough $\rho_0$, alignment events are frequent and efficient 
enough that global orientational order eventually emerges. 
Figures~\ref{FIG1}c and~\ref{FIG:NEMA}a are typical snapshots in this homogeneous but fluctuating ordered ``liquid" 
for the polar/Vicsek and the nematic classes. (Note that, naturally, this liquid is polarly ordered in the Vicsek class, and nematically
ordered in the nematic classes.)

The ordered liquid state is separated, in the $(\rho_0,\eta)$ plane, by a coexistence region composed of ordered, dense, 
 domains moving in a residual sparse disordered gas (see, e.g.,  Fig.~\ref{FIG1}b for the polar/Vicsek class, and 
Fig.~\ref{FIG:NEMA}b for the classes with nematic alignment). 
This phase coexistence comes from the interplay between density and order:
aligned particles tend to travel together; such ordered packets become more ordered and ``recruit" more particles, promoting 
density and order. 
As it will become clear from hydrodynamic theories (see Section~\ref{properties_classes_hydro}), 
the resulting feedback mechanism leads to an instability of homogeneous configurations close to the ordering transition. 
We will show below that increasing the global density and/or the system size in the coexistence region, 
the vapor density of the residual gas remains constant and the ``liquid fraction", i.e. the fraction of space covered by dense ordered domains, increases, a signature of phase separation.

The two lines delimiting the coexistence phase meet at the origin, 
and converge to each other when $\rho_0\to\infty$ (Fig.~\ref{FIG2}). They can be seen as the binodal lines of a liquid-gas phase separation scenario.
In contrast with equilibrium liquid-gas phase separation, they do not have the parabolic shape culminating at a critical point. 
Here, in some sense, the critical point is sent to the $\rho_0\to\infty$ limit 
(no jamming occurs for Vicsek-like models since particles have no physical size) because the rotational symmetry of the gas
is spontaneously broken in the ordered liquid.

\subsubsection{Anisotropic long-range correlations and giant number fluctuations}

The ordered liquid, in all 3 classes, is endowed with generic long-range algebraic correlations and anomalous fluctuations, 
as predicted by Toner, Tu, Ramaswamy, {\it et al.}. The long-range correlations of order fluctuations arise from the spontaneous
breaking of the continuous rotational invariance, like in the equilibrium XY model. In DADAM, these, as we will see below,
can be strongly enhanced, and they are coupled to density fluctuations, leading to new phenomena.

Toner \& Tu's landmark calculation for the Vicsek class, based on their phenomenological hydrodynamic equation, 
focused on the stability of the ordered phase with respect to fluctuations, leading to their proof of true long-range 
polar order for the Vicsek class, even in 2D. Beyond this crucial result, 
Toner \& Tu predicted that the two-point correlation functions
of density and velocity fluctuations display generic anisotropic algebraic decay:
\begin{equation}
C({\bf r}) = \left| {\bf r}_\perp \right|^{2\chi} f(r_\|/\left| {\bf r}_\perp \right|^{\zeta}) \,,
\end{equation}
where $\|$ and $\perp$ indices respectively refer to directions longitudinal and transverse to the mean motion of the flock
and the exponents $\chi$ and $\zeta$ as well as the function $f$ are universal.
For a long-time, the values of the scaling exponents $\chi$ and $\zeta$ 
predicted in the 1995 and 1998 papers \shortcite{toner1995long,toner1998flocks}
were believed to be `exact' in 2D. In 2012, it was realized that this is not true \shortcite{toner2012reanalysis}. 
As of today, there is no direct, quantitative,
numerical investigation of the Toner-Tu predictions, but a general consensus that they are qualitatively correct.
It was realized later that the Toner-Tu scaling of correlation functions is tantamount to
the existence of anomalously-strong, ``giant" number fluctuations (GNF) and of superdiffusive behavior of particles 
in the dimensions transverse to global order~\shortcite{ginelli2016physics}.

Similar conclusions were reached for active nematics by Ramaswamy and collaborators \shortcite{ramaswamy2003active}, 
predicting GNF and algebraic correlations
within the quasi-long-range, nematically-ordered liquid phase (see below). 

\begin{figure}[h!]
	\centering
	\includegraphics[scale=0.145]{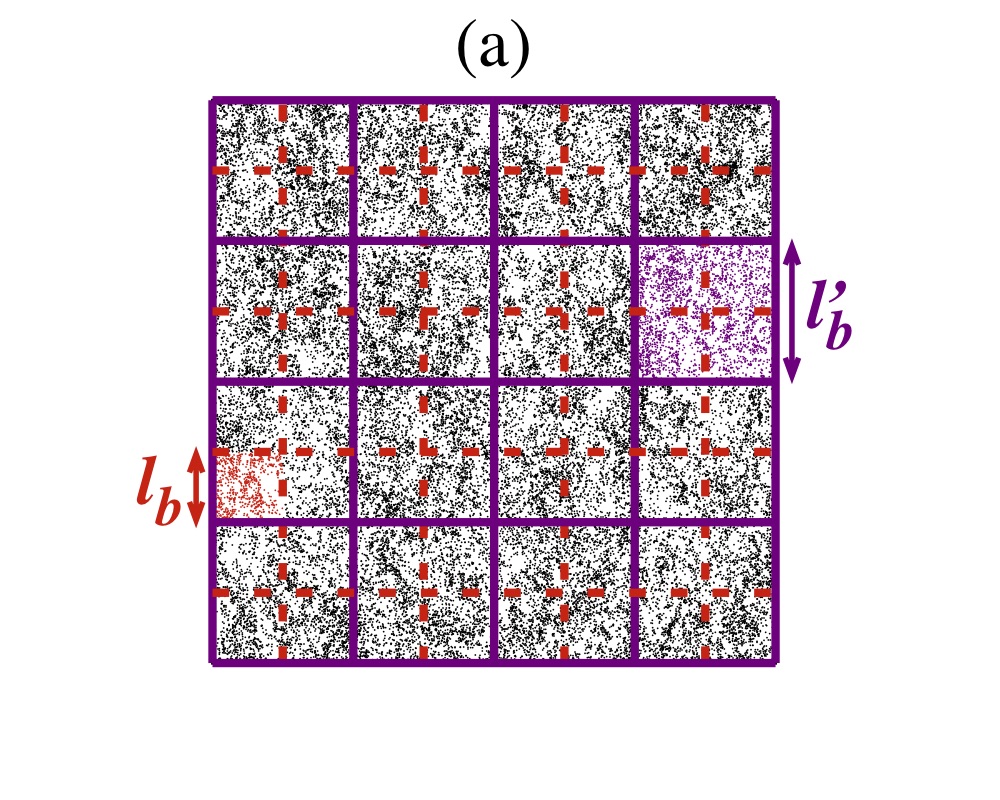}
	\includegraphics[scale=0.8]{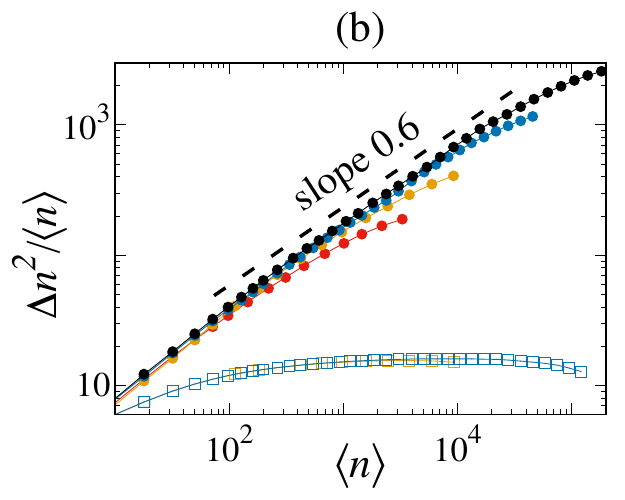}
	\caption{Giant number fluctuations in the orientationally-ordered liquid phase.
	(a): Schematic view showing a large system divided into sub-boxes of linear sizes $l_b$ and $l_b' > l_b$
	containing respectively $n_b(t)$ and $n_b'(t)$ particles at time $t$.
	The variance of $n_b(t)$ is given by $\Delta n^2 = \langle (n_b(t) - \langle n \rangle )^2 \rangle_{b,t}$,
	where $\langle \ldots \rangle_{b,t}$ stands for the average over all boxes of same size $l_b$ and time.
	$\langle n \rangle = \rho_0 l_b^2$, with $\rho_0$ the mean density of the system,
	denotes the mean number of particles in the sub-box of size $l_b$. 
	Computing $\Delta n^2$ and $\langle n \rangle$ for various values of $l_b$ 
	with $l_b \ll L$, we obtain the type of curves shown in (b).
	(b): Canonical Vicsek model: 
	$\Delta n^2 / \langle n \rangle$ vs.\  $\langle n \rangle$ showing GNF in the liquid phase (filled circles) with an exponent
	compatible with the Toner and Tu prediction $\phi = \frac{8}{5}$ (dashed line),
	 and normal fluctuations in the gas phase (hollow squares) for linear system sizes 
	 $L = 128$ (red), 256 (orange), 512 (blue), and 1024 (black).
	Note that the numerical effort required to obtain these results is rather large 
	(averaging over typically $10^7$ timesteps necessary for the largest size).
	Nevertheless, apparently-converged scaling is observed over 1-2 decades at best. 
	This difficulty renders all numerical estimates
	of the scaling exponent $\phi$ rather uncertain. 
	Thus, even though in all 3 DADAM classes $\phi=1.6-1.7$ has been reported,
	this ``super-universality" of $\phi$ should not be taken too seriously.
	Parameters: $\rho_0 = 2$, $\eta = 0.2$ (liquid) and $0.6$ (gas).
	 }
	\label{FIG:GNF}
\end{figure}

By giant number fluctuations, one usually means that the variance 
$\Delta n^2=\langle n^2\rangle - \langle n\rangle^2$ of the number of particles $n$ 
contained in a sub-box of a much larger system scales like 
\begin{equation}
\label{eq:GNF}
 \Delta n^2 \sim \langle n\rangle^\phi \;\; {\rm with} \;\; \phi>1 \;.
\end{equation}
(See Fig.~\ref{FIG:GNF}a for a schematic description of the procedure.)
In other words, number fluctuations are stronger than what the central limit theorem would predict 
for a system with exponentially-decaying density correlation functions (in that case, $\phi=1$).

%%%%%%%%%%%%%%%%%%%%%%%%%%%
%%%%%%%%%%%%%%%%%%%%%%%%%%%

GNF, because their investigation implies measuring quantities seemingly easier to extract from numerical or experimental data, 
have become a sort of landmark signature of the `new physics' displayed by active matter. 
Some confusion exists in the literature, where GNF have been claimed to exist for systems displaying some kind of clustering
dynamics: indeed, in systems without orientational order but which exhibit configurations 
made of dense clusters sitting in a sparse gas,
one can record some evidence of ``GNF'' on scales smaller or equal to the typical dense cluster size. 
But these are trivially related to the phase separation into clusters (one can easily calculate that $\phi=2$), 
and not asymptotic since they disappear for scales beyond the cluster size.

Giant number fluctuations have been found in numerical simulations in the ordered liquid phase of all 3 basic DADAM classes. 
Data obtained on the canonical Vicsek model are shown in Fig.~\ref{FIG:GNF}b. 
Surprisingly, numerical estimates of the associated scaling exponent have found that $\phi\simeq 1.6-1.7$ in all 3 cases
\shortcite{chate2008collective,ginelli2010large,ngo2014large}.
This is roughly in agreement with the Toner-Tu prediction that $\phi=\frac{8}{5}$ in 2D (for the polar class), but in disagreement 
with the prediction by Ramaswamy {\it et al.} that $\phi=2$ in the nematic classes 
(see however, \shortcite{shankar2018low} for a discussion of this point).

\subsection{Specific properties of the polar, Vicsek class}

\begin{figure}[h!]
	\centering
	\includegraphics[scale=0.65]{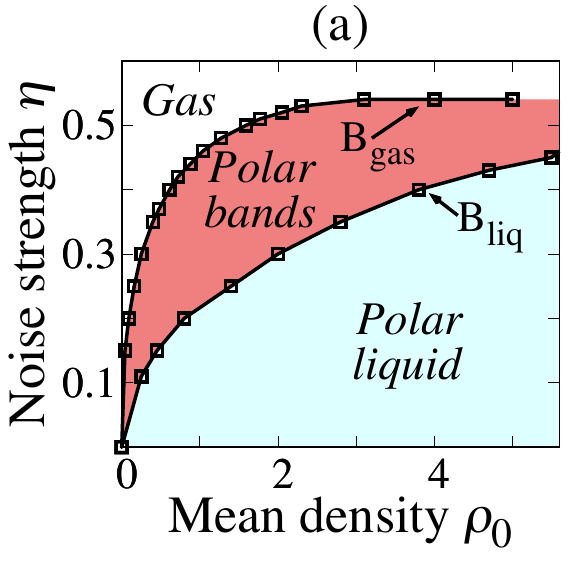}
	\includegraphics[scale=0.65]{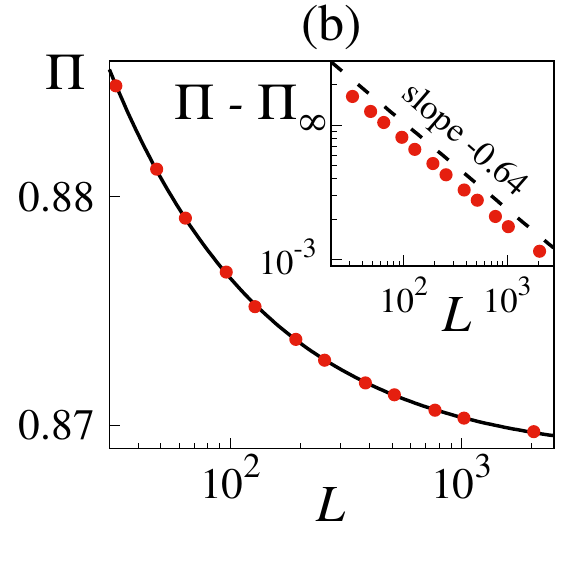}
	\includegraphics[scale=0.65]{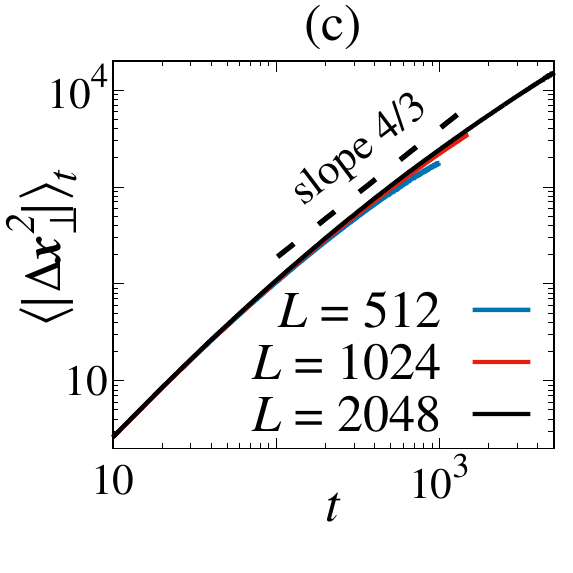}
	\caption{Global properties of the `canonical' Vicsek model (as defined by Eqs.~(\ref{eqcvm})).
	(a): Phase diagram in the density/noise plane showing the three phases characteristic of DADAM.
	(b): LRO in liquid phase: 
	Polar order parameter $\Pi = \langle|\langle \exp(i\theta_j^t)\rangle_j|\rangle_t$ as function of system size $L$
	showing a power law decay with an exponent $\simeq -0.64$ to an asymptotic value $\Pi_\infty = 0.8685(2)$ (dashed line). 
	Inset: $\Pi - \Pi_\infty$ as function of system size.
	(c): Mean square displacement of particles in the direction transverse to the order (labelled by $\perp$) as function of time
	exhibiting superdiffusion in the liquid phase with an exponent roughly compatible with the Toner \& Tu prediction $\frac{4}{3}$. 
	This measurement has been done in a channel configuration, i.e. by imposing reflective boundary conditions 
	on one direction, such that the orientation of the global polar order remains steady in time.
	Parameters for (b,c): $\rho_0 = 2$, $\eta = 0.2$.
	 }
	\label{FIG:VM}
\end{figure}

Extensive simulations of the Vicsek model have shown that its phase diagram conforms to the 
general picture sketched above (Fig.~\ref{FIG:VM}a).
True long-range polar order is observed in the liquid phase (Fig.~\ref{FIG:VM}b).
As mentioned above, giant number fluctuations have been found in that phase,
with a scaling exponent $\phi\simeq 1.6$, i.e. compatible with the Toner and Tu value $\frac{8}{5}$.
Superdiffusive behavior in the direction transverse to order, a property also following from the Toner-Tu calculation,
is indeed present, with again a scaling exponent compatible with the Toner-Tu calculation 
(Fig.~\ref{FIG:VM}c). 
However, a direct, accurate assessment of the Toner-Tu predictions has not yet been reported.

%%%%%%%%%%%%%%%%%%%%%%%%%%%
%%%%%%%%%%%%%%%%%%%%%%%%%%%

%%%%%%%%%%%%%%%%%%%%%%%%%%%
\begin{figure}[h!]
	\centering
	\includegraphics[scale=0.18]{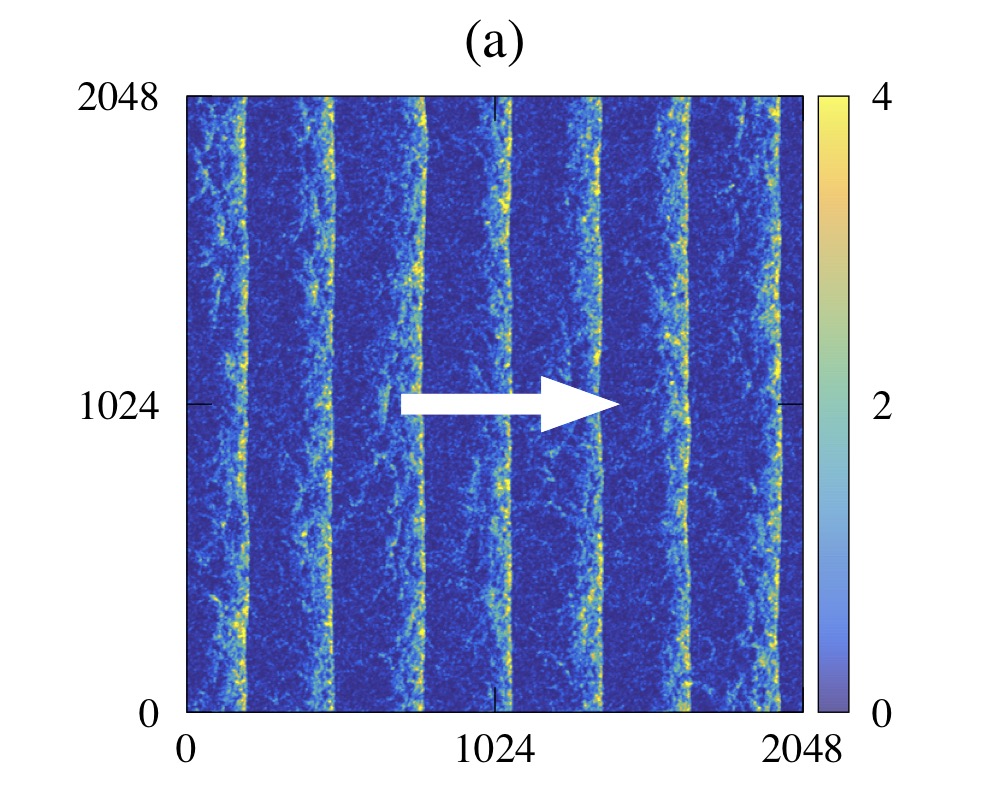}
	\includegraphics[scale=0.18]{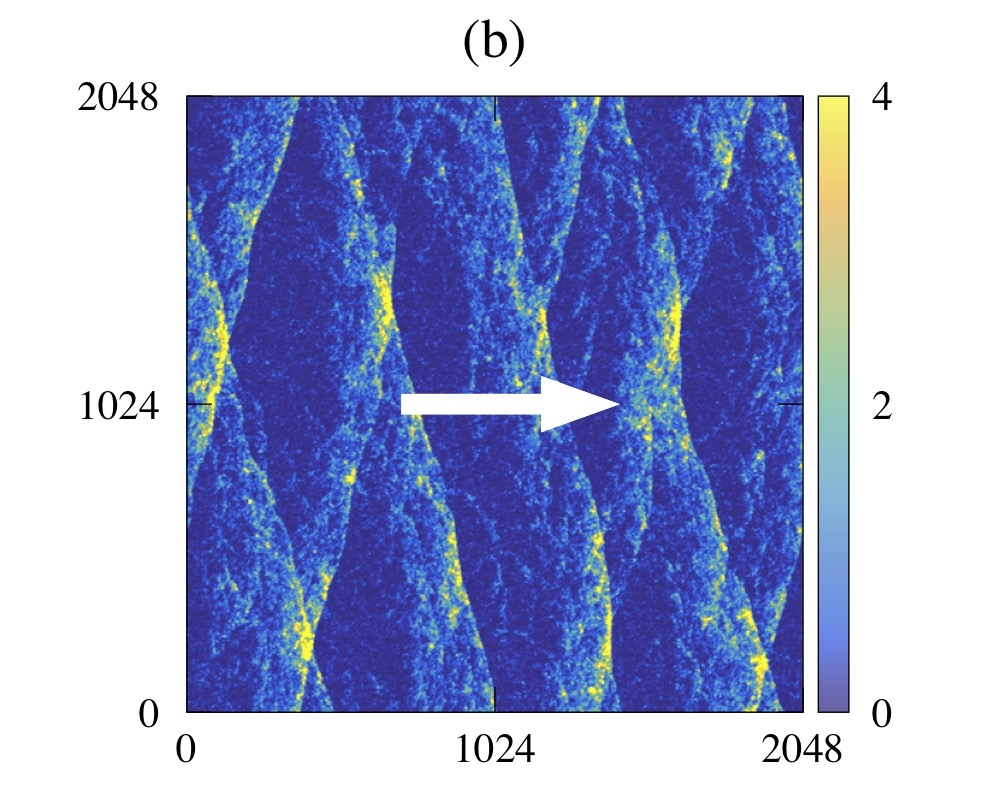}
	\includegraphics[scale=0.65]{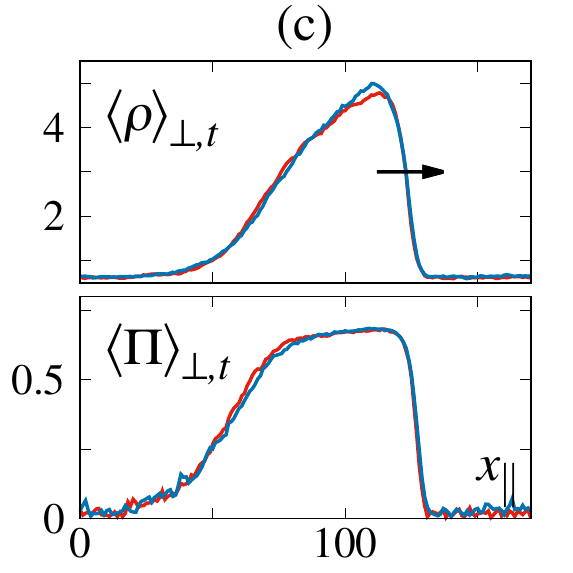}
	\includegraphics[scale=0.65]{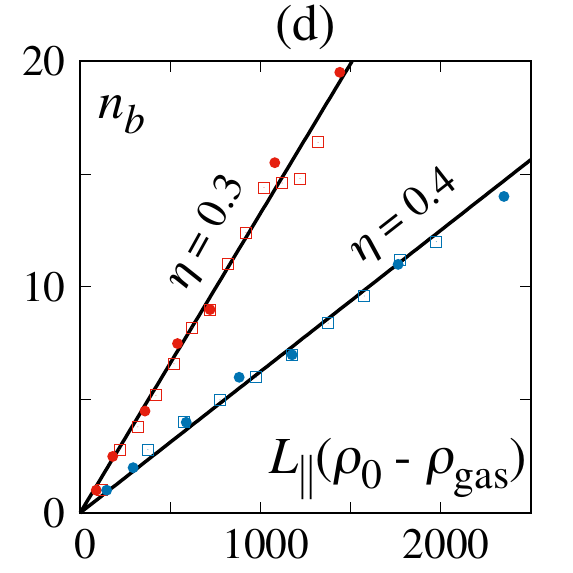}
	\includegraphics[scale=0.65]{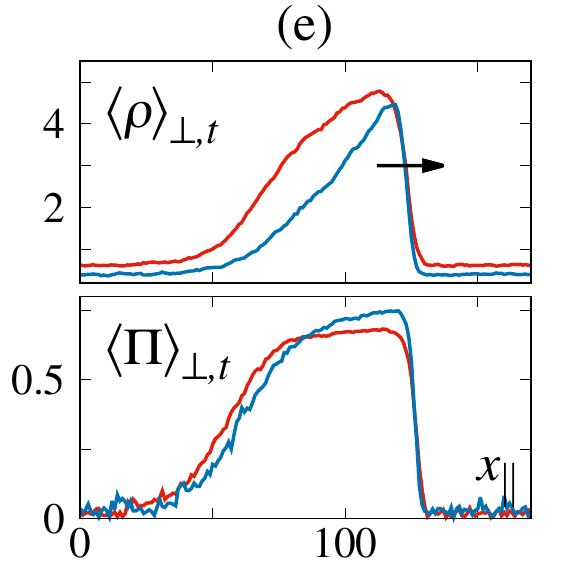}
	\caption{Coexistence phase of the `canonical' Vicsek model (as defined by Eqs.~(\ref{eqvm})).
	(a,b): Snapshots of coarse-grained density field in the regular smectic of bands and irregular one (near the liquid binodal).
	Parameters: $\rho_0 = \frac{1}{2}$, $\eta = 0.2$ \& $0.25$.
	(c): Density and order profiles averaged over the direction transverse to the order and time at densities 
	$\rho_0 = 1$ (red) and 2 (blue) for $\eta = 0.4$ and system size $2048 \times 256$.
	(d): Number of bands vs $L_\|(\rho_0 - \rho_{\rm gas})$ varying either the excess density for 
	$L_\| \times L_\perp = 2000 \times 100$ (hollow squares), or the system size along the direction of propagation (filled circles)
	for $\rho_0 = 0.6$ ($\eta = 0.3$) or $\rho_0 = 1.2$ (fixed noise strength $\eta = 0.4$).
	(e): Variation of the shape of individual bands with noise level: Density and order profiles averaged over the direction 
	transverse to the order and time at noise strengths 
	$\eta = 0.35$ (blue), 0.4 (red) for $\rho_0 = 1$ and system size $2048 \times 256$.
	 }
	\label{FIG:VM2}
\end{figure}
%%%%%%%%%%%%%%%%%%%%%%%%%%%

Large-enough systems simulated long enough in the coexistence phase display many traveling bands such as that of Fig.~\ref{FIG1}b.
Not too far from the gas binodal line ${\rm B}_{\rm gas}$, 
the bands are arranged in a smectic way: they form a train of identical, equally-spaced, equal-speed
objects (Fig.~\ref{FIG:VM2}a). 
Close to the liquid binodal ${\rm B}_{\rm liq}$, the bands are more numerous and interact strongly with each other, 
leading to disordered configurations (Fig.~\ref{FIG:VM2}b).

The bands are `quanta' of liquid, uniquely defined at fixed noise strength: bands observed at different
densities not too far from the gas binodal are identical objects, while the gas level (vapor density) remains constant 
(Fig.~\ref{FIG:VM2}c).
Their number increases linearly with the dimension of the system
along the polar order and/or with the global density $\rho_0$ (Fig.~\ref{FIG:VM2}d).
One is thus in presence of a microphase separation scenario,
one in which the coexisting liquid is quantized into ``microphases" (the bands).
The shape of these individual quanta of liquid (i.e. the density and order profiles of the bands) varies with the noise level, 
from rather sharp at low noise to wider and smoother at high noise (Fig.~\ref{FIG:VM2}e).
It is largely believed, but was never reported clearly, that true long-range polar order is also present in the coexistence phase.

The two binodal lines defining the phase diagram correspond to non-critical transitions. If the liquid binodal ${\rm B}_{\rm liq}$
(separating the coexistence phase from the liquid) remains hard to estimate numerically 
(see \shortcite{solon2015phase} for a reasonable attempt), the gas binodal ${\rm B}_{\rm gas}$ can be clearly defined by the 
density of the residual gas separating bands (which varies only with the noise strength $\eta$). 
At finite system size, the transition appears discontinuous since the system has to jump from zero to one band (or vice-versa). 
Tn the infinite size limit on the other hand, one has always an infinite number of bands
but the distance between bands diverges at the gas binodal, so that
the liquid fraction goes continuously to zero and the transition is actually continuous.

\subsection{Specific properties of the nematic classes}

\begin{figure}[h!]
	\centering
	\includegraphics[scale=0.65]{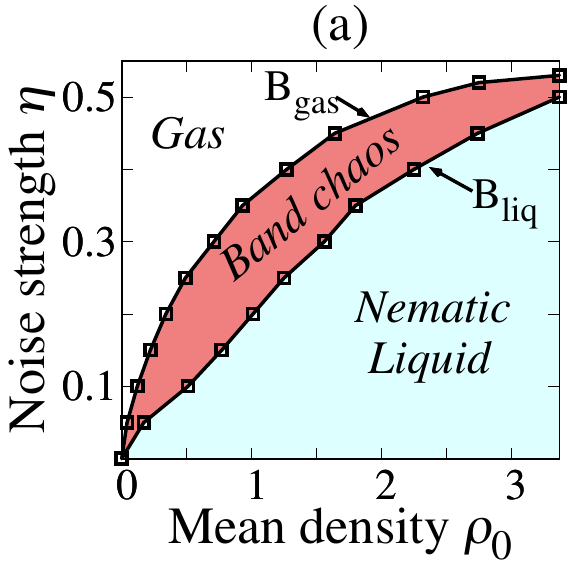} \hspace{1cm}
	\includegraphics[scale=0.65]{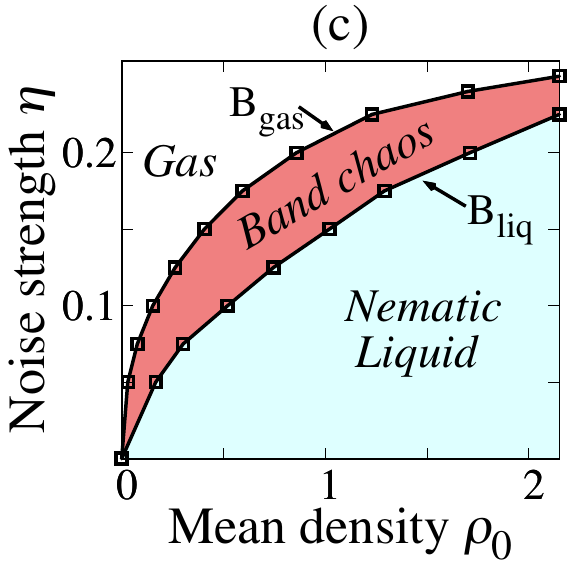}\\
	\includegraphics[scale=0.65]{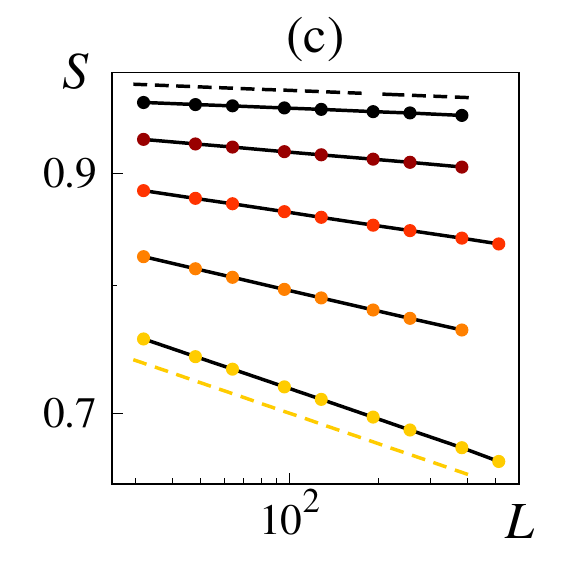}
	\includegraphics[scale=0.65]{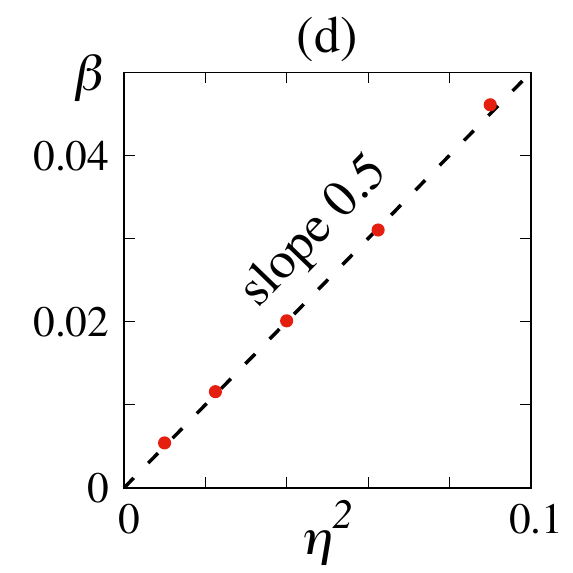}
	\includegraphics[scale=0.65]{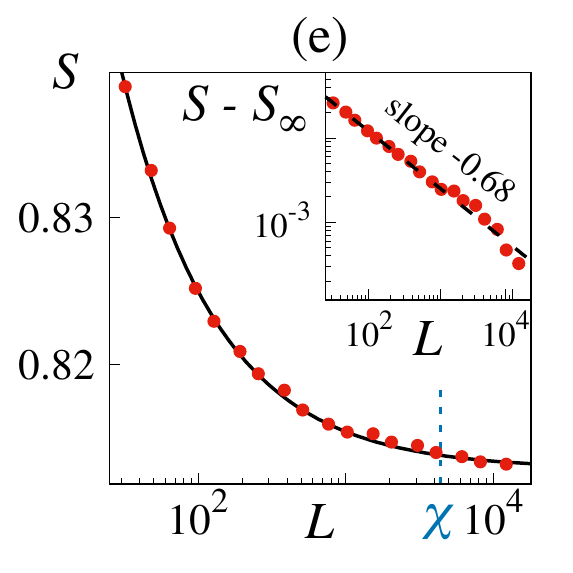}
	\caption{Collective properties of the Vicsek-style models with nematic alignment (as defined by Eqs.~(\ref{eqvm})).
	(a): Phase diagram of the active nematics model with fast velocity reversals ($\alpha= 0.5$).
	(b): Phase diagram of the model of self-propelled rods without velocity reversals ($\alpha= 0$).
	(c): Power law decay of the nematic order parameter $S = \langle |\langle \exp(2i\theta_j^t)\rangle_j|\rangle_t$
	with linear system size in the nematic liquid phase for active nematics with
	fast velocity reversals ($\alpha=0.5$); going downward, $\eta = 0.1$, 0.15, 0.2, 0.25, and 0.3 at fixed density $\rho_0 = 2$.
	The two dashed line respectively correspond to decay exponents $-0.005$ and $-0.045$.
	(d): Decay exponent $\beta$ of the curves of (c) showing a linear variation with the variance 
	of the noise $\eta^2$. Note that this is
	similar to the equilibrium $XY$ model for which the decay exponent increases linearly with temperature.
	(e): Decay of nematic order parameter with linear system size for Vicsek-style self-propelled rods 
	without velocity reversal ($\alpha=0$) in the liquid phase ($\rho = \frac{1}{8}$, $\eta = 0.048$). 
	This decay, slower than a power law, is best fitted by an algebraic decay to $S_{\infty} = 0.813(1)$ with an
	exponent $\approx -0.68$. 
	The dashed vertical line indicates the lengthscale $\chi$ corresponding to the characteristic flight time of particles 
	along one of the two directions defined by the global nematic order.
}
	\label{FIG:NEMA2}
\end{figure}

The two DADAM classes with nematic alignment, namely the active nematics and the  self-propelled rods classes, display very similar properties.
We thus present them together. Again the phase diagram in the density/noise strength plane contains two binodal lines delimiting
the coexistence phase situated between the disordered gas and the ordered liquid (Fig.~\ref{FIG:NEMA2}a,b).

Nematic alignment favors local nematic order. At finite system size, for weak enough noise and/or large enough global density, 
i.e. in the liquid phase, global nematic order arises, with typically one half of particles moving in each of the two directions 
determined by the global nematic director (Fig.~\ref{FIG:NEMA}a)). 
No segregation into oppositely-going polar lanes is observed, even in the  self-propelled rods case where 
particles do not reverse their velocity\footnote{This is in contrast with models of `explicit' self-propelled hard rods, which often show 
nematic ``laning". Here again, the pointwise nature of the Vicsek particles amounts to the possibility of what would be numerous overlaps
in explicit rods models. Such overlaps can prevent laning and allow for the globally ordered nematic liquid to remain `mixed' \shortcite{shi2018self}. Note also that laning is observed in Vicsek-style rods when memory/inertia is added \shortcite{nagai2015collective}}.

%%%%%%%%%%%%%%%%%%%%%%%%%%%
%%%%%%%%%%%%%%%%%%%%%%%%%%%

%%%%%%%%%%%%%%%%%%%%%%%%%%%
\begin{figure}[h!]
	\centering
	\includegraphics[width=\textwidth]{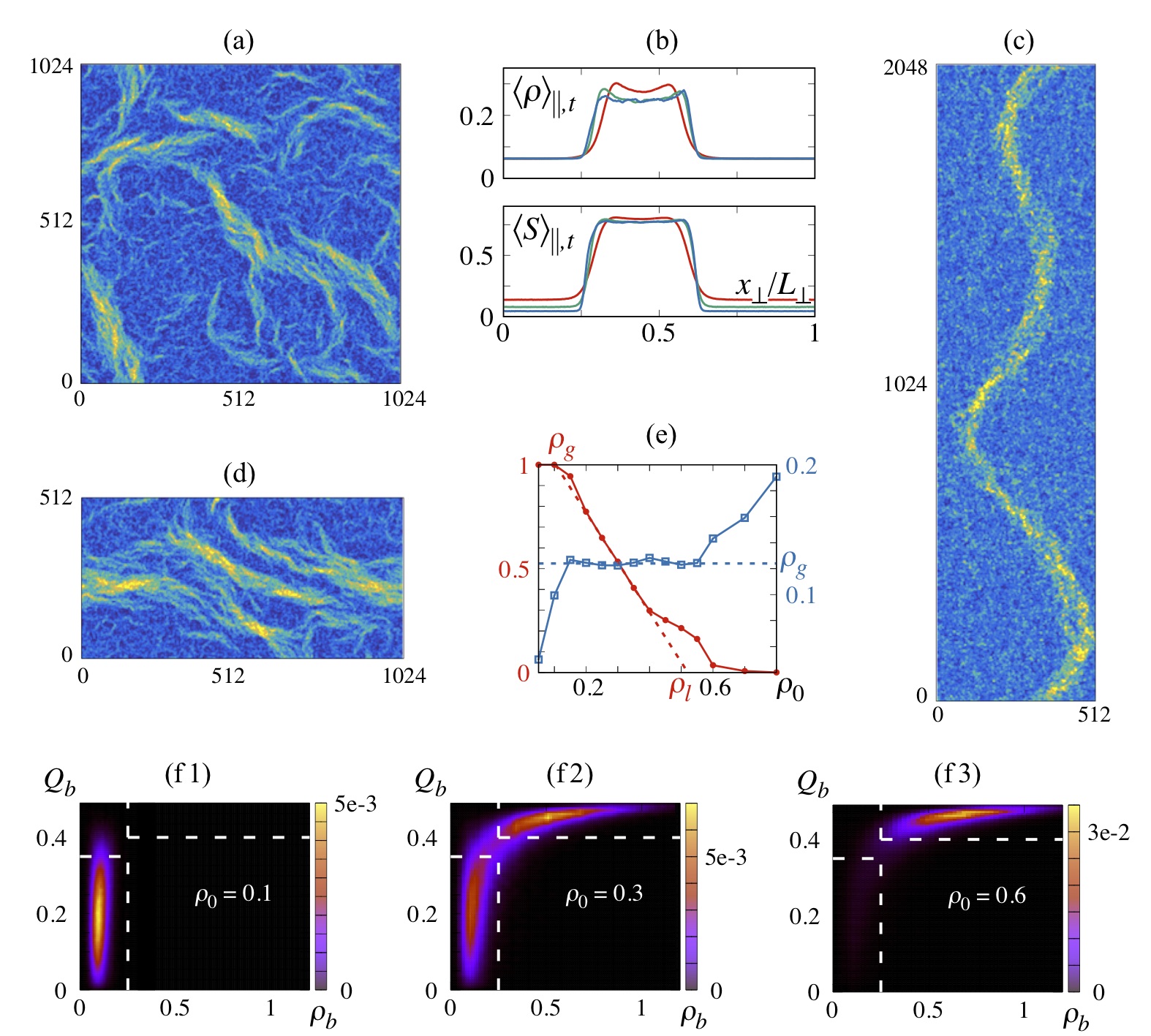}\\
	\caption{Coexistence phase of the Vicsek-style models with nematic alignment (as defined by Eqs.~(\ref{eqvm})).
	(a) Snapshot of the coarse grained density field in the band chaos regime ($\rho_0 = 0.2$, $\eta = 0.1$, $\alpha = 0.5$).
	(b) Density and nematic order profiles averaged over the direction longitudinal to the band and time in the self-propelled rods 
	model ($\alpha = 0$) as function of the rescaled system size. 
	The red, green and blue curves respectively correspond to linear system sizes $512$, $1024$ and $2048$ 
	($\rho = \frac{1}{8}$ and $\eta = 0.07$).
	(c): Same as (a) but showing the long-wavelength band instability for the rods case (
	$\alpha=0$, $\rho_0 = \frac{1}{2}$ and $\eta = 0.15$).
	(d): Same as (a) at same parameters but showing the transversal breakdown of the nematic band in the 
	fast-reversals active nematics.
	(e): Full red dots: gas fraction as function of the mean density and fixed noise $\eta = 0.1$. 
	Blue hollow squares: Gas density as function of $\rho_0$ showing a plateau in the coexistence phase.
	The red dashed line indicates the lever rule 
	from which the gas and liquid densities 
	($\rho_g$ and $\rho_l$) can be determined.
	(f): Probability distributions of density and nematic order computed in boxes of size $16 \times 16$ 
	at noise $\eta = 0.1$ in the gas (f1, $\rho_0 = 0.1$), coexistence (f2, $\rho_0 = 0.3$), and liquid (f3, $\rho_0 = 0.6$) phases.
	The white dashed lines mark the criteria for the definition of liquid and gas fractions.
	 }
	\label{FIG:NEMA3}
\end{figure}
%%%%%%%%%%%%%%%%%%%%%%%%%%%

The two nematic classes are only distinguished by the reversal rate $\alpha$. 
The major reported difference between the two classes concerns the nature of the long-range nematic order.
For active nematics with fast velocity reversals, one observes that the nematic order is only quasi-long-range:
the scalar nematic order parameter decays algebraically with system size, with typically a small decay exponent increasing 
continuously from zero when the noise strength is increased (Fig.~\ref{FIG:NEMA2}a,b). 
In contrast, without velocity reversals ($\alpha=0$), global nematic order appears truly long-range\footnote{For intermediate $\alpha$ values, one can observe a crossover to QLRO beyond a scale increasing when $\alpha\to 0$.} 
(Fig.~\ref{FIG:NEMA2}c)

Theoretical approaches (of the sort performed by Toner, Tu, Ramaswamy, et al.) predict asymptotic QLRO, and 
are currently unable to account for the true long-range nematic order observed for $\alpha=0$.
Caution is advised here: It could very well be that the difference is just a finite-size effect
and that quasi-long-range order is finally observed in the  self-propelled rods case at very large system sizes:
In the  self-propelled rods case, there is indeed a finite scale, 
independent of system size, which has been barely passed in numerical work: a given particle, in the nematic liquid phase, travels for some time along one of the two opposite global directions before collisions with other particles make it go along the other direction. 
These flight times are exponentially distributed, with a characteristic time typically in the thousands of timesteps. The corresponding lengthscale is indicated by the vertical dashed line in Fig.~\ref{FIG:NEMA2}c.
Simulations of systems much larger than this scale are needed. 
A recent publication~\shortcite{shankar2018low} also advocates a possibly very late crossover between LRO and QLRO behavior
at large system sizes.

The liquid phases of the two DADAM nematic classes also display long-range correlations and giant number fluctuations. 
At odds again with theoretical predictions, the scaling exponent of GNF in both nematic classes takes values $1.6-1.7$ very close to that observed in the polar class.
We note that in~\shortcite{shankar2018low} corrections to this scaling have been derived, however their amplitude remains too small to account for such numbers\footnote{It was found there that $\phi=2-\beta$, where $\beta$ is the decay exponent of nematic order with system size. Given the typically small values of $\beta$, this cannot account for the discrepancy between 2 and 1.6--1.7}.

The coexistence phase of the nematic DADAM classes is dominated by the emergence of high-density high-order bands, 
with the nematic direction along the band, spontaneously standing out of a sparse disordered gas 
(see Fig.~\ref{FIG:NEMA}b for an example). 
The nematic bands do not travel like their polar counterparts. 
Like in the polar class, one needs a large enough system to see these objects. In very large systems, many nematic bands
bend, stretch, split, and merge (Fig.~\ref{FIG:NEMA3}a). This is because the straight nematic band solution is in fact unstable.
It has been recently realized that this instability can take two different forms, depending on the velocity-reversal rate $\alpha$. 

For $\alpha$ small (including 0), the band is subjected to a long-wavelength instability along itself. 
In a system of moderate size,
one easily observes a single band occupying a growing fraction of space when the global density is increased, 
or a constant fraction of space when system size is increased at fixed parameters (Fig.~\ref{FIG:NEMA3}b). 
The surrounding gas remains roughly constant. One is thus, as announced, in the presence of 
a liquid-gas phase separation scenario. Taking the proportion of space occupied by the band as the liquid fraction, 
one can determine the binodal lines (corresponding to liquid fraction 0 and 1), which yields the phase diagram 
given in Fig.~\ref{FIG:NEMA2}b.
%The instability of the band is easiest to see near the gas binodal line ${\rm B}_{\rm gas}$, when it is relatively thin.
If one takes a single, stable band observed in a smallish system, and then collate many copies of this system to make an initially straight
long band, a long wavelength oscillation of the band develops, leading to its break-up, and finally to spatiotemporal chaos 
(Fig.~\ref{FIG:NEMA3}c). 

For fast velocity reversals (say $\alpha=0.5$), one can never observe a steady nematic band: 
it intermittently breaks up into many transversal pieces (Fig.~\ref{FIG:NEMA3}d) before eventually forming again. 
 The band interfaces are also rougher than in the small-$\alpha$ region.
All this makes it difficult to measure and 
even define the liquid fraction in the system.  One way out of this problem is to consider
coarse-grained density and nematic order fields, and build a 2D histogram of these field values: 
in the coexistence phase of interest here, one sees two well-separated regions that can be used to defining the gas and 
the nematic liquid and thus the liquid fraction (Fig.~\ref{FIG:NEMA3}f). Importantly, 
only the relative proportion of particles in each region varies
when the global density is increased. Using this methodology the phase diagram 
shown in Fig.~\ref{FIG:NEMA2}a can be built.
 
In all cases, the coexistence phase appears to be in fact a regime of spatiotemporal band chaos, with typically very large correlation 
lengths and times. Thus, the disorder-order transition, in the DADAM nematic classes, 
is relegated to the liquid binodal ${\rm B}_{\rm liq}$ separating the coexistence phase from the nematic liquid.

\section{Hydrodynamic theories for the 3 basic DADAM classes}
\label{hydro}

Numerous works on active matter are performed at the hydrodynamic level of a few partial differential equations (PDEs) 
governing the evolution of `slow' fields representing the large-scale dynamics of the system at stake. 
One can either consider stochastic or deterministic PDEs. Here, we mostly discuss deterministic hydrodynamic equations,
from which a lot can be learnt. However, it is clear that even though such deterministic PDEs do include some (mean-field) effects of 
the noise at play at the microscopic level, they cannot account for all fluctuations. In particular they cannot display
the anomalously strong fluctuations shown in the previous section to dominate the ordered liquid phases of DADAM systems.

There exist different ways to arrive at hydrodynamic equations. Here we first discuss this in general, 
before describing our favorite method, the Boltzmann-Ginzburg-Landau (BGL) approach, 
pioneered and developed by Eric Bertin and collaborators~\shortcite{bertin2006boltzmann,bertin2009hydrodynamic,peshkov2014boltzmann}. 
Finally, the rest of this section is devoted to studying the solutions of the hydrodynamic equations derived using the 
BGL method for the 3 basic classes of  DADAM.

\subsection{How to obtain hydrodynamic equations?}

Perhaps the easiest way to obtain hydrodynamic equations is to just write them down in the spirit of a 
gradient expansion (since one is interested in a large-scale description, 
only the lowest order terms should matter), identifying first the fields 
to be kept, then including all terms allowed by symmetries and conservation laws. 
In addition to these general principles, a good deal of experience/intuition and extra thinking is often needed, 
e.g. in deciding when to stop adding terms.

This is what Toner and Tu did in 1995 for the polar/Vicsek class. 
John Toner's contribution to this volume actually presents a detailed account of how to arrive 
at the Toner-Tu equations. These, superficially, look like the Navier-Stokes equations for a compressible fluid, governing 
a velocity field ${\bf v}$ coupled to the equation for the conservation of the local density field $\rho$:
\begin{subequations}
\begin{eqnarray}
& & \partial_t \rho + \nabla\cdot \left( \rho {\bf v} \right) = 0  \label{eqTT1} \\
& & \partial_t {\bf v} + \lambda_1 \left({\bf v}\cdot\nabla\right){\bf v} + \lambda_2\left(\nabla\cdot{\bf v}\right){\bf v}
+ \lambda_3 \nabla \left|{\bf v}\right|^2 = \left( \mu - \xi |{\bf v}|^2\right){\bf v} \nonumber \\
& & \quad\quad - \nabla P_1 - {\bf v}\left({\bf v}\cdot\nabla\right)P_2
+ D_0\Delta{\bf v} + D_1 \nabla\left(\nabla\cdot{\bf v}\right) + D_2\left({\bf v}\cdot\nabla\right)^2{\bf v}
+ {\bf f} \label{eqTT2}
\end{eqnarray}
\label{eqTT}
\end{subequations}
Note that these equations, describing situations with less symmetries and conservation laws than simple fluids, 
contain more terms than the Navier-Stokes equations. 
Whereas Eq.~(\ref{eqTT1}) expresses the advection of the conserved density by the velocity, Eq.~(\ref{eqTT2})
contains three `advection' terms with one $\nabla$ and two ${\bf v}$. Moreover,
the pressure ($P_1$ and $P_2$ terms) and the diffusion ($D_0$ and $D_1$ terms) are anisotropic
\footnote{As shown in the next sections, the $P_2$ and $D_2$ terms are usually absent of the same equations derived from kinetic theories. This is due to the fact that they are of higher oder in fields and gradient than the other terms. 
Moreover, it has been shown in~\shortcite{toner2012reanalysis} that they do not affect the scaling properties of the liquid phase. 
The main motivation for keeping them, as presented in J.Toner's contribution in this volume, 
is that they do not cancel in the linear theory.}.
The `potential' terms $\mu {\bf v}$ and $\xi |{\bf v}|^2 {\bf v}$ represent the small-scale injection 
and dissipation of energy leading to a
finite average speed $\bar{v}=\sqrt{\mu/\xi}$ when $\mu>0$.
Finally, the many `transport coefficients' ($\lambda_{1,2,3}$, $D_{0,1,2}$) are arbitrary 
and are in principle functions of $\rho$ and $|{\bf v}|$. Similarly, {\bf f} is an ad-hoc, additive, 
delta-correlated vectorial noise\footnote{Although we do not present it in details because this is not the purpose of these lecture notes, noise terms can also be derived from the microscopic dynamics~\shortcite{dean1996langevin,farrell2012pattern,bertin2013mesoscopic}. They are multiplicative in density and order, in contrast to the additive form considered by Toner and Tu, but the effects of these additional nonlinearities remain unexplored.}.

A major problem intrinsic to the above approach is that the connection to the microscopic level parameters is lost. 
How do the few parameters of the Vicsek model (or any model or experimental system in the same class) 
influence the many coefficients of the Toner-Tu equations? In the absence of such a connection, it is essentially
impossible to build phase diagrams corresponding to those established at the underlying level of microscopic models or experiments, and thus the general faithfulness of hydrodynamic equations cannot be assessed. 

Fortunately, methods exist which start from a given microscopic model and derive hydrodynamic equations from it. 
They provide a connection between microscopic parameters and hydrodynamic transport coefficients.
They have been applied in the context of DADAM, with various degrees of success.
Short of exact methods limited so far to very special classes of microscopic models \shortcite{kourbane2018exact}, 
all other approaches consist in building first a kinetic-level description of the problem at stake, 
and then derive hydrodynamic equations from the kinetic level. 
Most kinetic equations written for DADAM govern the one-body function $f({\bf r}, {\bf v}, t)$, i.e. the probability of finding particles
at position {\bf r}, with velocity {\bf v}, at time $t$.
To arrive at such an equation, it is necessary to factorize a multi-body function into a product of one-body functions, 
something legitimate under the strong ``molecular chaos" hypothesis that assumes decorrelation of particles' 
orientations from one interaction to the next.
Two avenues have been mostly followed in the context of dry active matter. The Smoluchowski/Fokker-Planck approach is probably
the most popular one, due to its `mean-field' character allowing a relatively easy implementation
\shortcite{degond2008continuum,baskaran2008enhanced,baskaran2008hydrodynamics,baskaran2010nonequilibrium,farrell2012pattern,romanczuk2012mean,grossmann2012active,grossmann2013self,barbaro2014phase}. 
The other approach consists in writing a Boltzmann equation, something that formally requires the assumption of diluteness, 
and is thus a priori suitable for DADAM~\shortcite{bertin2006boltzmann,bertin2009hydrodynamic,peshkov2014boltzmann}.

Note that in the context of aligning active matter, the molecular chaos hypothesis is a particularly strong assumption: indeed, particles 
that just interacted will remain aligned for some possibly long time, and thus remain correlated. 
As a result, quantitative agreement between the kinetic and microscopic levels is usually out of reach.
Given that further assumptions are needed to extract hydrodynamic equations from kinetic ones, the only reasonable goal
of hydrodynamic theories is to be qualitatively, possibly semi-quantitatively, faithful to the microscopic starting point. 

\subsection{Boltzmann-Ginzburg-Landau approach to DADAM}

\subsubsection*{Construction of the Boltzmann equation}

In the context of the Vicsek-style models representing the 3 basic DADAM classes, given the constant speed $v_0$ of particles, 
it is convenient to deal with the angular variable $\theta$ rather than with the velocity ${\bf v}=v_0 {\bf e}(\theta)$. 
Even though it is possible to treat discrete-time dynamics, we consider here the more traditional case of continuous time. 
To this end, we coarse-grain continuous time models, 
inspired by the discrete time dynamics defined by Eqs.~(\ref{eqvm}),
in which particles align their velocities in a Vicsek-style way and undergo tumbles at a finite rate $\lambda$.

Following \shortcite{peshkov2014boltzmann}, 
all 3 models are described by the following Boltzmann equation governing $f({\bf r},\theta,t)$:
\begin{eqnarray}
	 \partial_t f({\bf r},\theta,t) +  \bar{v}\, {\bf e}(\theta)\cdot \nabla f({\bf r},\theta,t) & = & D_0\Delta f({\bf r},\theta,t) + D_1q_{\alpha\beta}(\theta) \partial_\alpha\partial_\beta f({\bf r},\theta,t) \nonumber \\
	& & \!\!\!\! \!\!\!\! \!\!\!\! - a\left[f({\bf r},\theta,t) - f({\bf r},\theta+\pi,t)\right] + I_{{\rm sd}}[f]+I_{{\rm col}}[f]\,,
	\label{Boltzmann}
\end{eqnarray} 
where summation over repeated greek indices is assumed and 
$q_{\alpha\beta}(\theta) = e_{\alpha}(\theta) e_{\beta}(\theta)  - \frac{1}{2}$.

Eq.~(\ref{Boltzmann}) is nothing but a``master equation" counting gains and losses to $f({\bf r},\theta,t)$.
The first line corresponds to the general form of free motion dynamics: 
particles are moving at a mean speed $\bar{v}$ along their orientation, and they diffuse with isotropic diffusion constant $D_0$
and anisotropic diffusion constant $D_1$.
Coefficients $\bar{v}$, $D_0$ and $D_1$ are in general unknown, 
but can be computed for simple Vicsek-style microscopic dynamics.
Considering only the free motion of particles, the distribution $f$ obeys in general
\begin{equation}
f({\bf r},\theta,t + \Delta t) = \int_0^{2\pi} d\theta_v \, g(\theta_v - \theta) f({\bf r} - v_0 \, \Delta t \, {\bf e}(\theta_v),\theta,t)  \,,
\label{master_fm}
\end{equation}
where $g$ is a distribution of velocity angles $\theta_v$ coding motion along/against polarity.
Expanding the r.h.s.\ of Eq.~(\ref{master_fm}) to second order in $v_0\Delta t$ and sending $\Delta t \to 0$ we get
\begin{equation}
 \partial_t f({\bf r},\theta,t) = - v_0 \langle {\bf e}\rangle_v \cdot \nabla f({\bf r},\theta,t) + D_0 \Delta f({\bf r},\theta,t) 
+ D_1 \langle q_{\alpha\beta} \rangle_v \partial_\alpha \partial_\beta f({\bf r},\theta,t) \,,
\label{fm_general}
\end{equation}
where $\langle \ldots \rangle_v$ denotes the average over the distribution $g$, 
$D_0 = \lim_{\Delta t\to 0} v_0^2 \Delta t/4$, and $D_1 = 2D_0$.

In the polar Vicsek and self-propelled rods classes, $g^{\rm polar}(\theta_v - \theta) = \delta(\theta_v - \theta)$.
Therefore, the dominant term of the expansion is the drift at constant speed $v_0$ such that $\bar{v} = v_0$ and $D_0 = D_1 = 0$.
On the other hand, for active nematics particles that experience fast velocity reversals,
$g^{\rm apolar}(\theta_v - \theta) = \frac{1}{2}\left[ \delta(\theta_v - \theta) + \delta(\theta_v - \theta + \pi)\right]$.
That way the drift term in Eq.~(\ref{fm_general}) cancels and the dominant contribution to the free motion 
is given by the diffusion terms: $\bar{v} = 0$, $D_0 = \lim_{\Delta t\to 0} v_0^2\Delta t/4$ and $D_1 = 2D_0$.

The second line of Eq.~(\ref{Boltzmann}) describes the dynamics of particle orientations.
In the self-propelled rods class, 
particles can undergo velocity reversals at a slow rate $a$ without destroying the local order.
Finally the two last terms are the integrals accounting for angular self-diffusion and 
collision events, that we define now.
In order to account for the finite-time dynamics of Vicsek-style models,
the self-diffusion integral considers tumbling events occurring at a rate $\lambda$, 
thus
\begin{equation}
I_{\rm sd}[f] = -\lambda f({\bf r},\theta,t) + \lambda\int_0^{2\pi} d\theta' f({\bf r},\theta',t) P_\eta(\theta - \theta') \,,
\end{equation}
where $P_\eta$ is the angular noise distribution of variance $\eta^2$.
As for any Boltzmann approach, the collisional integral is derived under the binary collision assumption,
which states that in a dilute system collisions of more than two particles are rare and thus negligible\footnote{
We take the limit where the collisional mean free path is much larger than the radius
of interaction: $\rho_0^{-1/2} \gg r_0$.}.
For its derivation, we consider a particle $1$ located at position ${\bf r}$ with an orientation $\theta_1$
and compute the number of collisions $N_c$ per unit volume and unit time it experiences with
 a particle $2$ of orientation $\theta_2$.
This quantity is given by the collisional cross section
\begin{equation}
N_c = 2r_0|{\bf v}_{\rm rel}| f^{(2)}({\bf r},\theta_1,\theta_2,t) \,,
\end{equation}
where $r_0$ is the interaction radius,
$|{\bf v}_{\rm rel}| = v_0K(\theta_2-\theta_1)$ is the speed of particle $2$ in the reference frame of $1$,
%that is rotationally invariant, 
and $f^{(2)}$ is the two-body distribution.
In order to have a closed equation for the single-particle distribution, 
we now need to work under the molecular chaos hypothesis\footnote{
This approximation is supposedly valid when the typical flight distance $v_0/\lambda$ is much larger than
the radius of interaction $r_0$. However, as discussed earlier its validity needs to be nuanced due to the aligning 
nature of the interactions.}
which ensures that 
the states of particles $1$ and $2$ are uncorrelated prior to the collision, such that
 \begin{equation}
f^{(2)}({\bf r},\theta_1,\theta_2,t) = f({\bf r},\theta_1,t) f({\bf r},\theta_2,t)\,.
\end{equation}
The collisional integral gathers loss and gain terms that thus read 
\begin{eqnarray}
 I_{\rm col}[f] & = & - 2r_0v_0\int_{0}^{2\pi} d\theta' K(\theta'-\theta)  f({\bf r},\theta,t)  f({\bf r},\theta',t)  \nonumber \\
 & +& 2r_0v_0\! \! \int_{0}^{2\pi} \!\!\! d\theta_1 \int_{0}^{2\pi} \!\!\!d\theta_2 K(\theta_2 \!-\! \theta_1)f({\bf r},\theta_1,t) f({\bf r},\theta_2,t)P_\eta\left[\theta - \Psi(\theta_1,\theta_2)\right]  \label{Icoll}
\end{eqnarray}
where $\Psi(\theta_1,\theta_2)$ is the post collisional alignment rule
and the noise is drawn from the same distribution $P_\eta$ as in 
the self-diffusion integral\footnote{This is done only to simplify the presentation. In principle, the two noises can be chosen to be different, something probably necessary for some microscopic rules. In the context of the Vicsek-style models studied here, keeping just one noise distribution is fine.}.

Let us now specify the functions that enter in Eq.~(\ref{Icoll}) for the three models considered here.
In the polar/Vicsek and self-propelled rods classes, the kernel of interaction $K(\theta_2 - \theta_1)$ is simply given by 
\begin{subequations}
\begin{equation}
K^{\rm polar}(\theta_2 - \theta_1) = \left| {\bf e}(\theta_2) - {\bf e}(\theta_1) \right| 
= 2\left|\sin\left(\frac{\theta_2 - \theta_1}{2}\right)\right|\,,
\end{equation} 
while, because of velocity reversals, it takes a $\pi$-periodic form for active nematics:
\begin{eqnarray}
K^{\rm apolar}(\theta_2 - \theta_1) & = & \frac{1}{2}\left[ \left| {\bf e}(\theta_2) - {\bf e}(\theta_1) \right| + \left| {\bf e}(\theta_2) + {\bf e}(\theta_1) \right| \right] \nonumber \\
& = & \left|\sin\left(\frac{\theta_2 - \theta_1}{2}\right)\right| + \left|\cos\left(\frac{\theta_2 - \theta_1}{2}\right)\right|\,.
\end{eqnarray} 
\end{subequations}
The alignment rule $\Psi$, because of global rotational invariance, obeys $\Psi(\theta_1,\theta_2) = \theta_1 + H(\theta_2 - \theta_1)$.
In the case of polar (or ferromagnetic) alignment, the function $H$ thus reads
\begin{subequations}
\begin{equation}
H^{\rm ferro}(\Delta) = \frac{\Delta}{2} \quad \forall \Delta\in \; ( -\!\pi;\pi]\,,
\end{equation} 
and is $2\pi$-periodic.
On the contrary, if particles align with a nematic symmetry,
\begin{equation}
H^{\rm nema}(\Delta) = \frac{\Delta}{2} \quad \forall \Delta\in \,\, \left( -\!\frac{\pi}{2};\frac{\pi}{2} \right]\,,
\end{equation} 
\end{subequations}
and it is $\pi$-periodic.

For convenience, we now make the Boltzmann equation~(\ref{Boltzmann}) dimensionless by applying the rescalings
\begin{eqnarray*}
& & t \rightarrow \lambda^{-1} t \,, \\
& & \partial_{x,y} \rightarrow \lambda v_0^{-1} \partial_{x,y} \quad \mbox{(polar/Vicsek and self-propelled rods)} \,, \\
& & \partial_{x,y} \rightarrow  \sqrt{\frac{\lambda}{2D_0}} \partial_{x,y} \quad \mbox{(active nematics)} \,, \\
& & f({\bf r},\theta,t) \rightarrow \rho_0 f({\bf r},\theta,t) \,, 
\end{eqnarray*} 
such that the polar and active nematics classes have only two control parameters: 
the variance $\eta^2$ of the noise distribution and
the dimensionless density $\tilde{\rho}_0 \equiv 2r_0v_0\rho_0\lambda^{-1}$.
The self-propelled rods class shows an additional free parameter: the dimensionless reversal rate $\tilde{a} \equiv a\lambda^{-1}$.
In the following we will drop the tildes in order to lighten the notations.

\subsubsection*{Angular Fourier mode expansion}

Since the variable $\theta$ is $2\pi$-periodic, it is natural to expand the distribution $f$
in terms of angular Fourier modes,
\begin{equation}
f({\bf r},\theta,t) = \frac{1}{2\pi} \sum_{k=-\infty}^{\infty} f_k({\bf r},t)e^{-\imath k\theta} \,,
\end{equation}
with $f_{-k}({\bf r},t) = f_k^*({\bf r},t)$ (we use the $*$ superscript to denote complex conjugates).
The first modes correspond to the complex representations of the 
density $\rho$, momentum ${\bf w} = \rho{\bf v}$ and nematic ${\bf S} = \rho {\bf Q}$ fields:
\begin{subequations}
\begin{eqnarray}
f_0({\bf r},t) &=& \int_0^{2\pi} d\theta f({\bf r},\theta,t) = \rho({\bf r},t) \,, \\
f_1({\bf r},t) &=& \int_0^{2\pi} d\theta f({\bf r},\theta,t) e^{\imath \theta} = w_x({\bf r},t) + \imath w_y ({\bf r},t)\,, \\
f_2({\bf r},t) &=& \int_0^{2\pi} d\theta f({\bf r},\theta,t) e^{2\imath \theta} = 2\left[S_{xx}({\bf r},t) + \imath S_{xy} ({\bf r},t) \right]\,.
\end{eqnarray}
\end{subequations}
As we will see below, when working near the onset of orientational order we have smooth angular variations of $f({\bf r},\theta,t)$
and the modes $f_k$ beyond $f_1$ and $f_2$ quickly decay to zero as $k$ increases. 
This will be useful below to obtain hydrodynamic equations,
which are closed equations for the first few modes.

We now re-express the Boltzmann equation Eq.~(\ref{Boltzmann}) as a hierarchy of PDEs governing the evolution of the modes $f_k$. 
 In what follows and for simplicity, space and time dependencies of the functions will be implicit.
 The drift term in Eq.~(\ref{Boltzmann})  becomes
 \begin{eqnarray*}
% \int_0^{2\pi} d\theta \, e^{\imath k\theta} {\bf e}(\theta)\cdot \nabla f(\theta) = 
  \int_0^{2\pi} d\theta \, e^{\imath k\theta} \left(\begin{array}{c} \cos(\theta) \\ \sin(\theta) \end{array} \right) \cdot
 \left(\begin{array}{c} \partial_x \\ \partial_y \end{array} \right) f(\theta) & = & 
 \frac{1}{2} \left(\begin{array}{c} \partial_x \\ \partial_y \end{array} \right) \cdot 
 \left(\begin{array}{c} f_{k+1} + f_{k-1} \\ -\imath( f_{k+1} - f_{k-1} ) \end{array} \right) \\
 &=& \frac{1}{2}\left( \triangledown^* f_{k+1} + \triangledown f_{k-1} \right) \,,
 \end{eqnarray*}
 where we have used the complex gradient $\triangledown = \partial_x + \imath \partial_y$.
 
 The Laplacian operator in~(\ref{Boltzmann}) does not depend on $\theta$, 
 thus $\Delta f(\theta)$ simply transforms as $\Delta f_k$ with $\Delta = \triangledown\triangledown^*$.
 The anisotropic diffusion, on the other hand, becomes
 \begin{eqnarray*}
 & &  \int_0^{2\pi} d\theta \, e^{\imath k\theta} {{\rm Tr}}\left[ \left(\begin{array}{c c} \cos^2(\theta) - \frac{1}{2} & \cos(\theta)\sin(\theta) \\
  \cos(\theta)\sin(\theta) & \sin^2(\theta) - \frac{1}{2} \end{array} \right) 
   \left(\begin{array}{c c} \partial_{xx}^2 & \partial_{xy}^2 \\
 \partial_{xy}^2 & \partial_{yy}^2 \end{array} \right) \right] f(\theta) \\
& &\quad\quad\quad\quad\quad\quad\quad\quad = \frac{1}{2} \int_0^{2\pi} d\theta \, e^{\imath k\theta} {{\rm Tr}}\left[ \left(\begin{array}{c c} \cos(2\theta) & \sin(2\theta) \\
 \sin(2\theta) & -\cos(2\theta) \end{array} \right) 
   \left(\begin{array}{c c} \partial_{xx}^2 & \partial_{xy}^2 \\
 \partial_{xy}^2 & \partial_{yy}^2 \end{array} \right) \right] f(\theta) \\
& &\quad\quad\quad\quad\quad\quad\quad\quad = \frac{1}{4} {{\rm Tr}}\left[ \left(\begin{array}{c c} \partial_{xx}^2 & \partial_{xy}^2 \\
 \partial_{xy}^2 & \partial_{yy}^2 \end{array} \right) 
 \left(\begin{array}{c c} f_{k+2} + f_{k-2} & -\imath( f_{k+2} - f_{k-2} ) \\
  -\imath( f_{k+2} - f_{k-2} ) & -(f_{k+2} + f_{k-2}) \end{array} \right) \right] \\
& &\quad\quad\quad\quad\quad\quad\quad\quad =  \frac{1}{4} \left( {\triangledown^*}^2 f_{k+2} + \triangledown^2 f_{k-2} \right) \,,
 \end{eqnarray*}
 where ${\rm Tr}$ denotes the trace operator.
 
 On the r.h.s.\ of Eq.~(\ref{Boltzmann}), the exchange term modeling slow velocity reversals
 takes the form of a damping of odd modes:
 \begin{equation*}
- \int_0^{2\pi} d\theta \, e^{\imath k\theta} \left[ f(\theta) - f(\theta+\pi) \right] = -\left[1 + (-1)^{k+1}\right] f_k \,,
  \end{equation*}
while the self-diffusion integral, 
whose gain part is given by the convolution product of $f$ and the noise distribution $P_\eta$, 
simply becomes
  \begin{equation*}
\int_0^{2\pi} d\theta \, e^{\imath k\theta} I_{\rm sd}[f] = (P_k - 1)f_k \,,
  \end{equation*}
  with $P_k = \int_{-\infty}^\infty d\sigma P_\eta(\sigma) \exp{(ik\sigma)}$ is the $k^{\rm th}$ mode of $P_\eta$.

Finally, we derive the expression of the Fourier transform of the gain term of the collisional integral.
We first expand $P$ and $f(\theta_2)$ and apply the change of variable $\theta_2 \rightarrow \Delta + \theta_1$,
hence
 \begin{eqnarray*}
& &  \int_0^{2\pi} d\theta \, e^{\imath k\theta}  \int_{0}^{2\pi} d\theta_1 \int_{0}^{2\pi} d\theta_2 K(\theta_2-\theta_1)f(\theta_1) f(\theta_2)P_\eta\left[\theta - \theta_1 - H(\theta_2 - \theta_1))\right] \\ 
& & = \frac{1}{(2\pi)^2} \sum_{q,l} P_l f_q \int_0^{2\pi} d\theta \, e^{\imath (k-l)\theta} 
 \int_{0}^{2\pi} d\theta_1 \, e^{\imath(l-q)\theta_1} f(\theta_1)
 \int_{0}^{2\pi} d\Delta \, K(\Delta) e^{\imath[-q\Delta + lH(\Delta)]} \,,
 \end{eqnarray*}
 which, after evaluating the first two integrals, leads to
  \begin{eqnarray*}
& &  \int_0^{2\pi} d\theta \, e^{\imath k\theta}  \int_{0}^{2\pi} d\theta_1 \int_{0}^{2\pi} d\theta_2 K(\theta_2-\theta_1)f(\theta_1) f(\theta_2)P_\eta\left[\theta - \theta_1 - H(\theta_2 - \theta_1))\right] \\ 
& & \quad\quad\quad\quad\quad\quad\quad\quad\quad\quad\quad\quad\quad\quad\quad\quad = P_k\sum_{q}I_{k,q} f_q f_{k-q} \,,
  \end{eqnarray*}
with the mode coupling function 
  \begin{equation}
  I_{k,q} = \frac{1}{2\pi}  \int_{0}^{2\pi} d\Delta \, K(\Delta) e^{\imath[-q\Delta + kH(\Delta)]} \,.
  \end{equation}
The corresponding loss part can be computed in a similar way.  
 After collecting all the terms, we obtain the following hierarchy
    \begin{eqnarray}
 \partial_t f_k &+& \frac{1}{2}\left( \triangledown^* f_{k+1} + \triangledown f_{k-1} \right)  =  
\frac{1}{2}\Delta f_k + \frac{1}{4} \left( \triangledown^{2^*} f_{k+2} + \triangledown^2 f_{k-2} \right)  \nonumber \\
& & - a\left[1 + (-1)^{k+1}\right] f_k+  (P_k - 1)f_k + \rho_0\sum_{q} \left(P_kI_{k,q} - I_{0,q}\right) f_q f_{k-q} \,.
\label{fourier_hierarchy}
  \end{eqnarray}
  
  In the next sections we detail the procedure used to truncate and close this hierarchy
  in a controlled way, leading to hydrodynamic equations for the three classes.

\subsubsection*{Polar/Vicsek class}

For this case, the hierarchy reads
\begin{equation}
 \partial_t f_k + \frac{1}{2}\left( \triangledown^* f_{k+1} + \triangledown f_{k-1} \right) = 
 (P_k - 1)f_k + \sum_{q} J^{\rm polar}_{k,q} f_q f_{k-q} \,,
 \label{hierarchy_polar}
\end{equation}
with $J^{\rm polar}_{k,q} = \rho_0(P_k I^{\rm polar}_{k,q} - I^{\rm polar}_{0,q})$ and the coefficients
\begin{equation}
I^{\rm polar}_{k,q} = \left\{  \begin{array}{c} \frac{4}{\pi}\frac{ 1 - (k-2q)(-1)^{q}\sin\left(\frac{k\pi}{2}\right)}{ 1 - (k-2q)^2}  \quad {\rm if } \; \vert k-2q \vert \ne 1 \\ 
						\frac{2}{\pi} \quad {\rm otherwise} \end{array} \right. \,.
\end{equation}
Eq.~(\ref{hierarchy_polar}) admits the trivial disordered solution $\rho = 1$ and $f_{k>0} = 0$.
The linear stability of this solution to space-independent perturbations is given by 
\begin{equation}
 \partial_t \delta f_k  = \left[P_k - 1 + J_{k,0} + J_{k,k}\right] \delta f_k
\equiv \mu_k  \delta f_k \,,
\end{equation}
such that it will remain stable whenever all $\mu_k$'s are negative and unstable if
at least one of them is positive.
Computing these coefficients, it can be shown that only 
$\mu_1 = P_1 - 1 + \frac{4}{\pi}\left(P_1 - \frac{2}{3}\right)\rho_0$ can become positive 
at low noises and large densities.
Below the line $\mu_1 = 0$ in the $(\rho_0,\eta)$ plane (see Fig.~\ref{FIG:LINSTAB}), 
the polar order $|f_1|$ thus grows, 
and close to the transition we can thus assume that $|f_1| \approx \varepsilon$ with $\varepsilon \ll 1$.
In order to compute the associated scaling of the higher modes, 
let us write explicitly the first equations of the hierarchy,
 \begin{eqnarray*}
 \partial_t \rho & = & -\Re\left(\triangledown^* f_1\right) \,, \\
 \partial_t f_1 & = & -\frac{1}{2}\left( \triangledown^* f_2 + \triangledown \rho \right) + \mu_1[\rho] f_1 + (J_{1,2} + J_{1,-1})f_1^* f_2 + \ldots \,, \\
 \partial_t f_2 & = & -\frac{1}{2}\left( \triangledown^* f_3 + \triangledown f_1 \right) + \mu_2[\rho] f_2 + J_{2,1} f_1^2 + \ldots \,, \\
 \ldots \,, \\
 \partial_t f_k & = & -\frac{1}{2}\left( \triangledown^* f_{k+1} + \triangledown f_{k-1} \right) + \mu_k[\rho] f_k +   
(J_{k,1} + J_{k,k-1})f_1 f_{k-1} + \ldots \,, \\
 \ldots \,.
  \end{eqnarray*}
  Each $k>1$ mode takes non-zero values due to its nonlinear coupling with $f_1$ but is maintained small
     because of the negative linear coefficients $\mu_k$.
  From the above equations, we thus have $|f_2| \approx |f_1|^2$, ..., $|f_k| \approx |f_1 f_{k-1}|$ $\forall k > 1$
  such that $|f_k| \approx |f_1|^{k} \approx \varepsilon^k$ $\forall k \ge 1$ close to the 
  transition\footnote{This is easily confirmed
  by calculating the homogeneous-order solution of the Boltzmann equation.}.
  Moreover, the continuity equation imposes that $|\partial_t \rho| \approx |\triangledown^* f_1|$ 
  while the pressure in the polar field equation gives $|\partial_t f_1| \approx |\triangledown\rho|$.
  Therefore, $|\delta \rho| = |\rho - 1| \approx |f_1| \approx \varepsilon$.
  We then immediately obtain from the mass conservation that $\partial_t \approx \triangledown \approx \varepsilon^\alpha$,
  and finally $\alpha = 1$ is determined compensating $|\triangledown^* f_2| \approx |f_1^*f_2|$ in $f_1$'s equation. 
  
  Thus, there is a unique {\it scaling ansatz} that allows one to attribute an $\varepsilon$ order to each term of the hierarchy.
   Truncating it at the first nontrivial order $\varepsilon^3$, we obtain equations for $\rho$, $f_1$ and $f_2$.
  Moreover, in the $f_2$ equation, the $\partial_t f_2$ term is negligible compared to the leading order terms,
  such that $f_2$ can be enslaved:
  \begin{equation}
  f_2 = \frac{1}{\mu_2}\left[ \frac{1}{2}\triangledown f_1 - J_{2,1} f_1^2\right] \,.
  \end{equation}
  Replacing $f_2$ by this expression in the $f_1$ equation we finally get the closed hydrodynamic
  equations for the polar/Vicsek class
 \begin{subequations}
    \begin{eqnarray}
 \partial_t \rho & = & -\Re\left(\triangledown^* f_1\right) \,, \\
 \partial_t f_1 & = & -\frac{1}{2}\triangledown \rho + \left(\mu_1[\rho] - \xi |f_1|^2\right) f_1 + \nu \Delta f_1 
- \kappa_1 f_1 \triangledown^* f_1 - \kappa_2 f_1^* \triangledown f_1  \,, 
  \end{eqnarray}
  \label{hydro_polar}
\end{subequations}
with 
\begin{align*}
\mu_1[\rho] = & P_1 - 1 + \frac{4}{\pi}\left(P_1 - \frac{2}{3}\right)\rho_0\rho \,, &
\mu_2 = &P_2 - 1 - \frac{8}{15\pi}(7 + 5P_2)\rho_0 \,, \\
\xi = & -\frac{16(5P_1-2)(3P_2+1)\rho_0^2}{15\pi^2 \mu_2} \,, &
\nu = &- \frac{1}{4\mu_2} \,, \\
\kappa_1 = &-\frac{4(1+3P_2)\rho_0}{3\pi \mu_2} \,, &
\kappa_2 = &\frac{2(5P_1-2)\rho_0}{5\pi \mu_2} \,.
 \end{align*}
  
  \subsubsection*{Active nematics class}
  
  In this case particles reverse their velocities at a fast rate, 
  therefore the Boltzmann hierarchy only shows spatial diffusion.
  Moreover, since the problem has a full nematic symmetry, all the odd modes of the distribution $f$ can be set to zero
  \shortcite{bertin2013mesoscopic,ngo2014large}. One is left with:
  \begin{equation}
 \partial_t f_{2k}  = \frac{1}{2}\Delta f_{2k} + \frac{1}{4} ( {{\triangledown^*}^2} f_{2(k+1)} + \triangledown^2 f_{2(k-1)} ) 
+  (P_{2k} - 1)f_{2k} + \sum_{q} J^{\rm an}_{k,q} f_{2q} f_{2(k-q)} \,,
 \label{hierarchy_an}
\end{equation}
with $J^{\rm an}_{k,q} = \rho_0(P_{2k} I^{\rm an}_{k,q} - I^{\rm an}_{0,q})$ and the coefficients
\begin{equation}
I^{\rm an}_{k,q} = \frac{4}{\pi}\,\frac{[1 - 2\sqrt{2}(k-2q)(-1)^q\sin\left(\frac{k\pi}{2}\right)]}{1 - 4(k-2q)^2} \,.
\end{equation}
  From the symmetry of the alignment, the field responsible for the transition from disorder to nematic order
  is now the second mode of the distribution $f_2$.
  Indeed, the only linear coefficient that can be positive (at low noise and large densities) is now 
  $\mu_2 = P_2 - 1 + \frac{8}{15\pi}[5(2\sqrt{2}-1)P_2 - 7]\rho_0$.
  Using similar arguments as for the polar/Vicsek class, we can arrive at the following unique scaling ansatz close to 
  the onset of nematic order
  \begin{equation}
|f_{2k}| \approx \varepsilon^{k} \; \forall k > 0 \,, \; |\delta \rho| \approx \varepsilon \,, \; \triangledown^2 \approx \partial_t \approx \varepsilon^2 \,,
\end{equation}
such that space and time now scale diffusively.
Truncating the hierarchy~(\ref{hierarchy_an}) up to order $\varepsilon^3$, 
we get equations for $\rho$, $f_2$ and $f_4$. 
Similarly to the polar class, $f_4$ can be enslaved to $f_2$,
 which leads to the following hydrodynamic equations
 \begin{subequations}
    \begin{eqnarray}
 \partial_t \rho & = & \frac{1}{2} \Delta\rho + \frac{1}{2}\Re\left({{\triangledown^*}^2} f_2 \right) \,,  \label{hydro_an1}\\
 \partial_t f_2 & = &  \left(\mu_2[\rho] - \xi |f_2|^2\right) f_2 + \frac{1}{2} \Delta f_2 + \frac{1}{4}\triangledown^2\rho \,,  \label{hydro_an2}
  \end{eqnarray}
  \label{hydro_an}
  \end{subequations}
with 
    \begin{eqnarray*}
\mu_2[\rho] = P_2 - 1 + \frac{8}{15\pi}\,[5(2\sqrt{2}-1)P_2 - 7]\, \rho_0\, \rho \,, \\
\xi = \frac{32}{15\pi} \, \frac{[9(1 + 6\sqrt{2})P_2-13](15P_4 + 1)\rho_0^2}{ [315\pi(1-P_4)+ 8(155+21P_4)\rho_0]} \,.
  \end{eqnarray*}
  
    \subsubsection*{Self-propelled rods class} 
    
    This class also shows nematic alignment, 
    but no or slow velocity reversals.
    The free motion in the Boltzmann hierarchy thus takes the form of a drift at constant speed
    \footnote{ It can be shown that the active nematics hierarchy~(\ref{hierarchy_an}) 
    is recovered in the limit of large reversal rate $a$
    imposing $v_0^2 \sim a$~\shortcite{3D_hydro}.}
      \begin{equation}
 \partial_t f_k + \frac{1}{2}\left( \triangledown^* f_{k+1} + \triangledown f_{k-1} \right) = 
 - a\left[1 + (-1)^{k+1}\right] f_k \nonumber +  (P_k - 1)f_k + \sum_{q} J^{\rm rods}_{k,q} f_q f_{k-q} \,,
\label{hierarchy_rods}
  \end{equation}
with $J^{\rm rods}_{k,q} = \rho_0(P_{k} I^{\rm rods}_{k,q} - I^{\rm rods}_{0,q})$ and
\begin{equation}
 I^{\rm rods}_{k,q} = \left\{\begin{array}{c c c} \frac{4}{\pi}\frac{2 + \sqrt{2}\left[\cos\left(\frac{\pi(k-2q)}{4}\right) + (k-2q)\sin\left(\frac{\pi(k-2q)}{4}\right)\right]\left[\cos\left(\frac{k\pi}{2}\right) - 1 \right]}{1 - (k-2q)^2} & & \;\;\; {\rm if}\;\; \vert k - 2q\vert \ne 1 \\
 \frac{2}{\pi} & & {\rm otherwise} \,. \end{array}\right.
\label{eq:Boltzmann_rods_Fourier}
\end{equation}
The ordering field in this case remains $f_2$
as witnessed by the associated linear coefficient $\mu_2 = P_2 - 1 + \frac{16}{15\pi}\,[5(2\sqrt{2} - 1)P_2 - 7] \rho_0$
which becomes positive at large densities and low noises.
From the symmetry of the motion, the scaling ansatz necessary to truncate the hierarchy
for this class is then derived assuming a ballistic scaling for space and time: $\triangledown \approx \partial_t$.
It is unique and reads~\shortcite{peshkov2014boltzmann,peshkov2012nonlinear}
\begin{equation}
|f_{2k-1}| \approx |f_{2k}| \approx \varepsilon^{k} \; \forall k > 0 \,, \; |\delta \rho| \approx \varepsilon \,, \; \triangledown \approx \partial_t \approx \varepsilon \,.
\end{equation}
The hierarchy~(\ref{hierarchy_rods}) truncated at order $\varepsilon^3$ thus includes equations for 
$\rho$, $f_1$, $f_2$, $f_3$ and $f_{4}$.
Enslaving the two highest modes to the other, 
we finally get the rather complicated equations
  \begin{subequations}
    \begin{eqnarray}
 \partial_t \rho & = & -\Re\left(\triangledown^* f_1\right) \,,   \label{hydro_rods1}\\
 \partial_t f_1 & = & -\frac{1}{2}\left(\triangledown^* f_2 + \triangledown \rho\right) + \left(\mu_1[\rho] + \beta |f_2|^2\right) f_1 
+ \zeta f_1^*f_2 + \gamma f_2^*\triangledown f_2 \,,  \label{hydro_rods2}\\
 \partial_t f_2 & = & -\frac{1}{2}\triangledown f_1 + \left(\mu_2[\rho] - \xi |f_2|^2\right) f_2 + \omega f_1^2 + \tau |f_1|^2f_2 + \nu \Delta f_2
\nonumber \\
& & \quad\quad\quad\quad\quad\quad\quad\quad\quad\quad\quad\quad\quad\quad\quad
  + \kappa_1 f_1^*\triangledown f_2 + \kappa_2 \triangledown^*\left(f_1f_2\right) \,,  \label{hydro_rods3} 
  \end{eqnarray}
  \label{hydro_rods}
\end{subequations}
with 
\begin{align*}
 \mu_1[\rho] = &P_1 - 1 -2a + \frac{4}{\pi}\left(P_1 - \frac{4}{3}\right)\rho_0\rho \,, &
\beta = &-\frac{32(7P_1-2)(4 + 5P_3)\rho_0^2}{3\pi(35\pi(1-P_3 + 2a)+272\rho_0)}\,, \\
 \xi = &\frac{128(9(1+6\sqrt{2})P_2-13)(1+15P_4)\rho_0^2}{15\pi(315\pi(1-P_4)+16(155+21P_4)\rho_0)} \,, \;&
 \zeta = &\frac{16\rho_0}{5\pi}    \,, \\
 \mu_2[\rho] = &P_2 - 1 + \frac{16}{15\pi}\left(5(2\sqrt{2}-1)P_2 - 7\right)\rho_0\rho \,, \quad\quad&
\gamma = &\frac{20(7P_1-2)\rho_0}{3(35\pi(1-P_3 + 2a)+272\rho_0)} \,, \\
 \omega = &\frac{8}{3\pi}\left(1-3(\sqrt{2}-1)P_2\right)\rho_0\,, \;&
\tau = &\frac{64(7(1+\sqrt{2})P_2-19)(4+5P_3)\rho_0^2}{15\pi(35\pi(1-P_3 + 2a)+272\rho_0)} \,, \\
 \nu = &\frac{35\pi}{4(35\pi(1-P_3 + 2a)+272\rho_0)}\,, \;&
\kappa_1 = &\frac{8(7(1+\sqrt{2})P_2-19)\rho_0}{3(35\pi(1-P_3 + 2a)+272\rho_0)} \,, \\
 \kappa_2 = &-\frac{14(4+5P_3)\rho_0}{35\pi(1-P_3 + 2a)+272\rho_0} \,.
  \end{align*}

\subsection{Comments on the structure of the obtained equations}

First of all, as promised earlier, all transport coefficients of 
Eqs.~(\ref{hydro_polar}),~(\ref{hydro_an}) and~(\ref{hydro_rods})  depend explicitly on 
microscopic parameters (Fourier coefficients $P_k$ of the noise distribution, global density $\rho_0$,
explicit reversal rate $a$), so that
it is possible to study these PDEs by varying the same parameters as in the previous section, 
to confront their solutions to those observed at the microscopic level,
and to compare hydrodynamic-level and microscopic-level phase diagrams (the topics of the rest of this section).

Next, these equations are simpler and contain less terms than those written phenomenologically:
\begin{itemize}
\item Only the coefficients $\mu_1[\rho]$ and $\mu_2[\rho]$ of the linear terms in $f_1$ and $f_2$ depend 
explicitly on the {\it local} density $\rho$.
\item Some coefficients take simple numerical values because of the dimensionless form of the Boltzmann equation. 
     Note that it is thus impossible, in our context, to give them arbitrary values.
\item The `pressure' terms $\triangledown\rho$ and $\triangledown^2\rho$ 
in the order-fields equations are ``as simple as possible".
\item The density-conservation equation, in the classes with polar particles (polar/Vicsek class,  self-propelled rods), 
does not contain a diffusion term because the Vicsek-style models do not include positional diffusion of particles.
\item Diffusion in the ordering-field equations is isotropic, an effect of the 
pointwise nature of particles \footnote{It is only when repulsive interactions are taken into account that anisotropic diffusion arises, see, {\it e.g.}, \shortcite{dense-active-nema}.}.
\end{itemize}

Next, these equations are naturally written in terms of the $f_1$ and $f_2$ fields, 
which are not velocity/polarity and nematic tensor fields. Rather, $f_1$ represents ${\bf w}=\rho{\bf v}$,
and $f_2$ codes for ${\bf S}=\rho{\bf Q}$.
This is the outcome of any method where hydrodynamic equations are obtained by truncation/closure of a kinetic equation expressed in Fourier modes of the angular variables. 
Of course, equations for {\bf v} and {\bf Q} at order $\varepsilon^3$ can be obtained from our $f_1$ and $f_2$ equations.
This usually comes at the price of slightly more complicated pressure terms.

A point of particular interest here is the status of the equations derived above for the polar, Vicsek class with respect to
those written by Toner and Tu. 
Equations~(\ref{hydro_polar}) obtained at order $\varepsilon^3$ 
from the BGL method for the polar Vicsek class, can be easily rewritten 
in the more conventional vectorial notation:
\begin{subequations}
\begin{eqnarray}
& & \partial_t \rho + \nabla\cdot{\bf w} = 0  \label{eqpolar1v} \\
& & \partial_t {\bf w} + \lambda_1 \left({\bf w}\cdot\nabla\right){\bf w} + \lambda_2\left(\nabla\cdot{\bf w}\right){\bf w}
+ \lambda_3 \nabla |{\bf w}|^2 = - \frac{1}{2}\nabla\rho 
+ \left(\mu_1[\rho] - \xi|{\bf w}|^2\right){\bf w} + \nu\Delta{\bf w} \nonumber \\    \label{eqpolar2v} 
\end{eqnarray}
\end{subequations}
where ${\bf w}=\rho {\bf v} =({\rm Re}f_1,{\rm Im}f_1)$ is a momentum field,
$\lambda_1 = \kappa_1 + \kappa_2$, and $\lambda_2 = -2\lambda_3 = \kappa_1 - \kappa_2$.
%%%XXX change to eq in terms of polarity field p? 
%%---> We need to choose... In RMP MCM use p for v, I don't know which notation is more coherent with the literature.
As mentioned above, Eq.~(\ref{eqpolar2v}) contains less terms than Eq.~(\ref{eqTT2}): diffusion and pressure are isotropic, 
and moreover the pressure term is simply $\nabla\rho$. Let us note also that here the $\lambda_2$  and $\lambda_3$ coefficients
(in the notation of Toner and Tu)
are related: $\lambda_2= -2 \lambda_3$\footnote{
This is  a  relation satisfied if these two terms are derived from
a contribution $\int{\bf P}^2\nabla\cdot {\bf P}$ to a scalar free-energy functional governing dynamics at equilibrium 
%(where {\bf p} is an orientation vector)
\shortcite{marchetti2013hydrodynamics}. Note that this relation is not satisfied in the equivalent equations derived in 3D \shortcite{3D_hydro}}. 

Equations \eqref{hydro_an} obtained for active nematics are particularly simple when expressed in terms of $\rho$ and $f_2$. 
In tensorial notation, they read:
\begin{subequations}
\begin{eqnarray}
& &\partial_t \rho = \frac{1}{2}\Delta\rho + {\rm Tr}\left[\mathbf{\Gamma} {\bf S} \right]  \label{eqnema1Q} \\
& &\partial_t {\bf S} = \left(\mu_2[\rho] - 2\xi{\rm Tr}\left[{\bf S}^2\right] \right){\bf S}  
+ \frac{1}{2}\Delta{\bf S} + \frac{1}{4}\mathbf{\Gamma}\rho \label{eqnema2Q} 
\end{eqnarray}
\end{subequations}
where ${\bf S} = \rho{\bf Q}$ and $\Gamma_{ij} = \partial_i\partial_j - \frac{\delta_{ij}}{2}\Delta$.
The $\Re\left({\triangledown^*}^2 f_2\right)$ term of the conservation equation 
now reads ${\rm Tr}\left[\mathbf{\Gamma} {\bf S} \right]$, 
i.e. it is  the all-important ``nonlinear current" first written down by Ramaswamy and Simha~\shortcite{ramaswamy2003active}
and predicted to be at the origin of strong giant number fluctuations in active nematics.

In tensorial notations, the hydrodynamic equations \eqref{hydro_rods}  for  self-propelled rods  read
\begin{subequations}
\begin{eqnarray}
 & & \!\!\!\!\!\!\!\!\!\!\!\!\partial_t \rho + \partial_k w_k = 0\label{eqrods1Q} \,, \\
 & & \!\!\!\!\!\!\!\!\!\!\!\! \partial_t w_i = -\frac{1}{2}\partial_i \rho - \partial_k S_{ik} + \left(\mu_1[\rho] + 4\beta S_{kl}^2\right)w_i + 2\zeta w_kS_{ik} + 4\gamma S_{kl}\partial_l S_{ik} \label{eqrods2Q} \,,\\
& & \!\!\!\!\!\!\!\!\!\!\!\! \partial_t S_{ij} = -\frac{1}{2}\left[\partial_i w_j\right]_{\rm ST} + \left(\mu_2[\rho] - 2\xi S_{kl}^2\right)S_{ij} 
+ \omega \left[w_i w_j\right]_{\rm ST} + \tau w_k^2 S_{ij} + \nu\Delta S_{ij} \nonumber \\
& & \!\!\!\!\!\!\!\!\!\!\!\! + \kappa_1\left[ w_k\partial_k S_{ij} + (w_k\partial_i - w_i \partial_k) S_{kj} \right]
+ \kappa_2\left[ \partial_k( w_k S_{ij} + w_iS_{kj} ) - \partial_i(w_k S_{kj})\right] \,, \label{eqrods3Q}
\end{eqnarray}
\end{subequations}
where summation over repeated indices is assumed and we have kept explicit notations
for tensorial and inner products in order to avoid possible confusions between similar terms. 
The symmetric traceless part of the tensor ${\bf A}$ is defined by $[A_{ij}]_{\rm ST} = \frac{1}{2}\left(A_{ij} + A_{ji} - {\rm Tr}[{\bf A}]\delta_{ij} \right)$ with $\delta_{ij}$ being the Kronecker-delta. 
Because in this case the motion is propagative, 
the momentum field {\bf w} has been retained in the hydrodynamic description and the resulting equations
exhibit more nonlinearities than those describing active nematics.
As we will discuss later, both sets of equations for classes with nematic alignment show similar qualitative 
features at the deterministic level, 
their respective behavior at fluctuating level remains however unexplored.

\subsection{Properties common to all 3 classes}
\label{properties_classes_hydro}

The sets of hydrodynamic equations obtained above using the BGL approach share many properties, leading to the same
overall structure of their phase diagrams.

\begin{figure}[h!]
	\centering
	\includegraphics[scale=0.6]{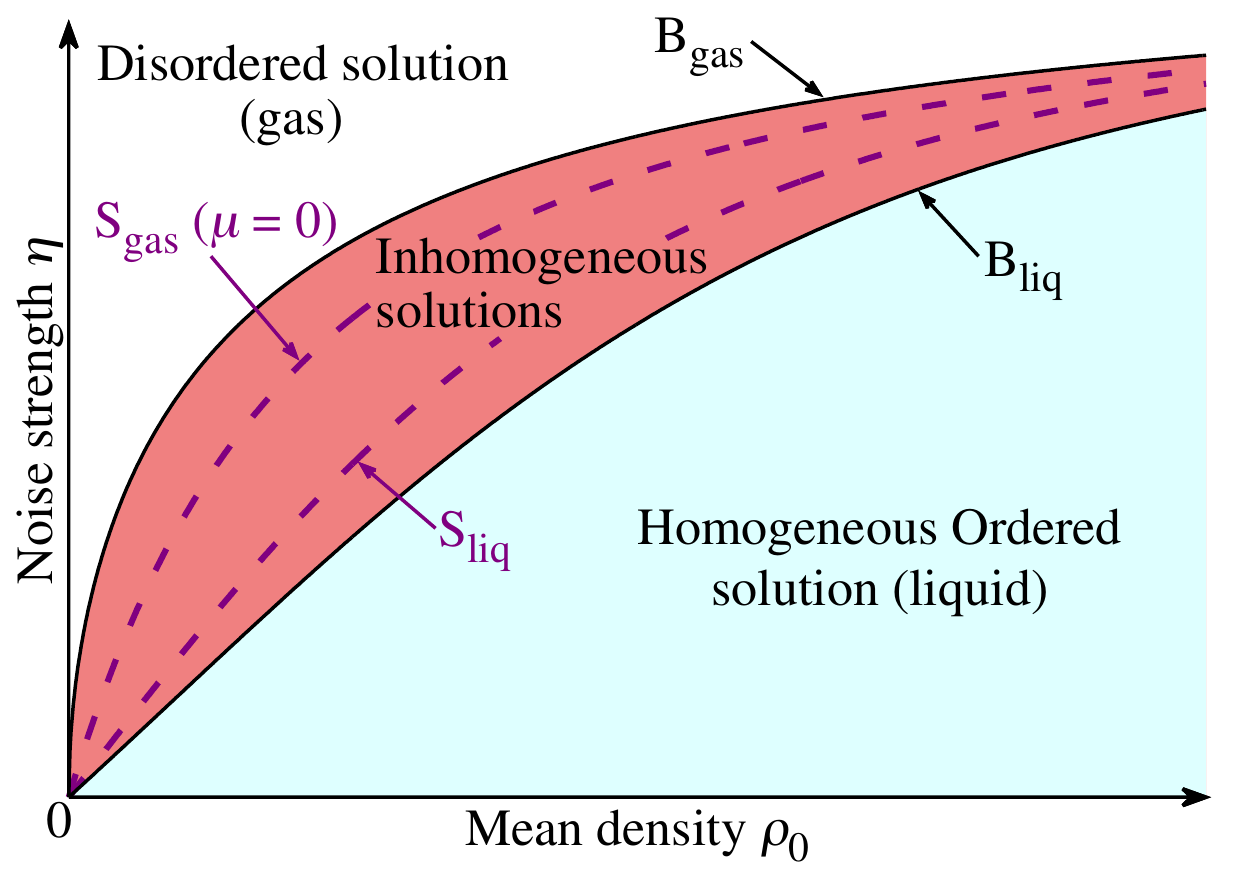}
	\caption{Schematic phase diagram of the hydrodynamic equations derived using the BGL method 
	in the (global density, noise intensity) plane. For all 3 DADAM classes, one finds 4 lines. 
	The innermost lines (${\rm S}_{\rm gas}$ and ${\rm S}_{\rm liq}$, dashed) 
	are, in the phase-separation language, the two spinodal lines. 
	They are given by the linear stability analysis of spatially-homogeneous solutions. 
	The ${\rm S}_{\rm gas}$ line marks the (lower) limit of stability of the disordered solution and is defined by $\mu=0$, where $\mu$ is the coefficient of the linear term in the ordering field ($\mu_1 f_1$ for the polar/Vicsek class, $\mu_2 f_2$ for the nematic classes).
	The ${\rm S}_{\rm liq}$ line marks the (upper) limit of stability of the homogeneous ordered solution, usually determined by the numerical resolution of a simple $3\times 3$ or $5\times 5$ ${\bf q}$-dependent matrix.
	The outermost lines (${\rm B}_{\rm gas}$ and ${\rm B}_{\rm liq}$) are the binodal lines usually corresponding to the lines delimiting the coexistence phase
	in the fluctuating context of microscopic models or experiments. They mark the limit of existence of inhomogeneous band solutions, and can only be determined by a detailed study of these. 
	 }
	\label{FIG:LINSTAB}
\end{figure}

First of all, the equation governing the ordering field ($f_1$ in the polar class, $f_2$ in the nematic classes) contains
the expected Ginzburg-Landau structure, 
$$\partial_t f = \mu f - \xi |f|^2 f + \ldots$$ 
In all cases $\xi>0$, which ensures that the cubic nonlinearity saturates the growth of order occurring when $\mu>0$. 
In all cases $\mu$ depends linearly on $\rho$: $\mu=a\rho+b$ with $a$ and $b$ depending on the noise coefficients.
This ``$\varphi^4$" or Mexican hat potential leads to the existence of a spatially-homogeneous ordered solution with $|f|^2=\mu/\xi$
as soon as $\mu>0$. In the ($\rho_0, \sigma$) plane, the line $\mu=0$ has the expected form, 
growing like a square root from the origin, saturating at high densities 
(${\rm S}_{\rm gas}$ line in Fig.~\ref{FIG:LINSTAB}). 
Above it, the homogeneous disordered solution is linearly stable, 
below it it is unstable\footnote{ This is obvious for homogeneous $q=0$ perturbations, 
but can also be shown for arbitrary perturbations ${\bf q}$.}. 
In this region, the homogeneous ordered solution exists. It is found to be linearly {\it unstable} near onset, 
i.e. below the $\mu=0$ line, down to a similar line (${\rm S}_{\rm liq}$ line in Fig.~\ref{FIG:LINSTAB}). 
This linear stability analysis can be performed semi-numerically 
for arbitrary perturbations of wavenumber ${\bf q}$\footnote{Linearizing the hydrodynamic equations around the 
homogeneous ordered solution, one has to solve, at each set of parameter values, 
a $3\times 3$ (polar and active nematic cases) or $5\times 5$ matrix (self-propelled rods case) that depends on ${\bf q}$.}
or analytically in the limit $|{\bf q}|\to 0$ of small wavenumbers. 
In the polar class for instance, the real part of the growth rate ${\mathcal G}(q)$ 
of the associated instability reads, in the limit of small longitudinal wavenumbers $q$~\shortcite{bertin2009hydrodynamic}:
\begin{equation}
{\mathcal G}(q) \underset{q\to0}{\approx} \left(\frac{(\partial \mu_1)^2}{2\mu_1} - \lambda\partial\mu_1 - \xi \right) q^2 \,, 
\end{equation}
where $\partial\mu_1 = d\mu_1/d\rho \; (> 0)$ and $\lambda = \kappa_1 + \kappa_2 \; (>0)$.
This growth rate will therefore always take positive values close to ${\rm S}_{\rm gas}$, where the linear coefficient $\mu_1 \to 0$.
Thus, one can easily see that the instability arises from the $\rho$-dependence
of the linear coefficient $\mu_1$ and, more precisely, from the fact that $\mu_1$ grows with $\rho$. 
A similar argument applies to the classes with nematic alignment, 
such that it constitutes a characteristic feature of DADAM.

 %%%XXX change label of liquid to "Homogeneous-order solution (liquid)"
 %%%%%%%%%%%%%%%%%%%%%%%%%%%
%%%%%%%%%%%%%%%%%%%%%%%%%%%

 The growth of $\mu$ with $\rho$ is the translation at hydrodynamic (and kinetic) level of the positive feedback 
 mechanism between local density and local order discussed at the microscopic level, 
and argued to be at the origin of phase separation. The ${\rm S}_{\rm gas}$ and 
${\rm S}_{\rm liq}$ lines delimiting the domain of linear instability 
of the homogeneous ordered solution are, in the language of phase-separation, the spinodal lines. 
In between them, perturbations around the ordered solution grow immediately. In this region, all our hydrodynamic equations
possess no stable homogeneous solutions, they only support inhomogeneous solutions. As we will see below, these closely resemble
the traveling wave trains observed in the Vicsek model, and the nematic bands observed in the nematic DADAM classes.
They exist beyond the region delimited by the spinodal lines. They are observed to compose the dominating attractor of the dynamics
up to a line (${\rm B}_{\rm gas}$ line in Fig.~\ref{FIG:LINSTAB}) located above the $\mu=0$ gas spinodal ${\rm S}_{\rm gas}$, 
and down to a line (${\rm B}_{\rm liq}$ line in Fig.~\ref{FIG:LINSTAB}) below the liquid spinodal ${\rm S}_{\rm liq}$.
The two  lines ${\rm B}_{\rm gas}$ and ${\rm B}_{\rm liq}$ are nothing but the binodal lines in the phase-separation language. They are the lines corresponding to those 
delimiting the coexistence region at the microscopic level. Locating them implies a fine study of the nonlinear, inhomogeneous
solutions of the hydrodynamic equations, something possibly delicate (see below). 
The spinodal lines, on the other hand, are easy to determine, but they have limited physical significance, 
and they do {\it not} correspond to the transitions found at the microscopic fluctuating level.

\subsection{Spurious instabilities, lack of positional diffusion}

To be honest, we have cheated (a bit) in the above presentation. In the polar/Vicsek and self-propelled rods classes, the 
linear stability analysis of the homogeneous ordered solution does not yield only the liquid spinodal line ${\rm S}_{\rm liq}$. 
Below this line, the ordered solution is {\it not} stable all the way down to the zero-noise axis. A second instability region is found, 
with growth rates typically a hundred times higher than those of the band instability, 
and rather large (polar/Vicsek class) or zero (rods class) most unstable wavenumber. 
%$(between $1$ to $10$ vs. $\approx 100$ for the band instability), 
This second instability does {\it not} correspond to anything known to happen at the microscopic level. 
It is not due to the particular choice of starting from a Boltzmann equation.
%%%XXX true? ---> I am not sure about the last sentence. For FP we also see an instability deep in the ordered phase, but its nature is different (it is present at the kinetic level, contrary to Boltzmann). Thus I don't think we can state that this instability could not also be due to the fact that we start from a Boltzmann equation.

\begin{figure}[h!]
	\centering
	\includegraphics[scale=0.6]{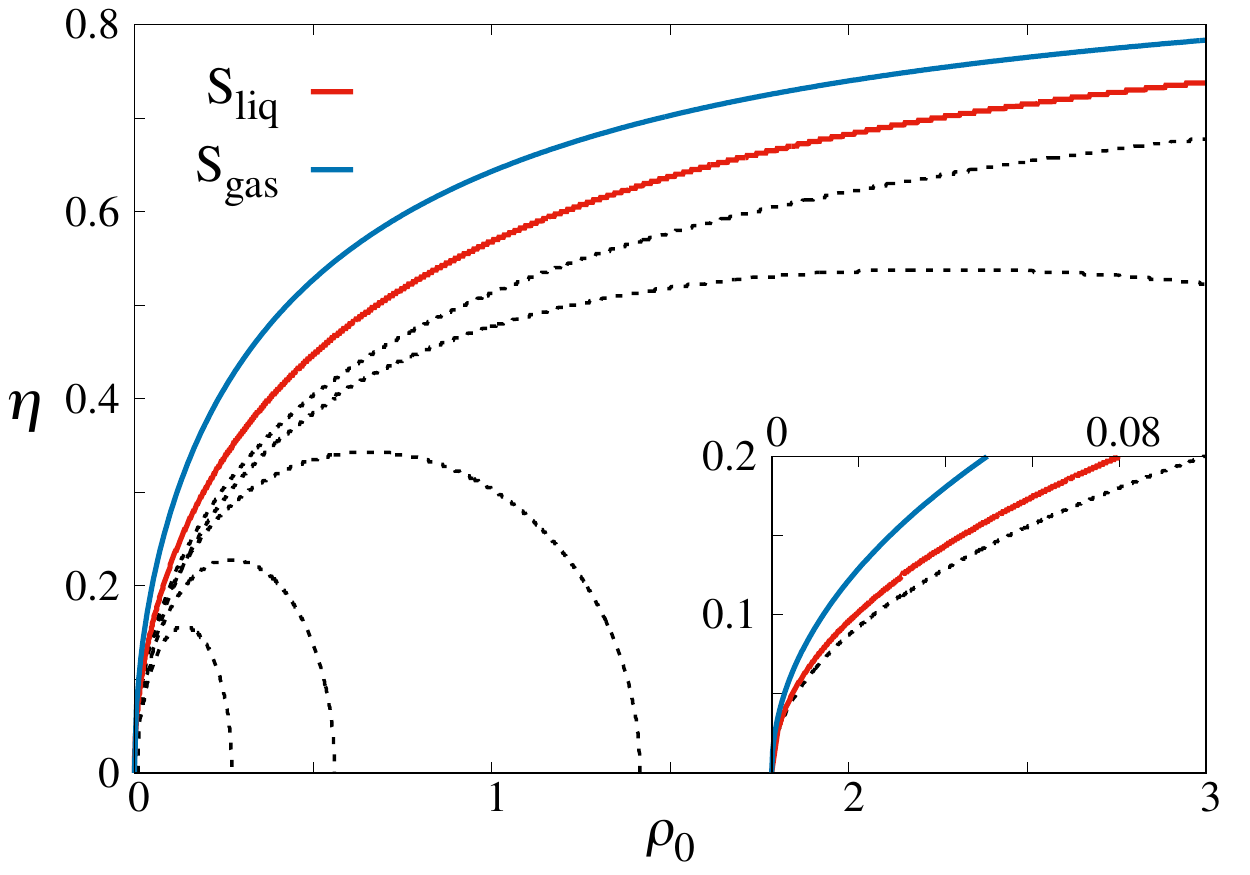}
	\caption{Linear stability of the homogeneous ordered solution 
	of the hydrodynamic equations~(\ref{hydro_polar}) for the Vicsek/polar class.
	In the region between the liquid and gas spinodals an instability w.r.t.\ perturbations 
	longitudinal to the order occurs, and leads to the inhomogeneous solutions characterized 
	in Sec.~\ref{Inhomogeneous_solutions_polar}.
	The black dashed lines correspond to the upper limit of the so-called spurious instability (see text)
	 considering different values of additional positional diffusion coefficient 
	 (of the form $D_{\rm add} \Delta f$ at the kinetic level).
	Going downwards, $D_{\rm add} = 0$, 0.05, 0.2, 0.5 and 1.
	The inset is a zoom showing how these curves reach the origin, 
	for clarity only the line corresponding to $D_{\rm add} = 0$ is represented.}
	\label{FIG:Spurious_hydro}
\end{figure}

For the polar/Vicsek class, it has been noticed that re-introducing some amount of positional diffusion in the dynamics 
(something Vicsek-style models formally lack) greatly reduces ---but does not eradicate--- the region of parameter space where this instability is present (Fig.~\ref{FIG:Spurious_hydro}). 
For the rods class, diffusion does not help since the instability is present for homogeneous perturbations.

%%%%%%%%%%%%%%%%%%%%%%%%%%%
%%%%%%%%%%%%%%%%%%%%%%%%%%%

This second instability can be considered spurious. It can be attributed to truncation effects 
since it is absent at the kinetic level: truncating the Boltzmann
hierarchies ~(\ref{hierarchy_polar}) and~(\ref{hierarchy_rods}) at some large number of modes $K$, 
performing the linear stability analysis of the homogeneous
ordered solution found numerically (solving this time large $(2K+1)\times(2K+1)$ matrices), 
one finds that this spurious instability region disappears as $K\to\infty$.
On the other hand, other weak instabilities can be found at kinetic level, but these are `cured' easily in the presence of
(additional) positional diffusion \shortcite{Kinetic_paper}. 

That such spurious instabilities should appear away from the onset of order 
is not too surprising since it is expected that higher-order fields, not retained at the hydrodynamic level,
may play an important role there. These remarks also stress that Vicsek-style models are somewhat formally singular
since they do not incorporate any positional diffusion. When simulated at discrete timesteps some effective diffusion arises, 
preventing any spurious instability deep in the fluid phase. On the other hand, deriving kinetic and hydrodynamic equations from them
yields continuous theories that are better behaved when complemented by some (effective) positional diffusion \shortcite{Kinetic_paper}.

\subsection{Inhomogeneous solutions in the polar class}
\label{Inhomogeneous_solutions_polar}

\begin{figure}[h!]
	\centering
	\includegraphics[scale=0.65]{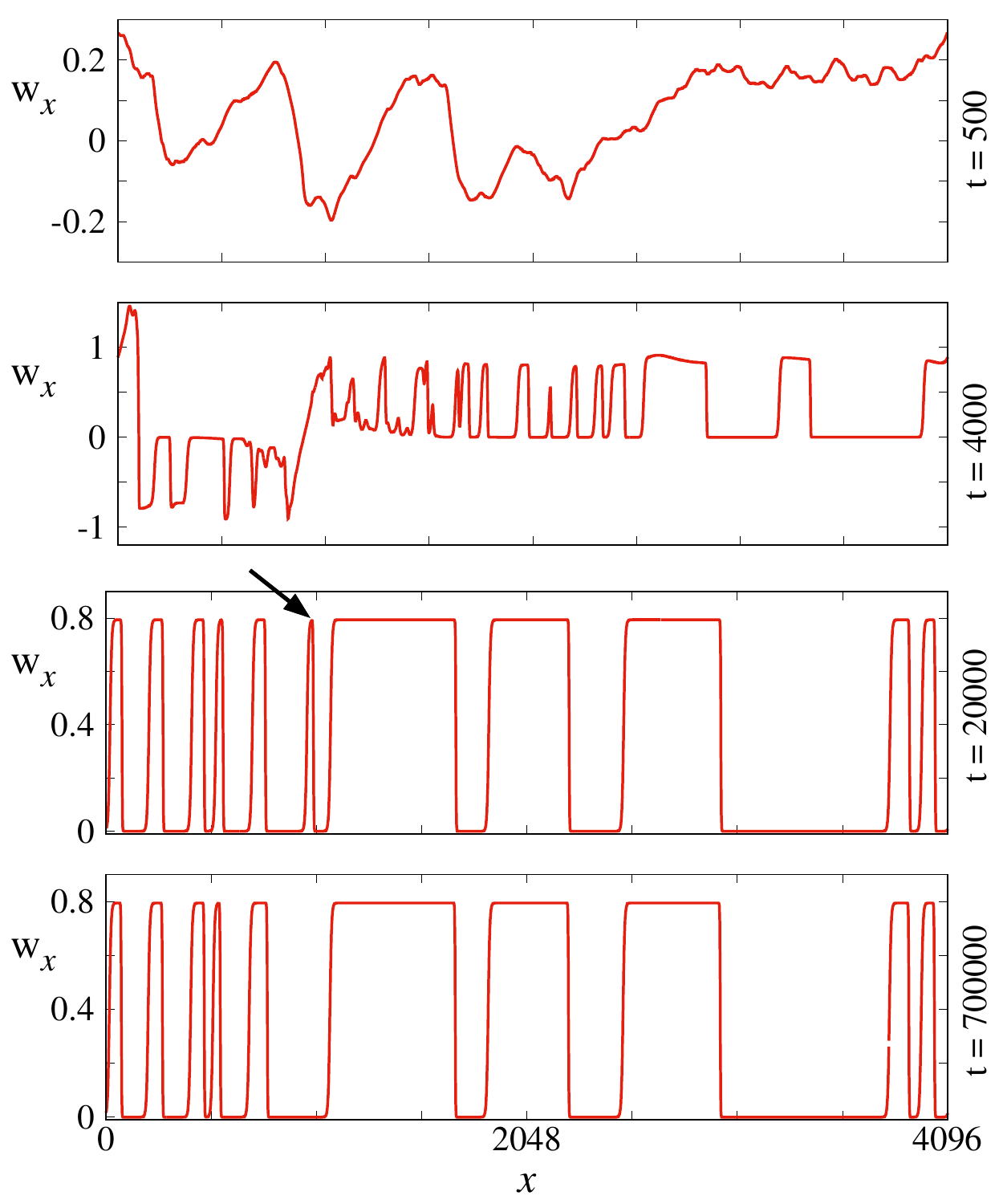}
	\caption{Direct numerical simulation of the (deterministic) Toner-Tu equations (Eqs.~(\ref{hydro_polar})) in the coexistence phase.
	Profiles of the $x$-component of ${\bf w}=\rho {\bf v}$ are shown at 4 different times following a 
	noisy disordered initial condition.
	The solution first remains ``two-dimensional''  ($t=500$).
	After a short transient ($\approx 1000$), it becomes homogeneous along one direction (here $y$) 
	and polar order is strictly along the other ($x$),
	with bands traveling in opposite directions	($t=4000$). Narrow bands may survive crossing each other, but eventually all remaining bands travel to the right ($t=20,000$). Their regularization into a periodic train of identical bands is extremely slow and
	numerically near-impossible due to pinning effects by the underlying simulation grid and the exponentially-weak interactions between bands. 
	At $t=700,000$ the solution indeed remains mostly unchanged, 
	except for only one thin band (indicated by the black arrow in the preceding snapshot) that has disappeared.
	Parameters: $\rho_0 = 0.5$, $\eta = 0.5$, $L = 4096$, $dx = 0.5$, $dt = 0.01$.
	 }
	\label{FIG:POLAR-COARSENING}
\end{figure}

In Figure~\ref{FIG:POLAR-COARSENING}, we show a typical
numerical simulation of Eqs.~(\ref{hydro_polar}) 
in two dimensions using parameter values taken 
between the two spinodal lines. It rapidly displays solutions homogeneous along one dimension
(usually along one of the axes of the simulation domain). 
These effectively one-dimensional solutions typically consist of bands of various shapes traveling in both possible directions. 
Eventually, one side wins and all remaining bands, after some time, have very slightly different speeds.
Then, a long ``coarsening" process follows during which the faster bands merge with slower ones. 
This is such a slow process that
often the ultimate state of a single propagating domain cannot be reached. 
(Moreover, space and time discretization effects are likely to ``interfere'', quantizing possible band speeds, de facto arresting the coarsening.)
As expected, this is observed not only in the region between the spinodals where no spatially-homogeneous solution is stable, 
but also beyond it, above the basic $\mu_1=0$ line, and also in the region of stability of the ordered homogeneous solution 
$|f_1|^2=\mu/\xi$.

%%%%%%%%%%%%%%%%%%%%%%%%%%%
%%%%%%%%%%%%%%%%%%%%%%%%%%%
%%%XXX are these profiles averaged over y? (the question makes sense for the first one) 
%%%---> These profiles are obtained in quasi-1D geometry in order to be able to reach long times, profiles not averaged over y.

This phenomenology is different from what happens
at the microscopic level in the Vicsek model, where a regular smectic arrangements of identical traveling bands is reached after transients. 

We now discuss the inhomogeneous solutions in some detail. As a matter of fact, the Toner-Tu equations do support, even at fixed parameter values, multiple coexisting, numerically observable (and thus linearly stable) solutions. 
This has been shown in detail in \shortcite{solon2015pattern}, and we sketch here the methodology followed and 
the main conclusions of this work.

Even though it is possible to work with the full equations (\ref{hydro_polar}), 
it is convenient to work on a simplified version, here written only in 
one space dimension $x$ for a scalar order parameter $m$ since we are looking for solutions homogeneous in $y$.
These simplified equations read:
\begin{subequations}
\begin{eqnarray}
\label{eq:hydroCC_rho1d} & & \partial_t \rho = -v_0\partial_x  m\\
\label{eq:hydroCC_m1d} & & \partial_t m +\lambda m\partial_x m = D \partial_x^2  m -P\partial_x\rho +(\rho-\varphi_{\rm g})  m -\xi  m^3
\end{eqnarray}
\label{eq:hydroCC_1d}
\end{subequations}
where the explicit coefficients $v_0$ and $P$ have been introduced to keep track of the corresponding terms in the following,
a single advection term $m\partial_x m$ appears in 1D,
and the linear term is expressed in a minimal form, 
all this without loss of generality at the qualitative level. (Note also that all explicit dependences on the global density $\rho_0$
and on the noise have been suppressed.)

The spinodal lines  marking the linear stability limits of homogeneous solutions 
are readily found: the disordered solution 
$m=0, \rho=\rho_0$ become unstable for  $\rho_0>\varphi_{\rm g}$, which defines the gas spinodal ${\rm S}_{\rm gas}$. 
The ordered solution $m^2=(\rho_0-\varphi_{\rm g})/\xi$
then exists, but is unstable until the liquid spinodal line ${\rm S}_{\rm liq}$ defined by 
$\rho_0>\varphi_\ell=\varphi_{\rm g}+\frac{1}{4\xi v_0+2P}$\footnote{ This expression for the liquid spinodal is only exact for longitudinal
perturbation in the small wavelength limit.}.

%%%XXX and what about spurious instability for these equations? ---> It has no longitudinal component, so these eqs. are exempt of spurious instability.

Solutions of Eqs.~(\ref{eq:hydroCC_1d}) propagating steadily at a speed $c$ only depend on $z=x-ct$.
They are solutions of the ODE system:
\begin{subequations}
\begin{eqnarray}
\label{eq:hydroCC_rhoz} & c\dot \rho-  v_0 \dot m=0\\
\label{eq:hydroCC_mz}   & D \ddot m+(c-\lambda m)\dot m-P\dot \rho+(\rho-\varphi_{\rm g}) m -\xi  m^3=0
\end{eqnarray}
\label{eq:hydroCC_z}
\end{subequations}
where the dots denote derivation with respect to $z$.  Solving Eq.~(\ref{eq:hydroCC_rhoz}) gives $\rho(z)=\rho_g+\frac{v_0}{c}m(z)$.
Since $\rho(z)=\rho_g$ when $m(z)=0$, the integration constant $\rho_g$ is the density in the
gaseous phase surrounding a localized polar excitation. 
Inserting the expression of $\rho$ in Eq.~(\ref{eq:hydroCC_mz}) we obtain the second-order ordinary differential equation
\begin{equation}
  \label{eq:ode}
  D \ddot m+(c-\frac{P v_0}{c}-\lambda m)\dot m-(\varphi_g-\rho_g) m+\frac{v_0}{c}m^2 -\xi  m^3=0
\end{equation}

\begin{figure}
\centering
  \includegraphics[width=0.8\textwidth]{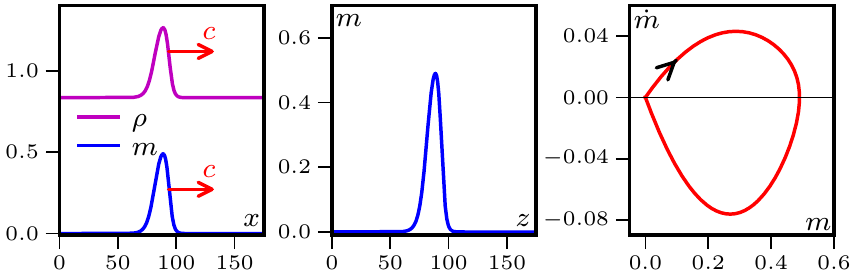}
  \caption{{\bf Left}: Density and magnetization profiles of a
    propagative solution of the hydrodynamic
    Eqs.~(\ref{eq:hydroCC_1d}). {\bf Center}:
    Magnetization profile in the comoving frame $z=x-ct$ or,
    equivalently, trajectory $m(z)$ of a point particle in the
    spurious time $z$. {\bf Right}: Phase portrait corresponding to
    the trajectory $m(z)$. Parameters:
    $D=v_0=\lambda=\xi=P=\varphi_g=1$.}
  \label{fig:first-trajectory}
\end{figure}

Equation~(\ref{eq:ode}) has a mechanical interpretation, in which it describes the position $m$, varying along time $z$,
 of a particle of mass $D$ in a potential $H(m)$, subjected to a (position-dependent) friction $f(m)$:
\begin{equation}
  \label{eq:ode_potential} D \ddot m=-f(m)\dot m-\frac{dH}{dm} \;.
\end{equation}
The potential and the friction read:
\begin{subequations}
\begin{eqnarray}
  \label{eq:potential}H(m)&= &-(\varphi_g-\rho_g)\frac{m^2}{2}+\frac{v_0}{3c}m^3-\frac{\xi}{4}m^4\\
  \label{eq:friction}f(m)&= &c-\frac{P v_0}{c}-\lambda m
\end{eqnarray}
\end{subequations}
%%%BEGINNING OF COPY OF PRE (not quite exact copy, but still)
The original problem of finding all the inhomogeneous propagative
solutions $m(x,t)$, $\rho(x,t)$ of the hydrodynamic equations is now
reduced to finding all the pairs ($c$, $\rho_g$) for which the
corresponding trajectories $m(z)$ exist. Mass conservation,
Eq.~(\ref{eq:hydroCC_rho1d}), imposes the boundary condition
$m(z=-\infty)=m(z=+\infty)$ so that we are looking for solutions
of~(\ref{eq:ode_potential}) which are closed in the $(m,\dot m)$ plane.
An example of propagative solutions $m(x,t)$, $\rho(x,t)$
together with the corresponding trajectory $m(z)$ and its phase
portraits is shown in Fig.~\ref{fig:first-trajectory}.

 %%%%%%%%%%%%%%%%%%%%%%%%%%%
 %%%%%%%%%%%%%%%%%%%%%%%%%%%

 %%%%%%%%%%%%%%%%%%%%%%%%%%%
\begin{figure}
\centering
  \includegraphics[width=0.7\textwidth]{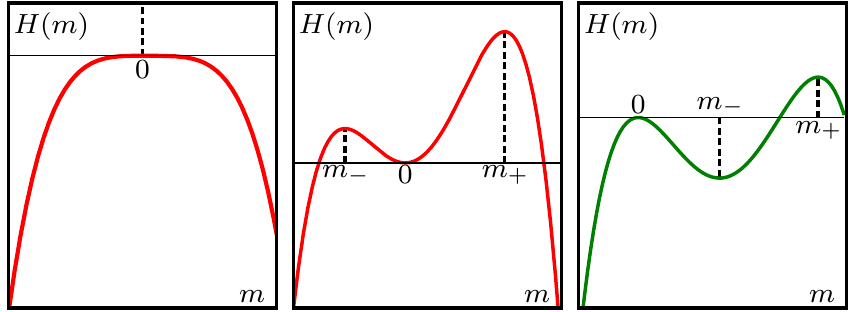}
  \caption{The green potential can give rise to physical (positive,
    non-exploding) solutions while the red ones are ruled out by our
    conditions (C1) (left) and (C2) (center).}
  \label{fig:bad_potential}
\end{figure}
 %%%%%%%%%%%%%%%%%%%%%%%%%%%

%%%XXX Eric doit avoir raison.. dans le panneau en haut a gauche, la partie en rouge ne doit pas atteindre le maximum, non?
%%% si c'est le cas, tu peux rectifier? ou bien tu demandes a  Alex?
 %%%%%%%%%%%%%%%%%%%%%%%%%%%
\begin{figure}
\centering
  \includegraphics[width=1\textwidth]{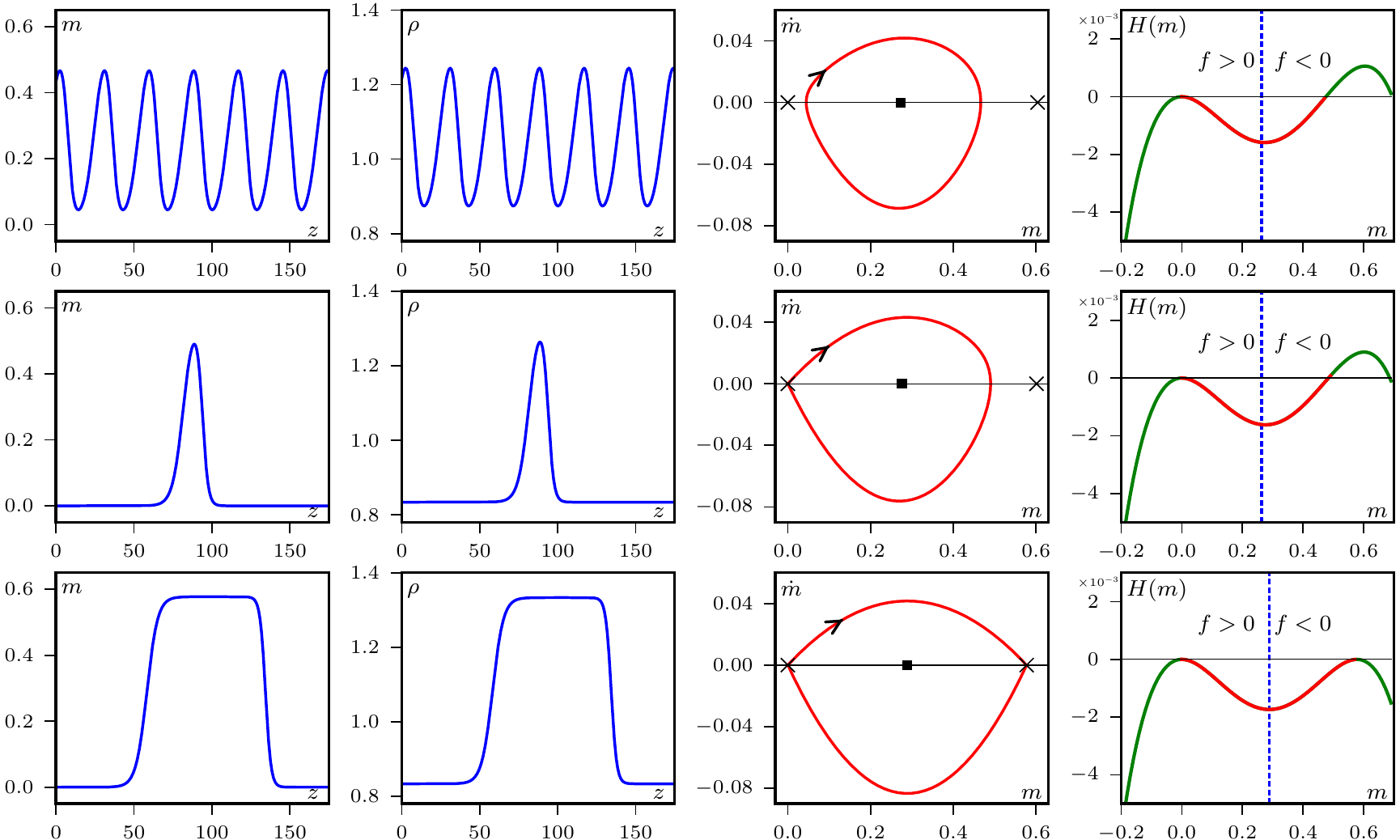}
 \caption{Example of the three types of trajectories. From left to
    right: magnetization and density profiles, phase portrait, and
    potential $H$. {\bf Top row}: Periodic trajectory, $\rho_g=0.835$,
    $c=1.14$. {\bf Center row}: Homoclinic trajectory,
    $\rho_g=0.83412$, $c=1.14$. {\bf Bottom row}: Heteroclinic
    trajectory, $\rho_g=0.83333$, $c=1.1547$. Phase portrait: Crosses
    indicate saddle points at $m=0$ and $m=m_+$. Squares indicate
    stable fixed points at $m=m_-$.  Potentials: The blue dashed line
    indicate where the friction changes sign. The red portion of the
    potential is the one visited by the trajectory. Parameters:
    $D=v_0=\lambda=\xi=P=\varphi_g=1$.}
  \label{fig:trajectories}
\end{figure}
 %%%%%%%%%%%%%%%%%%%%%%%%%%%

Some general constraints exist on $\rho_g$ and $c$.
Some potentials cannot give physical solutions. The extrema of
$H$, solutions of $H'(m)=0$, are located at $m=0$ and $m=m_\pm$.
We can already discard the case where $H'(m)$ has two complex roots
since $H$ then has a single maximum at $m=0$, and all trajectories
wander to $m=\pm \infty$ (see Fig.~\ref{fig:bad_potential},
left). This leads to a first condition on $c$, $\rho_g$:
\begin{equation}\label{eq:C1}%\tag{C1}
 {\rm C1:} \;\;\; (\varphi_g-\rho_g)c^2<\xi v_0^2
\end{equation}

Without loss of generality we can assume that $c>0$ and look only for
solutions with $m\ge 0$. This rules out the ($c$, $\rho_g$) values for
which $m_-<0$ and $m_+>0$ which give oscillations between negative and
positive values of $m$ (see Fig.~\ref{fig:bad_potential}, central
panel). 
%At the hydrodynamic equation level, such solutions would indeed
%correspond to different parts of the profiles moving in opposite direction. 
The corresponding condition
\begin{equation}
  \label{eq:C2}%\tag{C2}
{\rm C2:} \;\;\;  \rho_g<\varphi_g
\end{equation}
imposes $0<m_-<m_+$. The potential $H$ then has two maxima, at $m=0$
and $m=m_+$, and one minimum, at $m=m_-$. The typical shape of
potential which gives admissible solutions is shown in
Fig.~\ref{fig:bad_potential} along with examples of potentials ruled
out by conditions~(\ref{eq:C1}) and~(\ref{eq:C2}).

From the admissible shape of the potential $H$, we can now list all possible
trajectories $m(z)$ and the corresponding fields $m(x,t)$,
$\rho(x,t)$:
\begin{itemize}
\item Limit cycles, whose corresponding spatial profiles are
  periodic bands, as shown in the first row of
  Fig.~\ref{fig:trajectories}.
\item Homoclinic orbits, that start infinitely close to a maximum of
  $H$, hence spending an arbitrary large time there, before crossing
  twice the potential well in a finite time to finally return to the
  same maximum of $H$ at $z=\infty$. These trajectories correspond
  to isolated solitonic band profiles, as shown in the second row of
  Fig.~\ref{fig:trajectories}.
\item Heteroclinic orbits that spend an arbitrary large time close to
  a first maximum of $H$, cross the potential well in a finite time,
  spend an arbitrary large time close to the second maximum of $H$,
  before returning to the first maximum. These trajectories correspond
  to phase separated profiles. The arbitrary waiting times at the two
  maxima of $H$ then reflect the arbitrary sizes of two
  phase-separated domains (see the third row of
  Fig.~\ref{fig:trajectories}).
\end{itemize}

A third condition on $\rho_g,c$ arises from the non-linear friction
term. Following the classical mechanics analogy, we define an energy
function 
%\begin{equation}
 $E=\frac1 2 D \dot m^2+H$.
%\end{equation}
Multiplying the equation of motion~(\ref{eq:ode}) by $\dot m$, we get
%\begin{equation}
 % \label{eq:energy}
 $\frac{dE}{dz}=-f(m)\dot m^2$.
%\end{equation}
Energy is injected when $f(m)<0$ and dissipated when $f(m)>0$. On a
closed trajectory, the friction $f$ must thus change sign. Since $f$
is a decreasing function of $m$, this imposes $f(0)>0$ for
trajectories with $m(z)>0$, or equivalently
\begin{equation}\label{eq:C3}%\tag{C3}
{\rm C3:} \;\;\;  c>\sqrt{\lambda v_0}
\end{equation}

The conditions~(\ref{eq:C1}), (\ref{eq:C2}) and (\ref{eq:C3}) thus
provide loose bounds on the subspace of the ($c, \rho_g$) plane which
contains the three types of trajectories $m(z)$ described above. 
%These trajectories correspond to the three types of inhomogeneous profiles
%seen in the microscopic models. 

\begin{figure}
\centering
  \includegraphics[width=0.9\textwidth]{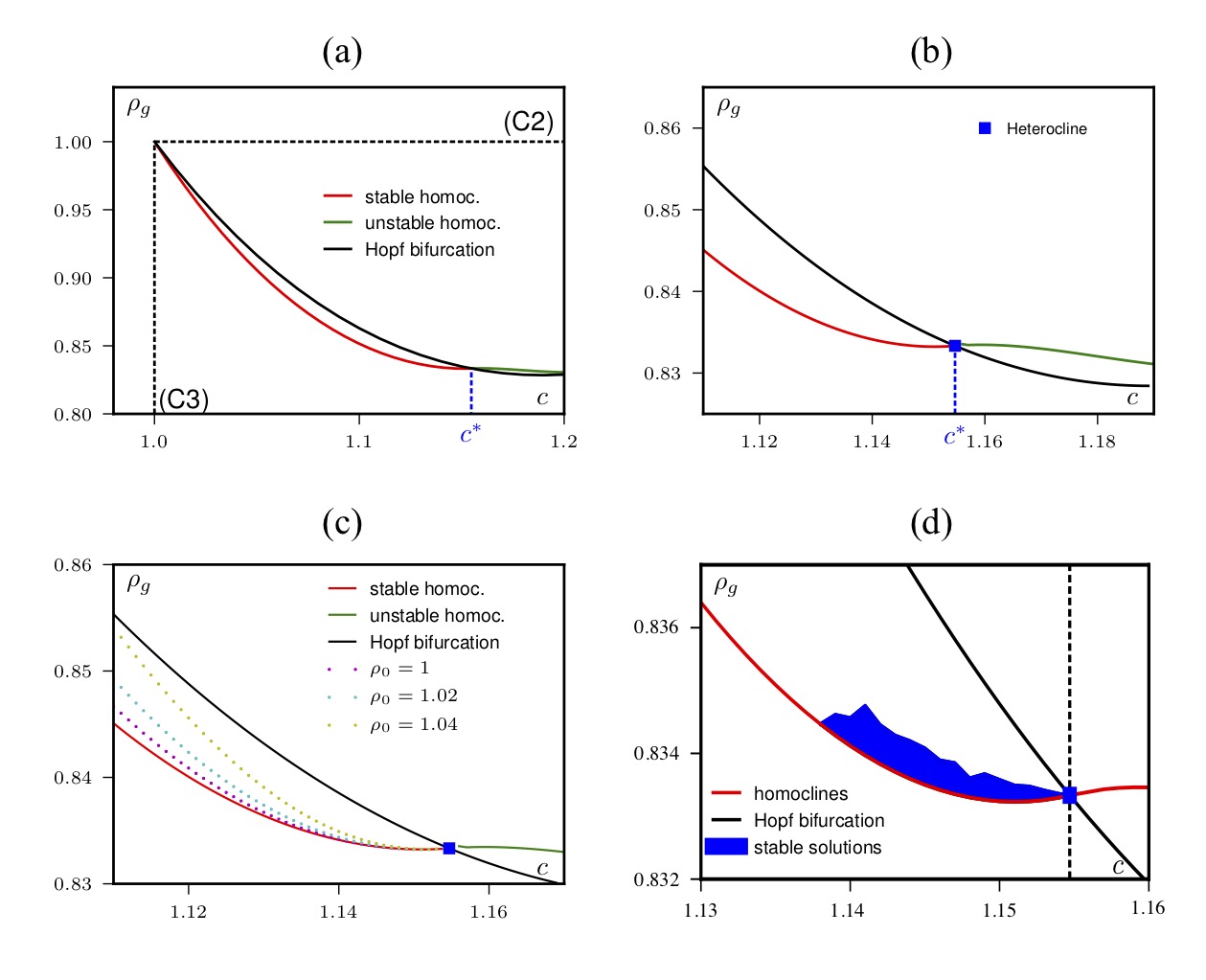}
  \caption{Space of all solutions. 
  	(a):  A 2-parameter family of periodic
    	orbits is found inside the band delimited by the black line (corresponding to a Hopf bifurcation in the dynamical system)
    	and the homoclinic trajectories. The constraints (C2) and (C3) are
    	indicated by the black dashed lines. The constraint (C1) lies out
    	the range of this plot.  
     	(b):   Zoom of (a)   around the point $c=c^*$ where the unique
    	heteroclinic trajectory is found.  
    	(c): Lines of solutions having a fixed average density $\rho_0$ in the space of all solutions. 
	(d): Subset of admissible propagative
    	solutions which are dynamically stable (blue region). None of the solutions for
    	$c>c^*$ is stable in the hydrodynamic
	equations. (The rugged boundary of the blue region is due to numerical difficulties in performing this study, 
	see~\protect\shortcite{solon2015pattern}.)
	Parameters: $D=v_0=\lambda=\xi=P=\varphi_g=1$.
    }
  \label{fig:rhog-c}
\end{figure}

A detailed study of the phase portrait of the simple dynamical system defined by Eq.~(\ref{eq:ode_potential}),
and how it evolves when $\rho_g$ and $c$ are varied,
is necessary to find the final, complete, domain of existence of solutions.
This study, presented in \shortcite{solon2015pattern}, is centered around the stability of the fixed points, something that can be done
analytically, but one cannot avoid the systematic numerical determination of the solutions, which is not a particularly
difficult task given all the known constraints on them.
The final results are presented in Fig.~\ref{fig:rhog-c}.
The domain of existence of our solutions is much smaller than that given by the 3 constraints above.
It is organized around the unique heteroclinic solution of speed $c^*$. 
A one-parameter family of homoclinic orbits with speed $c<c^*$ is found on the red line.
In the whole region between this red line and a black line located above, a two-parameter family of periodic orbits is found.
The black and the red lines meet at a point fixed by the third condition (C3), closing this 
region\footnote{A similar region exists for $c>c^*$, 
but these orbits are found to be unstable solutions at the PDE level.}.

 %%%%%%%%%%%%%%%%%%%%%%%%%%%

In the above study of the multiple solutions existing at fixed parameter values, the global density $\rho_0$ was not fixed.
For a given trajectory $m(z)$, $\rho_0$ is given by the average of $\rho(z)=\rho_g+v_0 m(z)/c$ over time.
Fixing $\rho_0$ adds a constraint that selects a line of solutions in
the ($c$, $\rho_g$) space, as shown in Fig.~\ref{fig:rhog-c}c.
These lines end at the heteroclinic trajectory. 

%\begin{figure}
 %\includegraphics[width=0.8\columnwidth]{figure/lines_cst-rho0_rhog.pdf}
  %\caption{Lines of solutions having a fixed average
   % density $\rho_0$ in the space of all solutions. 
% Parameters:
  %  $D=v_0=\lambda=\xi=a_4=\varphi_g=1$.}
  %\label{fig:fixed-rho0}
%\end{figure}

At this point, we still have multiple solutions of Eq.~(\ref{eq:ode_potential}) at fixed parameters and global density.
Each of them is a propagative solution of the original PDEs. However, we only know their existence, not their stability.
The discrepancy between the selection of a unique solution at the microscopic level and the multiple solutions 
found at hydrodynamic level could thus be lifted if only one of the PDE solutions were stable.
The linear stability of inhomogeneous solutions of PDEs cannot be performed analytically in general. 
In \shortcite{solon2015pattern}, a numerical procedure was adopted to estimate the linear stability of the ODE-found solutions. 
This was performed essentially in quasi-one dimensional domains. The result is presented in Fig.~\ref{fig:rhog-c}d:
%Fig.~\ref{fig:stabilityPDE}: 
at the PDE level, stable solutions form only a small subset of the existing ones.
However this subset still includes the
three possible types of trajectories: periodic bands, solitonic bands,
and phase-separated profiles. The dynamically stable solutions are all
found in the region of the ($c$, $\rho_g$) plane close to the
heteroclinic trajectory and close to the line of homoclinic orbits.

\begin{figure}
\centering
  \includegraphics[clip,width=\textwidth]{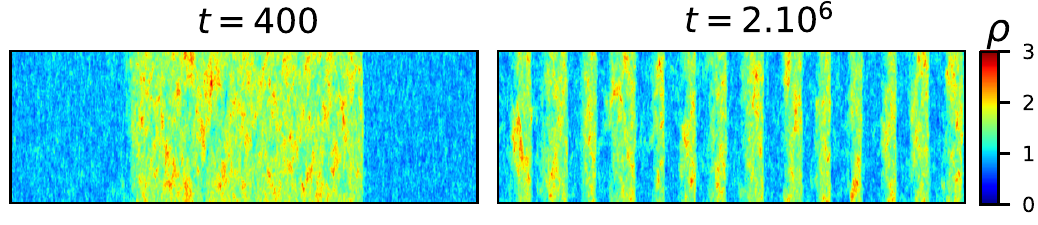}
  \caption{Illustration of solution selection in the stochastic Toner-Tu equations. Here a simple additive, delta-correlated white noise of variance 0.4 as been added to the Eqs.~(\ref{hydro_polar}). (Domain of size 2000$\times$100, parameters in the coexistence region, 
  figure adapted from~\protect\shortcite{solon2015pattern}) 
  Starting from the heteroclinic solution of the deterministic PDE, this unique propagating domains subsists for some time (left panel), but then gradually breaks and finally reaches a steady configuration with 14 bands (right panel). Similar experiments, starting with ``too many'' bands, also lead to the 14-band solution (not shown).
}
  \label{fig:noise_selection}
\end{figure}

Thus, the hydrodynamic equations for the polar/Vicsek class generically possess a multitude of linearly-stable, inhomogeneous
propagating solutions. Most of these solutions are periodic trains of bands, 
as observed in the Vicsek model, but the selection observed at the microscopic model is not occurring. 
It does occur, though, when the hydrodynamic equations are complemented by some `effective noise', 
be it an ad-hoc additive noise like in the works of Toner and Tu, or multiplicative noise terms determined by some
stochastic calculus (see \shortcite{bertin2013mesoscopic} for an example of such a calculation). 
Simulating such noisy hydrodynamic equations clearly shows
selection at play: starting with an initial solution with ``too many" bands, one observes, along a long transient, that their number
gradually decreases and stabilizes itself. Starting with too few bands, one observe that some of them split in two, until the same asymptotic number is reached (Fig.~\ref{fig:noise_selection}).

 %%%%%%%%%%%%%%%%%%%%%%%%%%%
 %%%%%%%%%%%%%%%%%%%%%%%%%%%

 %%%%%%%%%%%%%%%%%%%%%%%%%%%
\begin{figure}
\centering
  \includegraphics[width=0.5\textwidth]{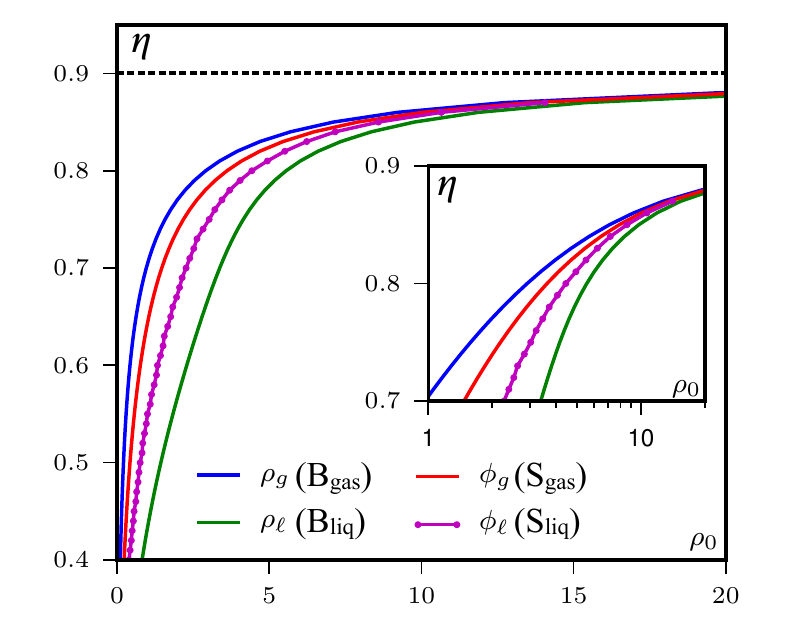}
  \caption{Phase diagram of the Vicsek/Toner-Tu
    hydrodynamic equations. The spinodal lines ${\rm S}_{\rm gas}$ and ${\rm S}_{\rm liq}$ are given by $\varphi_g$ and
    $\varphi_\ell$.
%     are known exactly except $\varphi_\ell$ in the
 %   Vicsek hydrodynamic equation which we computed numerically by
 %   looking at the stability of systems of size $50\times 50$. 
    The binodals ${\rm B}_{\rm gas}$ and ${\rm B}_{\rm liq}$ are estimated via the coexisting densities $\rho_g$ and $\rho_\ell$ 
    of the phase-separated (heteroclinic) solution as explained in the text.
    The dashed line indicate the asymptote above which only disordered
    solutions exist. Insets are close-ups on the high-density regions
    with a logarithmic scale on the $x$-axis. Parameters: $v_0=1$.}
  \label{fig:phase-diagram_PDE}
\end{figure}
 %%%%%%%%%%%%%%%%%%%%%%%%%%%

Finally, even though the deterministic Toner-Tu equations cannot account for the selection of a unique solution, the phase diagram
at hydrodynamic level can be estimated from them with some high accuracy. As we saw before, the spinodal lines 
${\rm S}_{\rm gas}$ and ${\rm S}_{\rm liq}$ are determined 
by the linear stability limits of the homogeneous solutions. The difficulty is with the binodal lines
${\rm B}_{\rm gas}$ and ${\rm B}_{\rm liq}$ that do correspond to the transitions lines observed at microscopic level. 
They can be defined by the minimum
and maximum densities beyond which inhomogeneous propagative profiles
cannot be observed in simulations of the hydrodynamic equations.
It was found in \shortcite{solon2015pattern} that no other stable solution has a larger
density than the liquid domain of the unique heteroclinic solution. This fixes the liquid binodal ${\rm B}_{\rm liq}$.
The situation is more subtle for the gas binodal ${\rm B}_{\rm gas}$. Depending
on the external parameters, the line of homoclinic trajectories
solution of the dynamical system is not always monotonous as a
function of $c$, so that it cannot in principle, be given by the heteroclinic solution (which has maximal speed $c^*$).
But this remains a very good approximation, and can be used to produce the final phase diagram
shown in shown in Fig.~\ref{fig:phase-diagram_PDE}.

%%%XXX some general conclusion here about the polar class?

\subsection{Inhomogeneous solutions in the nematic classes}

Once again, here we treat the two nematic classes (active nematics and self-propelled rods) together, 
all the more so since their dynamics
at the deterministic hydrodynamic level present no important differences.

\begin{figure}[h!]
	\centering
	\includegraphics[width=\textwidth]{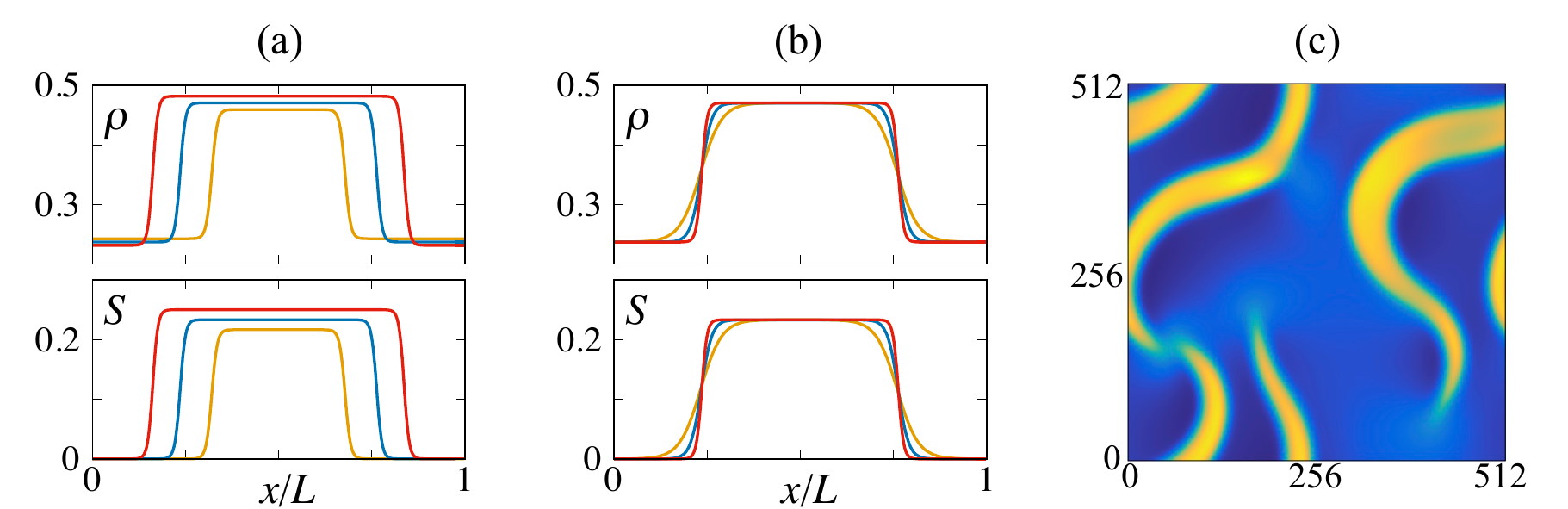}
	\caption{Coexistence phase of the hydrodynamic equations for the DADAM nematic classes.
(a) (Stable) band profiles obtained from simulations of Eqs.~(\ref{hydro_an}) in 1D at fixed noise $\eta = 0.2$, 
increasing the global density: $\rho_0 = 0.32$, 0.36, and 0.4 at system size $L = 512$.
(b) Rescaled (stable) band profiles at fixed noise $\eta = 0.2$ and density $\rho_0 = 0.36$, increasing system size: $L = 128$, 256 \& 512.
(c) Snapshot taken in the chaotic regime observed at large sizes (Parameters: $\rho_0 = 0.30$, $\eta = 0.20$, and $L = 512$).
For (a,b,c): $dx = \frac{1}{2}$ and $dt = 0.01$.
 }
	\label{FIG:HYDRO-NEMABAND}
\end{figure}

Numerical simulations of Eqs.~(\ref{hydro_an}) or Eqs.~(\ref{hydro_rods}) performed in moderate-size domains at parameter values 
between the two spinodal lines
usually lead to a stable, unique band with high-density and high nematic order standing out of a disordered sparse gas.
Increasing/decreasing the global density with other parameters fixed, eventually passing the spinodal lines,
the gas density, the nematic order and the density
inside the band all remain nearly constant, but the width of the band increases/decreases (Fig.~\ref{FIG:HYDRO-NEMABAND}a). 
Increasing system size at fixed parameters, the band occupies a near-constant fraction of space (Fig.~\ref{FIG:HYDRO-NEMABAND}b). 
All this leads to conclude a bona fide phase separation scenario with a single, macroscopic ordered liquid domain (the band)
occupying a fraction of space going from zero to one between two binodal lines.

Of course the above is true if the system size is big enough to accommodate the two finite-width interfaces delimiting the band, 
and/or as long as the band is wider than the width of its two interfaces: 
the band disappears near the gas binodal ${\rm B}_{\rm gas}$ (where it is expected to be very thin) due to finite-size effects. 
Moreover, for large-enough systems, 
%and/or thin enough bands, 
one can observe that the band is unstable, bends, elongates, 
and splits, leading to sustained chaotic dynamics (Fig.~\ref{FIG:HYDRO-NEMABAND}c).

%%%%%%%%%%%%%%%%%%%%%%%%%%%
%%%%%%%%%%%%%%%%%%%%%%%%%%%

These numerical observations are confirmed and better understood at the analytical level. First of all, it is 
relatively easy to find an explicit expression for the basic band solution. 
Looking for a solution of Eqs.~(\ref{hydro_an}) homogeneous in $x$, varying in $y$, one has 
${\rm Re}f_2=f_2(y)$, ${\rm Im}f_2=0$, $\rho=\rho(y)$. The conservation equation~(\ref{hydro_an1}) becomes:
\begin{equation}
\partial_y^2 \rho = \partial_y^2 f_2 \label{eq:band1}
\end{equation}
Integrating twice, imposing that the fields remain finite when $y\to\pm\infty$, we get 
\begin{equation}
\rho =  f_2 + \rho_g \label{eq:band2}
\end{equation}
Rewriting $\mu_2[\rho]= \mu' (\rho -\varphi_g)$ with $\varphi_g=\frac{1-P_2}{\mu'}$, using (\ref{eq:band1}) and (\ref{eq:band2}), 
Eq.~(\ref{hydro_an2}) reads:
\begin{equation}
\partial_y^2 f_2 = 4\mu'(\rho_g-\varphi_g) f_2 -4\mu' f_2^2 +4\xi f_2^3
\label{eq:band3}
\end{equation}
Multiplying by $\partial_y f_2$ and integrating, one gets:
\begin{equation}
\frac{1}{2}(\partial_y f_2)^2 = -2\mu'(\rho_g-\varphi_g) f_2^2 -\frac{4}{3}\mu' f_2^3 +\xi f_2^4 \;.
\label{eq:band4}
\end{equation}
Separating the variables, integrating under the condition $\lim_{y\to\pm\infty} f_2 =0$, one obtains, after some simplifications:
\begin{equation}
f_2(y) = \frac{3 (\varphi_g-\rho_g)}{1+a\cosh (2y\sqrt{\mu' (\varphi_g-\rho_g)})} \;\;{\rm with}\;\; 
a^2=1-\frac{9\xi}{4\mu'}(\varphi_g-\rho_g)
 \label{eq:band5}
\end{equation}
Finally, the gas density $\rho_g$ is fixed by the condition $\int_L \rho(y) dy = \rho_0 L$, where L is the length (in $y$) of the system.
Neglecting exponentially small terms, taking the $L\to\infty$ limit, one gets:
\begin{equation}
\rho_g \approx \varphi_g-\frac{2\mu'}{9\xi}\left(1-4e^{-KL}\right) \;\;{\rm with}\;\;
K=\frac{2\sqrt{2}\mu'}{9\sqrt{\xi}}\left[1+\frac{9\xi}{2\mu'}(\rho_0-\varphi_g)\right]
 \label{eq:band6}
\end{equation}
Substituting into (\ref{eq:band5})
%, taking the large $L$ limit, 
one finally gets
\begin{equation}
f_2(y) = \frac{f_2^{\rm band}}{1+2e^{-KL/2} \cosh (y \frac{2\sqrt{2}\mu'}{3\sqrt{\xi}})}
\;\;{\rm where}\;\; f_2^{\rm band}= \frac{2\mu'}{3\xi}
 \label{eq:band7}
\end{equation}
and
\begin{equation}
\rho_{\rm band} = f_2^{\rm band} + \rho_g=\varphi_g+\frac{4\mu'}{9\xi}\left(1+2e^{-KL}\right)
 \label{eq:band8}
\end{equation}
Note that, as expected, $\rho_{\rm band} > \varphi_g > \rho_g$, which guarantees the stability of both the ordered and
disordered parts of the solution, and $f_2>0$ which means that the nematic order is along $x$. 
It can be easily shown that this band solution, determined here for the active nematics equations~(\ref{hydro_an}),
 is also a solution of the self-propelled rods hydrodynamic equations~(\ref{hydro_rods}), 
 albeit with the polar field $f_1=0$.

The liquid fraction $\Omega$, i.e. the relative surface occupied by the band, is given, in the large $L$ limit where the interfaces are
negligibly thin, by $\Omega(\rho_{\rm band} -\rho_g)+ \rho_g=\rho_0$. 
Substituting the values of $\rho_{\rm band}$ and $\rho_g$, one gets:
\begin{equation}
\Omega= \frac{1}{3} +\frac{3\xi}{2\mu'}(\rho_0 - \varphi_g)
 \label{eq:band9}
\end{equation}
The binodal lines are then simply obtained by the conditions $\Omega=0,1$.

The band solution is found always stable in simulations of quasi-1D domains of short dimension along the band. 
However, we know that, at least in some cases, the band can develop long wavelength undulations leading to chaos.
As a matter of fact, with the expression of the band solution in closed form, it is possible to prove that it is {\it always} unstable 
to a long-wavelength instability. We do not reproduce here this calculation, first performed in \shortcite{ngo2014large} for the
active nematics case, 
then repeated in \shortcite{peshkov2013boltzmann} for the self-propelled rods case. We just note here that it is not a full linear stability analysis, 
what is obtained being a criterion for the presence of a long-wavelength instability that is found to be always satisfied.
The above results are valid in the infinite-size limit. At finite-size, naturally, one observes a more complicated scenario,
notably because the band is stable up to a certain parameter-dependent size. 
A detailed analysis of this can be found in \shortcite{grossmann2016mesoscale} 

The main conclusions of the above findings are that it is possible to determine analytically 
the binodal lines in the nematic DADAM classes and that the band solution is linearly unstable to a long-wavelength
instability in the whole coexistence phase. This leads to the spatiotemporal chaos of bands bending, elongating, splitting 
and merging illustrated in Fig.~\ref{FIG:HYDRO-NEMABAND}c, which is similar to what happens at the microscopic level. 
This chaotic regime is found to have large but finite correlation length- and time-scales. 
Asymptotically the coexistence phase is thus a disordered phase, with the order/disorder transition 
occurring at the liquid binodal line ${\rm B}_{\rm liq}$.

The band chaos observed is very similar in the two classes, active nematics and self-propelled rods. 
In the rods case, the polar field $f_1$ only takes non-zero, small values near the interfaces of the bending and moving bands. 
This indicates that $f_1$ can be then enslaved to $f_2$. As a matter of fact, doing this on the hydrodynamic equations for the rods class
(Eqs.~(\ref{hydro_rods})) yields the simple equations for the active nematics class (Eqs.~(\ref{hydro_an})).

A lot of the above properties of the hydrodynamic equations for the nematic DADAM classes are in agreement with
the microscopic level of Vicsek-style models (structure of phase diagram, existence and instability of the nematic band solution, chaotic coexistence phase, etc.). But some discrepancies remain. 

First, and rather obviously, these deterministic PDEs
cannot account for features intrinsic to the fluctuating level, such as GNF, or 
the main difference between the active nematics and self-propelled rods cases recorded at fluctuating level, 
i.e. the observation that nematic order is QLRO in the first case,
but possibly LRO in the second case. Whether this difference can be accounted for
at the level of fluctuating hydrodynamics, e.g. by adding back some noise terms to the deterministic hydrodynamic equations, 
is an important open question.

Second, the hydrodynamic equations studied above 
only capture the long-wavelength longitudinal instability of the nematic band, whereas
we have seen that this instability can be overcome by a short-wavelength transversal one for fast velocity reversals. 
It was shown recently that this second instability can be accounted for by higher-order hydrodynamic equations, keeping 
all fields until at least $f_4$ \shortcite{cai2019dynamical}. This is an indication that at the nonlinear level of inhomogeneous solutions the simplest
hydrodynamic theories may not be enough, and that higher-order terms may have to be introduced even at the bare level.

\section{Discussion and perspectives}
\label{discuss}

\subsection{Summary and remarks}

We have seen that in spite of its restricted scope, dry aligning dilute active matter can display a lot of 
unexpected, if not new, physics. This is true even for minimal Vicsek-style models, which were taken here as 
representatives of large classes. 

We have shown that rather than the original idea of order-disorder transitions
similar to magnetic systems, the emergence of orientational order in DADAM is best described as a phase separation scenario
between a microscopically-disordered gas, and a (quasi-) ordered liquid. 
The coexistence phases separating the gas from the ordered liquid involve mesoscale structures (usually in the form of high-density high-order bands). This renders numerical work aiming at uncovering asymptotic properties difficult. 

That is not to say that the proximity of DADAM systems with equilibrium magnetic models such as the XY model is
irrelevant. The long-range correlations observed in the liquid phase are primarily due to the spontaneous breaking  
of rotational symmetry, like in equilibrium. But the essential couplings between density and order and the advection modes
can modify substantially the long-range correlations and introduce anomalous number fluctuations, 
 in qualitative agreement with the early predictions by Toner, Tu, Ramaswamy and colleagues, even though a full understanding of the scaling laws involved is  still missing.

We have explained how one can derive continuous hydrodynamic theories from simple microscopic models of aligning active particles.
This is important if one wants to keep track of the microscopic-level parameters, something necessary in order to really estimate
to what extent the hydrodynamic equations are faithful to the `microscopic truth'. 
In particular, we have demonstrated that the Boltzmann-Ginzburg-Landau approach is systematic, and that it 
 yields, in spite of the strong assumptions made along the way,  good qualitative agreement with the particle-level
behavior. 

The simple hydrodynamic equations we obtained have been studied in depth, elucidating in particular the nature and the stability of 
the inhomogeneous band solutions governing the coexistence phase. The analysis involved typical nonlinear dynamics 
tools and concepts, and one general lesson here is of the limited role played by the linear stability analysis of the simple 
spatially-homogeneous solutions.
 
Further theoretical analysis requires to be working at some fluctuating continuous level. This is beyond the scope of these lecture notes,
which mostly considered deterministic hydrodynamic descriptions. Incorporating effective/dominant noise terms into them can be done
in various ways, starting by the addition of a simple additive noise, but also following systematic methods. These, however,
generically provide multiplicative noise terms, which are more difficult to handle, even though it is not clear at present whether this makes a difference. 
Given the difficulties of a numerical integration of stochastic PDEs, these issues will have to be resolved by a 
full-blown renormalisation group approach taking seriously into account the crucial couplings between density and order 
at play in DADAM.

\subsection{Expanding from the DADAM limit}

We conclude with some comments about possible directions into which knowledge of DADAM can be extended in order to
account for wider sets of active matter systems. 

{\it Dilute $\to$ dense:} 
First of all, one can dispense with the diluteness of the systems by introducing some kind of repulsive interaction in complement
of alignment. As a matter of fact, all the phenomena reported above are still observed when some degree of soft repulsion 
is added, or even, sometimes, in systems of hard particles \shortcite{deseigne2010collective}. This said, 
one can also expect to observe new collective phases in dense systems  even without reaching jamming conditions. 
In such cases, the prior knowledge of the behavior of the system in the dilute limit is a plus.
In dense dry active nematics, for instance,
a bending instability appears within the ordered liquid phase, 
leading to some incessant chaos of topological defects \shortcite{putzig2016instabilities,nema_repulsion}. There 
the density field is essentially constant, preventing the coupling between order and density to act as usual. 
It was also shown that repulsive interactions can lead to active smectic phases where translational invariance is broken, on top of the spontaneously broken rotation symmetry \shortcite{romanczuk2016emergent}.

{\it Dry $\to$ wet:} 
There exist many situations where elongated swimmers come in frequent close contact and align locally. This is the case of some dense
bacterial suspensions, but also of mixtures of bio-filaments and motor proteins fueled by ATP. These more or less wet systems can
be described by Vicsek-like models where particles exert forces on the surrounding fluid and generate a fluid flow that, in turn, advects and rotates the particles. Depending on the degree of confinement or friction, such systems may be `more or less wet', and their phenomenology might keep some features of the strictly dry situations studied here.

{\it Other directions:} 
Finally, one can explore entirely new directions, starting from DADAM. Here is a partial list.
\begin{itemize}
\item In the context of kinematic models for the
collective motion of animal groups (bird flocks, fish schools, etc.), spatial cohesion must be achieved. This can be achieved by 
adding some pairwise attractive interactions~\shortcite{gregoire2003moving}.
\item Another important direction is to incorporate memory effects, at play for example in those systems 
where the trajectories of active particles, even though stochastic, are smooth and differentiable.
\item The effect of quenched disorder in DADAM has just started to be investigated, and obviously deserves further study.
\item Similarly, chiral active matter has been the focus of recent works.
\end{itemize}

In general, adding one mechanism/ingredient to a DADAM problem yields a much richer phenomenology than that contained
in the simple phase diagram of DADAM systems. We believe it is useful to approach these more complicated phase diagrams 
such that they contain the reference strictly DADAM limit.

%%%XXX DISCUSS EXPERIMENTS??? HERE???

\vspace{12pt}
\noindent
{\bf Acknowledgements}

The results discussed here have been obtained in collaboration with a large number of people: 
Igor Aranson,
Markus B\"ar,
Denis Bartolo,
Eric Bertin,
Li-bing Cai,
Jean-Baptiste Caussin,
Francesco Ginelli,
Guillaume Gr\'egoire,
Xin-chen Jiang,
Yu-qiang Ma,
Shradha Mishra,
Sandrine Ngo,
Aurelio Patelli,
Fernando Peruani,
Anton Peshkov,
Sriram Ramaswamy,
Franck Raynaud,
Xia-qing Shi,
Alexandre Solon,
\& Julien Tailleur.
All are warmly thanked!

\bibliographystyle{OUPnamed}
\bibliography{Biblio}

\thebibliography{0}

\bibitem[\protect\citeauthoryear{Barbaro and Degond}{Barbaro and
  Degond}{2014}]{barbaro2014phase}
Barbaro, Alethea~BT and Degond, Pierre (2014).
\newblock Phase transition and diffusion among socially interacting
  self-propelled agents.
\newblock {\em Discrete \& Continuous Dynamical Systems-B\/},~{\bf 19}(5),
  1249--1278.

\bibitem[\protect\citeauthoryear{Baskaran and Marchetti}{Baskaran and
  Marchetti}{2008{\em a}}]{baskaran2008enhanced}
Baskaran, Aparna and Marchetti, M~Cristina (2008{\em a}).
\newblock Enhanced diffusion and ordering of self-propelled rods.
\newblock {\em Physical Review Letters\/},~{\bf 101}(26), 268101.

\bibitem[\protect\citeauthoryear{Baskaran and Marchetti}{Baskaran and
  Marchetti}{2008{\em b}}]{baskaran2008hydrodynamics}
Baskaran, Aparna and Marchetti, M~Cristina (2008{\em b}).
\newblock Hydrodynamics of self-propelled hard rods.
\newblock {\em Physical Review E\/},~{\bf 77}(1), 011920.

\bibitem[\protect\citeauthoryear{Baskaran and Marchetti}{Baskaran and
  Marchetti}{2010}]{baskaran2010nonequilibrium}
Baskaran, Aparna and Marchetti, M~Cristina (2010).
\newblock Nonequilibrium statistical mechanics of self-propelled hard rods.
\newblock {\em Journal of Statistical Mechanics: Theory and Experiment\/}~(04),
  P04019.

\bibitem[\protect\citeauthoryear{Bechinger, Di~Leonardo, L{\"o}wen, Reichhardt,
  Volpe and Volpe}{Bechinger {\em et~al.}}{2016}]{bechinger2016active}
Bechinger, Clemens, Di~Leonardo, Roberto, L{\"o}wen, Hartmut, Reichhardt,
  Charles, Volpe, Giorgio, and Volpe, Giovanni (2016).
\newblock {Active Particles in Complex and Crowded Environments}.
\newblock {\em Reviews of Modern Physics\/},~{\bf 88}(4), 045006.

\bibitem[\protect\citeauthoryear{Bertin, Chat{\'e}, Ginelli, Mishra, Peshkov
  and Ramaswamy}{Bertin {\em et~al.}}{2013}]{bertin2013mesoscopic}
Bertin, Eric, Chat{\'e}, Hugues, Ginelli, Francesco, Mishra, Shradha, Peshkov,
  Anton, and Ramaswamy, Sriram (2013).
\newblock Mesoscopic theory for fluctuating active nematics.
\newblock {\em New Journal of Physics\/},~{\bf 15}(8), 085032.

\bibitem[\protect\citeauthoryear{Bertin, Droz and Gr{\'e}goire}{Bertin {\em
  et~al.}}{2006}]{bertin2006boltzmann}
Bertin, Eric, Droz, Michel, and Gr{\'e}goire, Guillaume (2006).
\newblock Boltzmann and hydrodynamic description for self-propelled particles.
\newblock {\em Physical Review E\/},~{\bf 74}(2), 022101.

\bibitem[\protect\citeauthoryear{Bertin, Droz and Gr{\'e}goire}{Bertin {\em
  et~al.}}{2009}]{bertin2009hydrodynamic}
Bertin, Eric, Droz, Michel, and Gr{\'e}goire, Guillaume (2009).
\newblock Hydrodynamic equations for self-propelled particles: microscopic
  derivation and stability analysis.
\newblock {\em Journal of Physics A: Mathematical and Theoretical\/},~{\bf
  42}(44), 445001.

\bibitem[\protect\citeauthoryear{Cai, Chat\'e, Ma and Shi}{Cai {\em
  et~al.}}{2019}]{cai2019dynamical}
Cai, Li-bing, Chat\'e, Hugues, Ma, Yu-qiang, and Shi, Xia-qing (2019, Jan).
\newblock Dynamical subclasses of dry active nematics.
\newblock {\em Physical Review E\/},~{\bf 99}, 010601.

\bibitem[\protect\citeauthoryear{Cates and Tailleur}{Cates and
  Tailleur}{2015}]{cates2015motility}
Cates, Michael~E and Tailleur, Julien (2015).
\newblock Motility-induced phase separation.
\newblock {\em Annu. Rev. Condens. Matter Phys.\/},~{\bf 6}(1), 219--244.

\bibitem[\protect\citeauthoryear{Chat{\'e}, Ginelli, Gr{\'e}goire and
  Raynaud}{Chat{\'e} {\em et~al.}}{2008}]{chate2008collective}
Chat{\'e}, Hugues, Ginelli, Francesco, Gr{\'e}goire, Guillaume, and Raynaud,
  Franck (2008).
\newblock Collective motion of self-propelled particles interacting without
  cohesion.
\newblock {\em Physical Review E\/},~{\bf 77}(4), 046113.

\bibitem[\protect\citeauthoryear{Chen, Patelli, Chat{\'e}, Ma and Shi}{Chen
  {\em et~al.}}{2017}]{chen2017fore}
Chen, Qiu-shi, Patelli, Aurelio, Chat{\'e}, Hugues, Ma, Yu-Qiang, and Shi,
  Xia-Qing (2017).
\newblock Fore-aft asymmetric flocking.
\newblock {\em Physical Review E\/},~{\bf 96}(2), 020601.

\bibitem[\protect\citeauthoryear{Czir{\'o}k, Stanley and Vicsek}{Czir{\'o}k
  {\em et~al.}}{1997}]{czirok1997spontaneously}
Czir{\'o}k, Andr{\'a}s, Stanley, H~Eugene, and Vicsek, Tam{\'a}s (1997).
\newblock Spontaneously ordered motion of self-propelled particles.
\newblock {\em Journal of Physics A: Mathematical and General\/},~{\bf 30}(5),
  1375.

\bibitem[\protect\citeauthoryear{Dean}{Dean}{1996}]{dean1996langevin}
Dean, David~S (1996).
\newblock Langevin equation for the density of a system of interacting
  {Langevin} processes.
\newblock {\em Journal of Physics A: Mathematical and General\/},~{\bf 29}(24),
  L613.

\bibitem[\protect\citeauthoryear{Degond and Motsch}{Degond and
  Motsch}{2008}]{degond2008continuum}
Degond, Pierre and Motsch, S{\'e}bastien (2008).
\newblock Continuum limit of self-driven particles with orientation
  interaction.
\newblock {\em Mathematical Models and Methods in Applied Sciences\/},~{\bf
  18}(supp01), 1193--1215.

\bibitem[\protect\citeauthoryear{Deseigne, Dauchot and Chat{\'e}}{Deseigne {\em
  et~al.}}{2010}]{deseigne2010collective}
Deseigne, Julien, Dauchot, Olivier, and Chat{\'e}, Hugues (2010).
\newblock Collective motion of vibrated polar disks.
\newblock {\em Physical Review Letters\/},~{\bf 105}(9), 098001.

\bibitem[\protect\citeauthoryear{Doostmohammadi, Ign{\'e}s-Mullol, Yeomans and
  Sagu{\'e}s}{Doostmohammadi {\em et~al.}}{2018}]{Doostmohammadi2018active}
Doostmohammadi, Amin, Ign{\'e}s-Mullol, Jordi, Yeomans, Julia~M., and
  Sagu{\'e}s, Francesc (2018).
\newblock Active nematics.
\newblock {\em Nature Communications\/},~{\bf 9}(1), 3246.

\bibitem[\protect\citeauthoryear{Farrell, Marchetti, Marenduzzo and
  Tailleur}{Farrell {\em et~al.}}{2012}]{farrell2012pattern}
Farrell, FDC, Marchetti, MC, Marenduzzo, D, and Tailleur, J (2012).
\newblock Pattern formation in self-propelled particles with density-dependent
  motility.
\newblock {\em Physical Review Letters\/},~{\bf 108}(24), 248101.

\bibitem[\protect\citeauthoryear{Ginelli}{Ginelli}{2016}]{ginelli2016physics}
Ginelli, Francesco (2016).
\newblock {The Physics of the Vicsek model}.
\newblock {\em The European Physical Journal Special Topics\/},~{\bf
  225}(11-12), 2099--2117.

\bibitem[\protect\citeauthoryear{Ginelli, Peruani, B{\"a}r and
  Chat{\'e}}{Ginelli {\em et~al.}}{2010}]{ginelli2010large}
Ginelli, Francesco, Peruani, Fernando, B{\"a}r, Markus, and Chat{\'e}, Hugues
  (2010).
\newblock Large-scale collective properties of self-propelled rods.
\newblock {\em Physical Review Letters\/},~{\bf 104}(18), 184502.

\bibitem[\protect\citeauthoryear{Gr{\'e}goire and Chat{\'e}}{Gr{\'e}goire and
  Chat{\'e}}{2004}]{gregoire2004onset}
Gr{\'e}goire, Guillaume and Chat{\'e}, Hugues (2004).
\newblock Onset of collective and cohesive motion.
\newblock {\em Physical Review Letters\/},~{\bf 92}(2), 025702.

\bibitem[\protect\citeauthoryear{Gr{\'e}goire, Chat{\'e} and Tu}{Gr{\'e}goire
  {\em et~al.}}{2003}]{gregoire2003moving}
Gr{\'e}goire, Guillaume, Chat{\'e}, Hugues, and Tu, Yuhai (2003).
\newblock Moving and staying together without a leader.
\newblock {\em Physica D: Nonlinear Phenomena\/},~{\bf 181}(3), 157 -- 170.

\bibitem[\protect\citeauthoryear{Gro{\ss}mann, Peruani and
  B{\"a}r}{Gro{\ss}mann {\em et~al.}}{2016}]{grossmann2016mesoscale}
Gro{\ss}mann, Robert, Peruani, Fernando, and B{\"a}r, Markus (2016).
\newblock Mesoscale pattern formation of self-propelled rods with velocity
  reversal.
\newblock {\em Physical Review E\/},~{\bf 94}(5), 050602.

\bibitem[\protect\citeauthoryear{Grossmann, Schimansky-Geier and
  Romanczuk}{Grossmann {\em et~al.}}{2012}]{grossmann2012active}
Grossmann, Robert, Schimansky-Geier, Lutz, and Romanczuk, Pawel (2012).
\newblock {Active Brownian} particles with velocity-alignment and active
  fluctuations.
\newblock {\em New Journal of Physics\/},~{\bf 14}(7), 073033.

\bibitem[\protect\citeauthoryear{Grossmann, Schimansky-Geier and
  Romanczuk}{Grossmann {\em et~al.}}{2013}]{grossmann2013self}
Grossmann, Robert, Schimansky-Geier, Lutz, and Romanczuk, Pawel (2013).
\newblock Self-propelled particles with selective attraction--repulsion
  interaction: from microscopic dynamics to coarse-grained theories.
\newblock {\em New Journal of Physics\/},~{\bf 15}(8), 085014.

\bibitem[\protect\citeauthoryear{Kourbane-Houssene, Erignoux, Bodineau and
  Tailleur}{Kourbane-Houssene {\em et~al.}}{2018}]{kourbane2018exact}
Kourbane-Houssene, Mourtaza, Erignoux, Cl{\'e}ment, Bodineau, Thierry, and
  Tailleur, Julien (2018).
\newblock {Exact Hydrodynamic Description of Active Lattice Gases}.
\newblock {\em arXiv preprint arXiv:1801.08952\/}.

\bibitem[\protect\citeauthoryear{Mahault, Patelli and Chat{\'e}}{Mahault {\em
  et~al.}}{2018}]{3D_hydro}
Mahault, Beno{\^\i}t, Patelli, Aurelio, and Chat{\'e}, Hugues (2018).
\newblock Deriving hydrodynamic equations from dry active matter models in
  three dimensions.
\newblock {\em Journal of Statistical Mechanics: Theory and Experiment\/},~{\bf
  2018}(9), 093202.

\bibitem[\protect\citeauthoryear{Mahault, Patelli, Shi and Chat{\'e}}{Mahault
  {\em et~al.}}{2019}]{Kinetic_paper}
Mahault, Beno{\^\i}t, Patelli, Aurelio, Shi, Xiaqing, and Chat{\'e}, Hugues
  (2019).
\newblock {\em {On the Faithfulness of the Boltzmann Equation for Dry,
  Aligning, Dilute Active Matter. To be published}}.

\bibitem[\protect\citeauthoryear{Marchetti, Joanny, Ramaswamy, Liverpool,
  Prost, Rao and Simha}{Marchetti {\em
  et~al.}}{2013}]{marchetti2013hydrodynamics}
Marchetti, M~Cristina, Joanny, Jean-Fran{\c{c}}ois, Ramaswamy, Sriram,
  Liverpool, Tanniemola~B, Prost, Jacques, Rao, Madan, and Simha, R~Aditi
  (2013).
\newblock Hydrodynamics of soft active matter.
\newblock {\em Reviews of Modern Physics\/},~{\bf 85}(3), 1143.

\bibitem[\protect\citeauthoryear{Mishra, Baskaran and Marchetti}{Mishra {\em
  et~al.}}{2010}]{mishra2010fluctuations}
Mishra, Shradha, Baskaran, Aparna, and Marchetti, M~Cristina (2010).
\newblock Fluctuations and pattern formation in self-propelled particles.
\newblock {\em Physical Review E\/},~{\bf 81}(6), 061916.

\bibitem[\protect\citeauthoryear{Nagai, Sumino, Montagne, Aranson and
  Chat{\'e}}{Nagai {\em et~al.}}{2015}]{nagai2015collective}
Nagai, Ken~H, Sumino, Yutaka, Montagne, Raul, Aranson, Igor~S, and Chat{\'e},
  Hugues (2015).
\newblock Collective motion of self-propelled particles with memory.
\newblock {\em Physical Review Letters\/},~{\bf 114}(16), 168001.

\bibitem[\protect\citeauthoryear{Ngo, Peshkov, Aranson, Bertin, Ginelli and
  Chat{\'e}}{Ngo {\em et~al.}}{2014}]{ngo2014large}
Ngo, Sandrine, Peshkov, Anton, Aranson, Igor~S, Bertin, Eric, Ginelli,
  Francesco, and Chat{\'e}, Hugues (2014).
\newblock Large-scale chaos and fluctuations in active nematics.
\newblock {\em Physical Review Letters\/},~{\bf 113}(3), 038302.

\bibitem[\protect\citeauthoryear{{Patelli}, {Djafer-Cherif}, {Aranson},
  {Bertin} and {Chat{\'e}}}{{Patelli} {\em et~al.}}{2019}]{dense-active-nema}
{Patelli}, Aurelio, {Djafer-Cherif}, Ilyas, {Aranson}, Igor~S., {Bertin}, Eric,
  and {Chat{\'e}}, Hugues (2019, Apr).
\newblock {Understanding dense active nematics from microscopic models}.
\newblock {\em arXiv e-prints\/}, arXiv:1904.12708.

\bibitem[\protect\citeauthoryear{Patelli, Djafer-Cherif, S.~Aranson, Bertin and
  Chat{\'e}}{Patelli {\em et~al.}}{2019}]{nema_repulsion}
Patelli, Aurelio, Djafer-Cherif, Ilyas, S.~Aranson, Igor, Bertin, Eric, and
  Chat{\'e}, Hugues (2019).
\newblock {Understanding active nematics from microscopic models. To be
  published}.

\bibitem[\protect\citeauthoryear{Peshkov}{Peshkov}{2013}]{peshkov2013boltzmann}
Peshkov, Anton (2013).
\newblock {\em {Boltzmann-Ginzburg-Landau} approach to simple models of active
  matter}.
\newblock Ph.D. thesis, Universit{\'e} Pierre et Marie Curie-Paris VI.

\bibitem[\protect\citeauthoryear{Peshkov, Aranson, Bertin, Chat{\'e} and
  Ginelli}{Peshkov {\em et~al.}}{2012}]{peshkov2012nonlinear}
Peshkov, Anton, Aranson, Igor~S, Bertin, Eric, Chat{\'e}, Hugues, and Ginelli,
  Francesco (2012).
\newblock Nonlinear field equations for aligning self-propelled rods.
\newblock {\em Physical Review Letters\/},~{\bf 109}(26), 268701.

\bibitem[\protect\citeauthoryear{Peshkov, Bertin, Ginelli and
  Chat{\'e}}{Peshkov {\em et~al.}}{2014}]{peshkov2014boltzmann}
Peshkov, A, Bertin, E, Ginelli, F, and Chat{\'e}, H (2014).
\newblock {Boltzmann}-{Ginzburg}-{Landau} approach for continuous descriptions
  of generic {Vicsek}-like models.
\newblock {\em The European Physical Journal. Special topics\/},~{\bf 223}(7),
  1315--1344.

\bibitem[\protect\citeauthoryear{Putzig, Redner, Baskaran and Baskaran}{Putzig
  {\em et~al.}}{2016}]{putzig2016instabilities}
Putzig, Elias, Redner, Gabriel~S., Baskaran, Arvind, and Baskaran, Aparna
  (2016).
\newblock Instabilities, defects, and defect ordering in an overdamped active
  nematic.
\newblock {\em Soft Matter\/},~{\bf 12}(17), 3854--3859.

\bibitem[\protect\citeauthoryear{Ramaswamy}{Ramaswamy}{2010}]{ramaswamy2010mechanics}
Ramaswamy, Sriram (2010).
\newblock {The Mechanics and Statistics of Active Matter}.
\newblock {\em Annual Review of Condensed Matter Physics\/},~{\bf 1}(1),
  323--345.

\bibitem[\protect\citeauthoryear{Ramaswamy, Simha and Toner}{Ramaswamy {\em
  et~al.}}{2003}]{ramaswamy2003active}
Ramaswamy, Sriram, Simha, R~Aditi, and Toner, John (2003).
\newblock Active nematics on a substrate: Giant number fluctuations and
  long-time tails.
\newblock {\em EPL (Europhysics Letters)\/},~{\bf 62}(2), 196.

\bibitem[\protect\citeauthoryear{Romanczuk, Chat{\'{e}}, Chen, Ngo and
  Toner}{Romanczuk {\em et~al.}}{2016}]{romanczuk2016emergent}
Romanczuk, Pawel, Chat{\'{e}}, Hugues, Chen, Leiming, Ngo, Sandrine, and Toner,
  John (2016, jun).
\newblock Emergent smectic order in simple active particle models.
\newblock {\em New Journal of Physics\/},~{\bf 18}(6), 063015.

\bibitem[\protect\citeauthoryear{Romanczuk and Schimansky-Geier}{Romanczuk and
  Schimansky-Geier}{2012}]{romanczuk2012mean}
Romanczuk, Pawel and Schimansky-Geier, Lutz (2012).
\newblock Mean-field theory of collective motion due to velocity alignment.
\newblock {\em Ecological Complexity\/},~{\bf 10}, 83--92.

\bibitem[\protect\citeauthoryear{Shankar, Ramaswamy and Marchetti}{Shankar {\em
  et~al.}}{2018}]{shankar2018low}
Shankar, Suraj, Ramaswamy, Sriram, and Marchetti, M~Cristina (2018).
\newblock Low-noise phase of a two-dimensional active nematic system.
\newblock {\em Physical Review E\/},~{\bf 97}(1), 012707.

\bibitem[\protect\citeauthoryear{Shi and Chat{\'e}}{Shi and
  Chat{\'e}}{2018}]{shi2018self}
Shi, Xia-qing and Chat{\'e}, Hugues (2018).
\newblock {Self-Propelled Rods: Linking Alignment-Dominated and
  Repulsion-Dominated Active Matter}.
\newblock {\em arXiv preprint arXiv:1807.00294\/}.

\bibitem[\protect\citeauthoryear{Solon and Tailleur}{Solon and
  Tailleur}{2013}]{solon2013revisiting}
Solon, AP and Tailleur, Julien (2013).
\newblock Revisiting the flocking transition using active spins.
\newblock {\em Physical Review Letters\/},~{\bf 111}(7), 078101.

\bibitem[\protect\citeauthoryear{Solon, Caussin, Bartolo, Chat{\'e} and
  Tailleur}{Solon {\em et~al.}}{2015{\em a}}]{solon2015pattern}
Solon, Alexandre~P, Caussin, Jean-Baptiste, Bartolo, Denis, Chat{\'e}, Hugues,
  and Tailleur, Julien (2015{\em a}).
\newblock Pattern formation in flocking models: {A} hydrodynamic description.
\newblock {\em Physical Review E\/},~{\bf 92}(6), 062111.

\bibitem[\protect\citeauthoryear{Solon, Chat{\'e} and Tailleur}{Solon {\em
  et~al.}}{2015{\em b}}]{solon2015phase}
Solon, Alexandre~P, Chat{\'e}, Hugues, and Tailleur, Julien (2015{\em b}).
\newblock From phase to microphase separation in flocking models: The essential
  role of nonequilibrium fluctuations.
\newblock {\em Physical Review Letters\/},~{\bf 114}(6), 068101.

\bibitem[\protect\citeauthoryear{Solon and Tailleur}{Solon and
  Tailleur}{2015}]{solon2015flocking}
Solon, Alexandre~P and Tailleur, Julien (2015).
\newblock {Flocking with discrete symmetry: The two-dimensional active Ising
  model}.
\newblock {\em Physical Review E\/},~{\bf 92}(4), 042119.

\bibitem[\protect\citeauthoryear{Toner}{Toner}{2012}]{toner2012reanalysis}
Toner, John (2012).
\newblock Reanalysis of the hydrodynamic theory of fluid, polar-ordered flocks.
\newblock {\em Physical Review E\/},~{\bf 86}(3), 031918.

\bibitem[\protect\citeauthoryear{Toner and Tu}{Toner and
  Tu}{1995}]{toner1995long}
Toner, John and Tu, Yuhai (1995).
\newblock Long-range order in a two-dimensional dynamical {XY} model: how birds
  fly together.
\newblock {\em Physical Review Letters\/},~{\bf 75}(23), 4326.

\bibitem[\protect\citeauthoryear{Toner and Tu}{Toner and
  Tu}{1998}]{toner1998flocks}
Toner, John and Tu, Yuhai (1998).
\newblock {Flocks, herds, and schools: A quantitative theory of flocking}.
\newblock {\em Physical Review E\/},~{\bf 58}(4), 4828.

\bibitem[\protect\citeauthoryear{Vicsek, Czir{\'o}k, Ben-Jacob, Cohen and
  Shochet}{Vicsek {\em et~al.}}{1995}]{vicsek1995novel}
Vicsek, Tam{\'a}s, Czir{\'o}k, Andr{\'a}s, Ben-Jacob, Eshel, Cohen, Inon, and
  Shochet, Ofer (1995).
\newblock Novel type of phase transition in a system of self-driven particles.
\newblock {\em Physical Review Letters\/},~{\bf 75}(6), 1226.

\endthebibliography

\end{document}